\documentclass[aps,prd,amsmath,floats,floatfix,twocolumn,
  superscriptaddress,nofootinbib,showpacs]{revtex4-1}

\pdfoutput=1 
\usepackage{diagbox}
\usepackage{multirow}
\usepackage{braket}
\usepackage{amsfonts}
\usepackage{amsmath}
\usepackage{amssymb}
\usepackage{mathtools}
\usepackage{amsthm}
\usepackage{bm}
\usepackage{dcolumn}
\usepackage{epsfig}
\usepackage{commath}
\usepackage{graphicx}
\usepackage{graphics}
\usepackage[latin1]{inputenc}
\usepackage{latexsym}
\usepackage{rotating}
\usepackage[dvipsnames]{xcolor}
\usepackage{mathrsfs}
\usepackage{microtype}
\usepackage{verbatim}
\usepackage{url}

\usepackage[caption=false]{subfig}
\usepackage[normalem]{ulem}
\usepackage{tikz}

\usepackage[breaklinks=true]{hyperref}
\hypersetup{
    colorlinks=true,
    linkcolor=NavyBlue,
    filecolor=Magenta,
    urlcolor=NavyBlue,
    citecolor=NavyBlue
}

\usepackage{yfonts}
\usepackage{float}
\usepackage{xspace} 
\usepackage{mathrsfs}
\usepackage[toc,page]{appendix}
\usepackage{siunitx}
\usepackage{array}
\usepackage[normalem]{ulem}

\newcommand{\Hertz}{{\mbox{\tiny H}}}
\newcommand{\Kerr}{{\mbox{\tiny Kerr}}}

\newcommand{\slow}{{\mbox{\tiny slow}}}
\newcommand{\be}{\begin{equation}}
\newcommand{\ee}{\end{equation}}

\newcommand{\RW}{{\mbox{\tiny RW}}}
\newcommand{\metric}{{h}}
\newcommand{\RWmod}{{\mbox{\tiny RWmod}}}

\newcommand{\ZM}{{\mbox{\tiny ZM}}}
\newcommand{\ZMmod}{{\mbox{\tiny ZMmod}}}

\newcommand{\geo}{{\mbox{\tiny geo}}}

\newcommand{\ba}{\begin{align}}
\newcommand{\ea}{\end{align}}

\makeatletter
\newcommand*{\rom}[1]{\expandafter\@slowromancap\romannumeral #1@}
\makeatother
\allowdisplaybreaks

\AtBeginDocument{%
    \newwrite\bibnotes
    \def\bibnotesext{Notes.bib}
    \immediate\openout\bibnotes=\jobname\bibnotesext
    \immediate\write\bibnotes{@CONTROL{REVTEX41Control}}
    \immediate\write\bibnotes{@CONTROL{%
    apsrev41Control,author="08",editor="1",pages="1",title="0",year="1"}}
    \if@filesw
    \immediate\write\@auxout{\string\citation{apsrev41Control}}%
    \fi
}

\newcommand{\UIUC}{Illinois  Center  for  Advanced  Studies  of  the  Universe \&
Department of Physics, University of Illinois at Urbana-Champaign, Urbana, Illinois 61801, USA}
\newcommand{\CAL}{Theoretical Astrophysics 350-17, California Institute of Technology, Pasadena, CA 91125, USA}
\newcommand{\AEI}{Max Planck Institute for Gravitational Physics (Albert Einstein Institute), D-14476 Potsdam, Germany}

\begin{document}

    \title{Perturbations of spinning black holes in dynamical Chern-Simons gravity: \\Slow rotation quasinormal modes}

    \author{Dongjun Li}
    \email{dongjun@illinois.edu}
    \affiliation{\CAL}
    \affiliation{\UIUC}
    
    \author{Pratik Wagle}
    \email{pratik.wagle@aei.mpg.de}
    \affiliation{\AEI}

    \author{Yanbei Chen}
    \affiliation{\CAL}
    
    \author{Nicol\'as Yunes}
    \affiliation{\UIUC}

    \date{\today}
    \begin{abstract}
    Gravitational waves offer new ways to test general relativity (GR) in the strong-field regime, including tests involving the ringdown phase of binary black hole mergers, characterized by oscillating and quickly decaying quasinormal modes (QNMs). Recent advances have extended QNM calculations to several theories beyond GR through the development of the modified Teukolsky formalism, including higher derivative gravity and dynamical Chern-Simons (dCS) gravity. Using the modified Teukolsky formalism, we previously derived radial second-order differential equations governing curvature and scalar field perturbations in dCS gravity at leading order in spin. In this work, we compute the QNM frequency shifts for slowly rotating black holes in dCS gravity from these modified Teukolsky equations, and (1) show that the radial equations for Weyl scalars $\Psi_{0,4}$ can be separated into even- and odd-parity parts, confirming that the scalar field couples only to the odd-parity sector; (2) extend the eigenvalue perturbation method to coupled fields; (3) compute the QNM spectrum, obtaining consistent results across independent calculations using different radiation gauges; (4) calculate the overtones in the QNM spectra for the first time in dCS gravity; (5) show that our findings align with previous metric perturbation studies and mark the first QNM spectrum calculation in a non-minimally coupled scalar-tensor theory via the modified Teukolsky formalism. This work lays the foundation for studying fast-rotating black holes in dCS gravity, advancing black hole spectroscopy in beyond-GR contexts.
    \end{abstract}
    \maketitle

\section{Introduction}
\label{sec:introduction}

The detection of gravitational waves (GWs) emitted by over a hundred binary black hole (BH) mergers \cite{KAGRA:2021vkt} has opened up a new avenue to study gravity in the simultaneously strong, non-linear, and highly dynamical regime. Although Einstein's general relativity (GR) has passed numerous tests in the Solar System \cite{Will:2014kxa}, it is still undergoing tests in strong gravity, especially using GW observations. The waveform of the GW sourced by a binary black hole (BBH) coalescence is comprised of three distinct stages: inspiral, merger, and ringdown. Each of these stages can be used separately or in combination to test Einstein's theory of GR. One such important test is to look at the ringdown part of GW signals, where GWs are dominated by quickly decaying and oscillating quasinormal modes (QNMs), as the remnant BH settles down~\cite{Maggiore:2018sht, Vishveshwara:1970cc, Vishveshwara:1970zz}. In GR, QNMs encode the mass, spin, and charge of the remnant BH, fully characterizing the BH spacetime, as dictated by the ``no-hair'' theorems~\cite{Carter:1971zc, Robinson:1975bv}, whose assumptions have now been tested with observational data from the LIGO-Virgo-KAGRA collaboration~\cite{Cardoso:2016ryw, Isi:2019aib, Islam:2021pbd, Capano:2021etf, Isi:2021iql}. In beyond Einstein theories, however, QNMs can also depend on the coupling constants of the theory, as these generically modify the black hole background and the dynamical perturbation equations. GW observations of the ringdown stage can therefore be used to detect or constrain such beyond-Einstein effects if we develop a more detailed understanding of QNMs in such theories. 

Why would we even consider beyond Einstein theories when GR has passed all tests with flying colors~\cite{Will:2014kxa}? Einstein's theory faces several theoretical and observational anomalies, as it is incompatible with quantum mechanics and struggles to explain certain observational puzzles, such as the asymmetry of matter and antimatter in our universe \cite{Canetti:2012zc}, the late-time acceleration of the universe~\cite{Perlmutter:1998np,Riess:1998cb}, and the anomalous galaxy rotation curves~\cite{Sofue:2000jx,Bertone:2016nfn}. For these reasons, modifications to GR are introduced either by constructing a unified theory for quantum gravity, such as string theory \cite{Schwarz:1982jn, Damour:1994zq, Banks:1996vh, Seiberg:1999vs, Aharony:1999ti, Mukhi:2011zz} and loop quantum gravity \cite{Birrell:1982ix, Ashtekar:1987gu, Rovelli:1989za, Rovelli:1997yv, Ashtekar:1997yu, Ashtekar:2004eh}, or by explaining certain observational puzzles through the inclusion of additional scalar, vector, or tensor fields or considering higher-order terms in the Einstein-Hilbert action. Such modifications to the action usually lead to modifications in the background BH spacetime, such as additional scalar or vector hair \cite{Kanti:1995qe, Kanti:1995vq, Jacobson:2007veq, Horava:2009uw, Yunes:2009hc, Yunes:2011we, Yagi:2012ya}. Since QNMs encode information about the properties of the remnant BH, studying QNM spectra in beyond-GR theories allows us to extract information about these additional hairs and their strong-field dynamics, as well as the length scale or the coupling constant associated with these beyond-GR corrections, which is the core idea behind BH spectroscopy \cite{Dreyer:2003bv, Berti:2005ys, Berti:2018vdi}. 

To compute the QNM spectrum of a non-rotating BH in GR, Regge and Wheeler~\cite{Regge:1957td} first presented a formalism using metric perturbations that are separated into even and odd parity. Focusing on odd-parity perturbations, they constructed a master function, known today as the Regge-Wheeler (RW) function, to obtain a second-order radial differential equation that describes the evolution of these perturbations in GR. Later, Zerilli and Moncrief extended this approach to study even-parity perturbations of a non-rotating BH in GR \cite{Zerilli:1971wd, Moncrief:1974am}. The QNM spectrum of a non-rotating BH in GR was initially computed in \cite{Chandrasekhar:1975zza} by numerical integration and can be more systematically calculated using the continuous fraction method developed by Leaver in \cite{Leaver:1985ax}. 

For rotating BHs in GR, directly solving for metric perturbations is much more challenging due to the lack of spherical symmetry of the background. Spinning BH backgrounds are axially symmetric, and in this case, one cannot easily decouple the even- and odd-parity metric perturbation modes to find two master functions characterizing all the metric components, and reduce their equations of motion into purely-radial, ordinary differential equations. Building upon the Newman-Penrose (NP) formalism \cite{Newman:1961qr}, Teukolsky found a solution to this problem in \cite{Teukolsky:1973ha, Press:1973zz, Teukolsky:1974yv} by instead solving for curvature perturbations represented by two Weyl scalars, $\Psi_0$ and $\Psi_4$. In this case, Teukolsky was able to reduce the NP equations describing the evolution of $\Psi_{0,4}$ into two decoupled and separable, partial differential equations. The solutions to the angular equations are called spin-weighted spheroidal harmonics, and the radial equations are generalized, spheroidal wave equations \cite{Leaver:1985ax}. Both the angular and radial parts of $\Psi_{0,4}$, as well as their QNM frequencies, can be calculated using Leaver's method in \cite{Leaver:1985ax} and extensions of it, such as the Mano-Suzuki-Takasugi method \cite{Mano:1996gn, Mano:1996vt, Fujita:2004rb, Fujita:2005kng}.

Many efforts have been made over the past ten years to study QNMs in beyond-GR theories. For non-rotating BHs, one can still apply the standard metric perturbation approach developed by \cite{Regge:1957td, Zerilli:1971wd, Moncrief:1974am}. Beyond-GR QNMs in the non-rotating case were computed in, for example, dynamical Chern-Simons (dCS) gravity \cite{Cardoso:2009pk, Molina:2010fb, Pani:2011xj}, Einstein-dilaton-Gauss-Bonnet (EdGB) theory \cite{Pani:2009wy, Blazquez-Salcedo:2016enn, Blazquez-Salcedo:2017txk}, higher-derivative gravity without extra fields \cite{Cardoso:2018ptl, deRham:2020ejn, Cardoso:2019mqo}, and Einstein-Aether theory \cite{Konoplya:2006rv, Konoplya:2006ar}. For rotating BHs, the calculations of QNMs are much more challenging, and thus, the field was stuck for many years. Up until very recently, there were only a few, very special examples of calculations that could be done, all of which relied heavily on the slow-rotation expansion, such as in dCS gravity \cite{Wagle:2021tam, Srivastava:2021imr}, EdGB gravity \cite{Pierini:2021jxd, Pierini:2022eim}, and higher-derivative gravity \cite{Cano:2020cao, Cano:2021myl}. Although QNMs for a rotating BH with a general spin could be extracted from full numerical relativity simulations, for example, in dCS gravity \cite{Okounkova:2019dfo, Okounkova:2019zjf}, this approach has several issues, such as the emergence of secularly growing terms and other numerical difficulties in the numerical extraction of small QNM shifts.

\begin{figure}[t]
    \centering
    \includegraphics[width=0.95\linewidth]{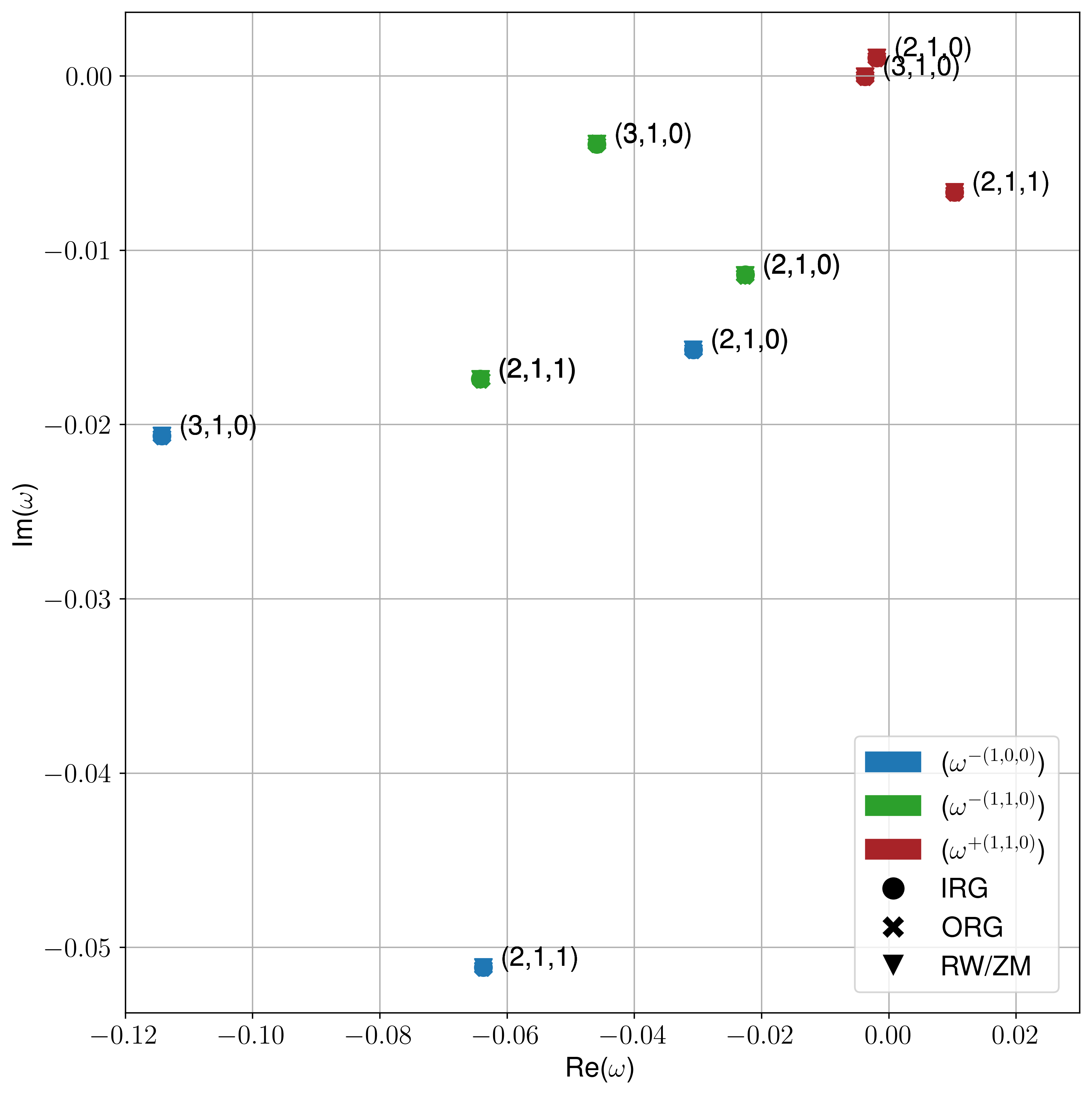}
    \caption{The QNM frequency shifts for three, selected, dominant, even- and odd-parity modes of a slowly rotating BH in dCS gravity up to $\mathcal{O}(\chi^1)$ for $\{\ell,m,n\}=\{(2,1,0),(2,1,1),(3,1,0)\}$ modes, as labeled at each point. Here we plot $\omega_{\ell 1}^{-(1,0,0)}$ (the frequency shift due to the dCS modification at $\mathcal{O}(\chi^0)$ order for odd-parity modes) in blue, while $\omega_{\ell 1}^{-(1,1,0)}$ (the frequency shift due to the dCS modification at $\mathcal{O}(\chi^1)$ for odd-parity modes) is plotted in in green, and $\omega_{\ell 1}^{+(1,1,0)}$ (the frequency shift due to the dCS modification at $\mathcal{O}(\chi^0)$ for even-parity modes) is presented in red, all as computed in Secs.~\ref{sec:QNM_nonrotating}, \ref{sec:QNM_slow_rotation_odd}, and \ref{sec:QNM_slow_rotation_even} with their values listed in Tables~\ref{tab:QNM_results_nonrotating}, \ref{tab:QNM_results_rotating_odd}, and \ref{tab:QNM_results_rotating_even}, respectively. The ``IRG'' marker labels the results computed from the equations of $\Psi_0^{(1,1)}$ [Eq.~\eqref{eq:master_eqn_Psi0_radial_simplify}] and  $\vartheta^{(1,1)}$ [Eq.~\eqref{eq:scalar_radial_IRG_2_lm}] in the IRG, while the ``ORG'' marker labels results computed from the equations for $\Psi_4^{(1,1)}$ [Eq.~\eqref{eq:master_eqn_Psi4_radial_simplify}] and $\vartheta^{(1,1)}$ [Eq.~\eqref{eq:scalar_radial_ORG_lm}] in the ORG, and ``RW/ZM'' labels the results computed from the RW equation [Eq.~\eqref{eq:RW_expand}] and the scalar field equation [Eq.~\eqref{eq:SF_expand}] for the odd-parity modes, or the ZM equation [Eq.~\eqref{eq:ZM_expand}] for the even-parity modes in the RW gauge, found by \cite{Wagle:2021tam, Srivastava:2021imr}. We have set $M=1/2$ in this plot.}
    \label{fig:summary_scatter}
\end{figure}

Recently, an important extension of the Teukolsky formalism was developed in~\cite{Li:2022pcy, Hussain:2022ins}, which now allows us to tackle a wide class of beyond-GR theories admitting an effective-field-theory description. This \emph{modified Teukolsky formalism} allows, for the first time, a systematic semi-analytical calculation of QNMs for BHs with a general spin, including fast-rotating ones, in beyond-GR theories. Besides the modified Teukolsky formalism, significant progress has also been recently made in applying spectral expansions to solve the linearized Einstein equations directly in both GR \cite{Chung:2023zdq, Chung:2023wkd} and beyond-GR theories \cite{Chung:2024ira, Chung:2024vaf, Blazquez-Salcedo:2024oek} in a framework dubbed \emph{METRICS}. As most of the remnant BHs of binary mergers are fast rotating \cite{Buonanno:2007sv, LIGOScientific:2021djp}, applying the modified Teukolsky formalism or the METRICS approach to specific beyond-GR theories is crucial for conducting BH spectroscopy.

The first attempt to do so was made by \cite{Cano:2023tmv, Cano:2023jbk, Cano:2024ezp}, which focused on computing the QNM frequency shifts in certain higher-derivative gravity that have no additional scalar or vector fields beyond the usual metric tensor. These works considered both the parity-preserving and parity-violating corrections to the Einstein-Hilbert action up to eight derivatives of the metric. Although the modified Teukolsky formalism works for BHs with a general spin, Refs.~\cite{Cano:2023tmv, Cano:2023jbk, Cano:2024ezp} performed a slow-rotation expansion up to $\mathcal{O}(\chi^{18})$, where $\chi=a/M$ is the dimensionless spin, since the background BH metric is known analytically only in the slow-rotation expansion \cite{Cano:2019ore}. For six-derivative gravity, their results for the fundamental modes are accurate for spins  $a\lesssim 0.7M$, with relative errors $\lesssim 1\%$ for most of the $(\ell,m)$ modes, and the errors mostly come from the slow-rotation expansion at high spins. Another attempt was recently made by \cite{Wagle:2023fwl} on a more complicated case, dCS gravity, where a pseudoscalar field $\vartheta$ is coupled to a quadratic term in the Riemann tensor and its dual, the so-called Pontryagin density. In \cite{Wagle:2023fwl}, we successfully implemented the Chrzanowski-Cohen-Kegeles (CCK) \cite{Cohen_Kegeles_1975, Chrzanowski:1975wv, Kegeles:1979an} procedure for metric reconstruction to compute all the source terms in the modified Teukolsky equation derived in \cite{Li:2022pcy}. After a projection to spin-weighted spheroidal harmonics, we obtained two sets of radial, ordinary differential equations in two different gauges, the ingoing radiation gauge (IRG) and the outgoing radiation gauge (ORG), for the radial part of the dCS correction to $\Psi_{0,4}$. For each set, we got an equation for the curvature perturbations ($\Psi_0$ in the IRG and $\Psi_4$ in the ORG) and another equation for the (pseudo)scalar field perturbations. In \cite{Wagle:2023fwl}, we performed a slow-rotation expansion up to $\mathcal{O}(\chi)$ since we wanted to compare our results to those in \cite{Cardoso:2009pk, Molina:2010fb, Pani:2011xj, Wagle:2021tam, Srivastava:2021imr}. In this work, we numerically solve the radial equations obtained in \cite{Wagle:2023fwl} to find the QNM frequency shifts for a slowly-rotating BH up to $\mathcal{O}(\chi)$ in dCS gravity. 

\subsection*{Executive Summary}
To calculate the QNM frequency shifts, we first apply the analysis of isospectrality breaking in \cite{Li:2023ulk} to simplify the radial master equations in \cite{Wagle:2023fwl}. The simplified equations share a similar structure to the RW/ZM equations and the scalar field equation found by \cite{Wagle:2021tam, Srivastava:2021imr}. Specifically, the even- and odd-parity modes are still solutions to the modified Teukolsky equation, while the scalar field perturbation only couples to the odd-parity modes. For each $(\ell, m)$, we use the ansatz for $\Psi_{0,4}$ and $\vartheta$ in Eqs.~\eqref{eq:solution_ansatz} and \eqref{eq:scalar_separation}. Then, for each $(\ell, m)$ mode, after solving the scalar field perturbation from Eq.~\eqref{eq:scalar_radial_IRG_2_lm} in the IRG or Eq.~\eqref{eq:scalar_radial_ORG_lm} in the ORG, we only need to solve one equation for the Weyl scalar perturbations, given by Eq.~\eqref{eq:master_eqn_Psi0_radial_simplify} for $\Psi_0$ in the IRG or Eq.~\eqref{eq:master_eqn_Psi4_radial_simplify} for $\Psi_4$ in the ORG.

To extract the QNM frequency shifts from these equations, we apply the eigenvalue perturbation (EVP) method of \cite{Zimmerman:2014aha, Mark:2014aja, Hussain:2022ins} and extend it, for the first time, to a set of coupled, gravitational and scalar field perturbation equations, with the validity of this extension formally shown in \cite{Dong:2025abc}. We compute the QNM frequency shifts of both even- and odd-parity prograde modes at $\ell=2,3,4$ and overtone numbers $n=0,1,2$ using both the modified Teukolsky equations of $\Psi_0$ in the IRG and of $\Psi_4$ in the ORG, respectively. The QNM frequency shifts obtained from these two independent sets of equations agree well, with relative differences $\lesssim10^{-4}$ at $\ell=2,n=0$ for all the corrections up to $\mathcal{O}(\chi^1)$ (displayed in Figs.~\ref{fig:even_scatter} and \ref{fig:odd_scatter}), verifying the self-consistency of our approach based on the modified Teukolsky formalism.

We further apply the EVP method to the modified RW/ZM equations previously obtained in \cite{Wagle:2023fwl, Srivastava:2021imr} and find that these results agree well with our modified Teukolsky formalism results, with relative differences between our IRG results and the RW/ZM results $\lesssim10^{-4}$ at $\ell=2,n=0$ for all the corrections up to $\mathcal{O}(\chi^1)$ (displayed in Figs.~\ref{fig:even_scatter} and \ref{fig:odd_scatter}). This comparison makes the first demonstration of the consistency across the curvature and the metric perturbation approaches for slowly-rotating BHs in dCS gravity up to $\mathcal{O}(\chi^1)$. Using our knowledge in GR, we estimate that these slow-rotation results using the EVP method should be valid for spins $a\lesssim0.22M$ (see Fig.~\ref{fig:slowroterror}).

For the first time, we successfully obtain the QNM frequency shifts of overtones in dCS gravity. By varying the overtone number $n$, we observe novel changes of dCS QNM frequencies with respect to $n$: (i) for non-rotating odd-parity QNM frequency shifts, both their real and imaginary parts decrease with $n$ for all $\ell$ modes we study, and they remain negative (see Table~\ref{tab:QNM_results_nonrotating} and Fig.~\ref{fig:odd_scatter}); (ii) for slowly-rotating, odd-parity shifts, their real part remains negative and decreases with $n$, while the imaginary part generally increases with $n$ (see Table~\ref{tab:QNM_results_rotating_odd} and Fig.~\ref{fig:odd_scatter}); (iii) for slowly-rotating, even-parity shifts, the real and the imaginary part of them increases and decreases with $n$, respectively (see Table~\ref{tab:QNM_results_rotating_even} and Fig.~\ref{fig:even_scatter}). All these features generally differ from their GR counterparts, resulting in a rich behavior of QNM frequencies in dCS gravity when one varies the spin and the coupling constant, as displayed in Figs.~\ref{fig:even_overtone} and \ref{fig:odd_overtone}. 

The remainder of this paper presents the details of the calculations described above and it is organized as follows. In Sec.~\ref{sec:review_master_eqn}, we briefly review the procedures for computing the modified Teukolsky equation in dCS gravity in \cite{Wagle:2023fwl}. In Secs.~\ref{sec:scalar_eqn} and \ref{sec:modified_teuk_eqn}, we apply the procedures in \cite{Li:2023ulk} to simplify the master equations for the scalar field $\vartheta$ and for the Weyl scalars $\Psi_{0,4}$, respectively. We also discuss the parity features of these equations and their implications on the structure of isospectrality breaking. In Sec.~\ref{sec:QNM_freq}, we review the EVP method in \cite{Zimmerman:2014aha, Mark:2014aja, Hussain:2022ins}, show how to apply it to the specific case of dCS gravity when an extra scalar field is present, and then compute the QNM frequency shifts. We present our numerical results for the QNM frequency shifts in Sec.~\ref{sec:QNM_freq} and compare them to previous results \cite{Cardoso:2009pk, Molina:2010fb, Pani:2011xj, Wagle:2021tam, Srivastava:2021imr} in Sec.~\ref{sec:comparison_to_previous_results}. In Sec.~\ref{sec:conclusions}, we discuss future directions to pursue. Henceforth, we use the following conventions unless stated otherwise. We work in 4-dimensions with metric signature $(-,+,+,+)$ as in~\cite{Misner:1973prb}. We set $M=1/2$ when plotting and tabulating the QNM frequencies following the convention in \cite{Leaver:1985ax}.

\section{Review of the master equations}
\label{sec:review_master_eqn}

Using the modified Teukolsky formalism developed in \cite{Li:2022pcy}, we computed in \cite{Wagle:2023fwl} the modified Teukolsky equation that describes the evolution of gravitational perturbations, characterized by the perturbations of the $\Psi_{0,4}$ Weyl scalars, and (pseudo)scalar field perturbations for a slowly-rotating BH in dCS gravity. Using the radiation gauges, we found four radial ordinary differential equations governing the perturbations of $\Psi_0$ and $\vartheta$ in the IRG and the perturbations of $\Psi_4$ and $\vartheta$ in the ORG. This section briefly reviews the procedure for obtaining these master equations.

In dCS gravity, the equations of motion for the scalar field and the metric field are \cite{Jackiw:2003pm, Alexander:2009tp, Yunes:2009hc, Yagi:2012ya}
 \begin{align}
    \square\vartheta=
    & \;-\frac{\alpha}{4}R_{\nu\mu\rho\sigma}{ }^{*} 
    R^{\mu\nu\rho\sigma} \,, \label{eq:EOM_theta}\\
    R_{\mu\nu}=
    & \;-\frac{\alpha}{\kappa_{g}}C_{\mu\nu}
    +\frac{1}{2\kappa_{g}}\bar{T}_{\mu\nu}^{\vartheta}\,, 
    \label{eq:EOM_R}
\end{align}
respectively, where $\kappa_{g}=(16\pi)^{-1}$, $\alpha$ is the dCS coupling constant with dimensions of ${\left[\textrm{Length}\right]}^2$, $\square=\nabla_{\mu}\nabla^{\mu}$ is the D'Alembertian operator, $^*\!R^{\mu\nu\rho\sigma}$ is the dual of the Riemann tensor,
\begin{equation} \label{eq:Riemann_dual}
    ^*\!R^{\mu\nu\rho\sigma}
    =\frac{1}{2}\epsilon^{\rho\sigma\alpha\beta}
    R^{\mu\nu}{}_{\alpha\beta}\,,
\end{equation}
and
\begin{align} 
    C^{\mu\nu}\equiv
    & \;\left(\nabla_{\sigma}\vartheta\right)
    \epsilon^{\sigma\delta\alpha(\mu}\nabla_{\alpha}R^{\nu)\delta}
    +\left(\nabla_{\sigma}\nabla_{\delta}\vartheta\right)^*\! R^{\delta(\mu\nu)\sigma}\,, 
    \label{eq:Ctensor} \\
    \bar{T}_{\mu\nu}^{\vartheta}\equiv
    & \;\left(\nabla_{\mu}\vartheta\right)
    \left(\nabla_{\nu}\vartheta\right)\,. 
    \label{eq:T_theta}
\end{align}
For a detailed review of dCS gravity, one can refer to \cite{Jackiw:2003pm, Alexander:2009tp, Yunes:2009hc, Yagi:2012ya}. 

To study perturbations of BHs in dCS gravity, one can, in principle, follow one of the following two approaches:
\begin{itemize}
    \item \textbf{Metric perturbation approach:} This technique involves directly examining metric perturbations, denoted as $h_{\mu\nu}$. By decomposing $h_{\mu\nu}$ into even- and odd-parity components within the RW gauge \cite{Regge:1957td, Zerilli:1971wd, Moncrief:1974am}, one can then linearize Eqs.~\eqref{eq:EOM_theta} and \eqref{eq:EOM_R} to obtain a set of evolution equations for the scalar and metric fields. This method has been utilized in prior studies of QNMs in dCS gravity \cite{Cardoso:2009pk, Molina:2010fb, Pani:2011xj, Wagle:2021tam, Srivastava:2021imr} up to leading order in the spin, and very recently using spectral expansions in the METRICS framework for larger spins~\cite{Chung:2025abc} 
    \item \textbf{Curvature perturbation approach:} This approach concentrates on curvature perturbations by projecting all geometric quantities onto the NP tetrad, a null tetrad that meets specific orthogonality conditions~\cite{Newman:1961qr}. Similar to the well-known Teukolsky formalism~\cite{Teukolsky:1973ha}, one can derive a set of evolution equations (known as the modified Teukolsky equations \cite{Li:2022pcy, Hussain:2022ins, Cano:2023tmv}) for the perturbations to the Weyl scalars $\Psi_{0,4}$ and the scalar field $\vartheta$. This `modified Teukolsky formalism' has been applied to obtain the master equations describing the perturbations of rotating BH to leading order in spin in dCS gravity~\cite{Wagle:2023fwl}.
\end{itemize}
The prime objective of this paper is to compute the QNM frequencies of a slowly-rotating BH in dCS gravity by solving the modified Teukolsky equation obtained in \cite{Wagle:2023fwl}. Furthermore, we will compare our results with those obtained using the metric perturbation approach in previous work \cite{Wagle:2021tam, Srivastava:2021imr}.

To apply the modified Teukolsky formalism in \cite{Li:2022pcy, Wagle:2023fwl}, it is convenient to define a two-parameter expansion in $\zeta$ and $\epsilon$ \cite{Wagle:2023fwl}, i.e.,
\begin{align}
    & \vartheta
    =\zeta\vartheta^{(1,0)}+\zeta\epsilon\vartheta^{(1,1)}\,, 
    \label{eq:expansion_scalar} \\
    & \Psi_i
    =\Psi_i^{(0,0)}+\zeta\Psi_{i}^{(1,0)}+\epsilon\Psi_{i}^{(0,1)}
    +\zeta\epsilon\Psi_{i}^{(1,1)}\,,
    \label{eq:expansion_NP}
\end{align}
where we take the Weyl scalars $\Psi_i$ as an example, and the other NP quantities follow the same expansion scheme. The dimensionless constant $\zeta$ is related to the dCS coupling constant $\alpha$ by
\begin{equation} \label{eq:zeta}
    \zeta\equiv\frac{\alpha^2}{\kappa_{g}M^4}
\end{equation}
with $M$ being the typical mass of the system. The $\zeta$ expansion parameter controls the strength of deviations from GR and it is connected to the cut-off scale of the effective field theory.  The other expansion parameter, $\epsilon$, represents the strength of GW perturbations of a binary merger's remnant BH. For example, for an EMRI, $\epsilon$ is proportional to the mass ratio between the stellar-mass object and the supermassive BH, while for a comparable-mass post-merger, $\epsilon$ is related to the ratio of the GW amplitude to the background's gravitational potential. We do not assume any hierarchy between $\zeta$ and $\epsilon$, since their relative size depends on the details of the beyond Einstein theory and the binary system [see \cite{Li:2023ulk, Wagle:2023fwl} for more details]. In our study of slowly-rotating BHs in dCS gravity, the terms at $\mathcal{O}(\zeta^0,\epsilon^0)$ and $\mathcal{O}(\zeta^1,\epsilon^0)$ are computed using the slowly-rotating Kerr metric and the dCS correction to it \cite{Yunes:2009hc}, respectively. The terms at $\mathcal{O}(\zeta^0,\epsilon^1)$ correspond to GWs in GR, while the terms at $\mathcal{O}(\zeta^1,\epsilon^1)$ are additional GWs driven by the dCS corrections, with the latter being the perturbations of GWs we intend to study in this work. Notice that, although $\vartheta$ enters at $\mathcal{O}(\zeta^{1/2})$ according to Eq.~\eqref{eq:EOM_theta}, we have followed the convention in \cite{Wagle:2021tam} to redefine $\zeta^{1/2}\vartheta \rightarrow\vartheta$ such that the expansion in Eq.~\eqref{eq:expansion_scalar} is valid.

Using the expansion in Eq.~\eqref{eq:expansion_scalar}, the scalar field equation [Eq.~\eqref{eq:EOM_theta}] becomes
\begin{align} \label{eq:EOM_scalar_11}
    \square^{(0,0)}\vartheta^{(1,1)}
    =-\frac{M^2}{16\pi^\frac{1}{2}}\left[R^*\!R\right]^{(0,1)}
    -\square^{(0,1)}\vartheta^{(1,0)}\,,
\end{align}
where $R^*\!R$ is shorthand notation for $R_{\nu\mu\rho\sigma}{ }^{*}R^{\mu\nu\rho\sigma}$. For the metric field, instead of solving the trace-reversed Einstein equation in Eq.~\eqref{eq:EOM_R}, we will solve the modified Teukolsky equation found by~\cite{Li:2022pcy}, i.e.,
\begin{align}
    & H_{0}^{(0,0)}\Psi_0^{(1,1)}
    =\mathcal{S}_{\geo}^{(1,1)}+\mathcal{S}^{(1,1)}\,,
    \label{eq:master_eqn_non_typeD_Psi0} \\
    & H_{4}^{(0,0)}\Psi_4^{(1,1)}
    =\mathcal{T}_{\geo}^{(1,1)}+\mathcal{T}^{(1,1)}\,.
    \label{eq:master_eqn_non_typeD_Psi4}
\end{align}
Here, $H_{0,4}^{(0,0)}$ are the Teukolsky operators in GR for $\Psi_{0,4}$, respectively, and hold the form presented by Teukolsky in his seminal paper~\cite{Teukolsky:1973ha}. The source terms $\mathcal{S}_{\geo}^{(1,1)}$ and $\mathcal{T}_{\geo}^{(1,1)}$ only depend on the dCS correction to the background spacetime $h_{\mu\nu}^{(1,0)}$, as well as the gravitational perturbation $h_{\mu\nu}^{(0,1)}$ in GR, so they are regarded as purely ``geometrical.'' In contrast, $\mathcal{S}^{(1,1)}$ and $\mathcal{T}^{(1,1)}$ directly depend on the effective stress-energy tensor [i.e., Eq.~\eqref{eq:EOM_R}], and are driven by both the GR GW perturbation $h_{\mu\nu}^{(0,1)}$ and the scalar field perturbation $\vartheta^{(1,1)}$. Since $\vartheta^{(1,1)}$ actually enters at $\mathcal{O}(\zeta^{1/2},\epsilon^1)$, one can solve for $\vartheta^{(1,1)}$ first using Eq.~\eqref{eq:EOM_scalar_11} and then plug it into Eqs.~\eqref{eq:master_eqn_non_typeD_Psi0} and \eqref{eq:master_eqn_non_typeD_Psi4} to solve for $\Psi_{0,4}^{(1,1)}$. For the complete expressions of Eqs.~\eqref{eq:EOM_scalar_11}--\eqref{eq:master_eqn_non_typeD_Psi4} in terms of NP quantities, one can refer to \cite{Li:2022pcy, Wagle:2023fwl}.

To evaluate certain contributions in the source terms that are of $\mathcal{O}(\zeta^0,\epsilon^1)$ in Eqs.~\eqref{eq:EOM_scalar_11}--\eqref{eq:master_eqn_non_typeD_Psi4}, one needs to reconstruct the metric perturbation $h_{\mu\nu}^{(0,1)}$ in GR using the Teukolsky equation \cite{Teukolsky:1973ha}. We choose to use the CCK metric reconstruction approach, which was explored previously in \cite{Cohen_Kegeles_1975, Chrzanowski:1975wv, Kegeles:1979an, Lousto:2002em, Ori:2002uv, Whiting:2005hr, Yunes:2005ve, Keidl:2006wk, Keidl:2010pm}, and \cite{Wagle:2023fwl} provides a quick summary of relevant equations used. Within the CCK approach, one either chooses the IRG or the ORG:
\begin{subequations}
\begin{align} \label{eq:IRG-def}
    \textrm{IRG:}
    & \quad h_{\mu\nu}l^\nu=0\,,\; h=0\,, \\
    \label{eq:ORG-def}
    \textrm{ORG:}
    & \quad h_{\mu\nu}n^\nu=0\,,\; h=0\,,
\end{align}   
\end{subequations}
where $l^\mu$ and $n^\mu$ are two null tetrad basis vectors in the NP formalism. Under these gauges, $h_{\mu\nu}^{(0,1)}$ can be expressed in terms of an intermediate function, known as the Hertz potential $\Psi_{\Hertz}$, 
\begin{equation}
    h_{\mu\nu}^{(0,1)}
    =\mathcal{O}_{\mu\nu}\bar{\Psi}_{\Hertz}+
    \bar{\mathcal{O}}_{\mu\nu}\Psi_{\Hertz}\,,
\end{equation}
where we use $\bar{f}$ to denote the complex conjugate of $f$. The operator $\mathcal{O}_{\mu\nu}$ in the IRG and the ORG can be found in \cite{Keidl:2006wk, Keidl:2010pm, Wagle:2023fwl} [i.e., Eqs.~(50) and (51) in \cite{Wagle:2023fwl}]. Furthermore, using separation of variables, we can decompose $\Psi_{0,4}^{(0,1)}$ into spin-weighted spheroidal harmonics ${}_{s}\mathcal{Y}_{\ell m}(\theta,\phi)$, such that
\begin{subequations}
\begin{align}
    \Psi_0^{(0,1)}
    =& \;\sum_{\ell,m}{}_{2}R_{\ell m}^{(0,1)}(r)
    {}_{2}\mathcal{Y}_{\ell m}(\theta,\phi)e^{-i\omega_{\ell m} t}\,, \\
    \rho^{-4}\Psi_4^{(0,1)}
    =& \;\sum_{\ell,m}{}_{-2}R_{\ell m}^{(0,1)}(r)
    {}_{-2}\mathcal{Y}_{\ell m}(\theta,\phi)e^{-i\omega_{\ell m} t}\,,
\end{align}
\end{subequations}
where ${}_{s}\mathcal{Y}_{\ell m}(\theta,\phi)={}_{s}S_{\ell m}(\theta)e^{im\phi}$ and $\rho$ is one of the spin coefficients~\cite{Teukolsky:1973ha}. Notice, ${}_{s}S_{\ell m}(\theta)$ also depends on $a\omega$, but we hide this dependence for simplicity. Similarly, decomposing the Hertz potential $\Psi_{\Hertz}$ using separation of variables , we have
\begin{subequations} \label{eq:Hertz_decompose}
\begin{align}
    \textrm{IRG}:\quad &\bar{\Psi}_{\Hertz}
    =\sum_{\ell,m}{}_{2}\hat{R}_{\ell m}(r)\,{}_{2}\mathcal{Y}_{\ell m}(\theta,\phi)e^{-i\omega_{\ell m}t}\,, \\
    \textrm{ORG}:\quad &\bar{\Psi}_{\Hertz}
    =\sum_{\ell,m}{}_{-2}\hat{R}_{\ell m}(r)\,{}_{-2}\mathcal{Y}_{\ell m}(\theta,\phi)e^{-i\omega_{\ell m}t}\,,
\end{align}
\end{subequations}
where one can compute the radial parts ${}_{s}\hat{R}_{\ell m}(r)$ of $\Psi_{\Hertz}$ from ${}_{s}R_{\ell m}^{(0,1)}(r)$ using that~\cite{Ori:2002uv}
\begin{subequations} \label{eq:hertztoteukall}
\begin{align}
    {}_{2}\hat{R}_{\ell m}(r)
    & =-\frac{2}{\mathfrak{C}}\Delta^2(r)(D^\dagger_{\ell m})^4 \left[\Delta^2(r)\,{}_{2}R_{\ell m}^{(0,1)}(r) \right]\,, \label{eq:hertztoteuk_IRG}\\
    {}_{-2}\hat{R}_{\ell m}(r)
    & =\frac{32}{\mathfrak{C}}(D_{\ell m})^4
    {}_{-2}R_{\ell m}^{(0,1)}(r)\,. \label{eq:hertztoteuk_ORG}
\end{align}
The operators $D_{\ell m}$ and $D_{\ell m}^{\dagger}$ are defined via
\end{subequations}
\begin{equation} \label{eq:reducedop}
\begin{aligned}
    & D_{\ell m}
    =\partial_r+i\frac{am-(r^2 +a^2)\omega_{\ell m}}{\Delta(r)}\,, \\
    & D^\dagger_{\ell m}
    =\partial_r-i\frac{am-(r^2 +a^2)\omega_{\ell m}}{\Delta(r)}\,,
\end{aligned}
\end{equation}
where $a$ is the (dimensional) spin parameter of the background BH, and $\Delta(r)=r^2-2Mr+a^2$. The coefficient $\mathfrak{C}$ is the mode-dependent Teukolsky-Starobinsky constant \cite{Teukolsky:1974yv, Starobinsky:1973aij, Starobinskil:1974nkd, Ori:2002uv, Cano:2023tmv, Cano:2023jbk},
\begin{widetext}
\begin{align} \label{eq:TS_constant}
    \mathfrak{C}
    =& \;144M^2\omega_{\ell m}^2
    +\left(8+6{}_{s}B_{\ell m}+{}_{s}B_{\ell m}^2\right)^2
    -8\left[-8+{}_{s}B_{\ell m}^2
    \left(4+{}_{s}B_{\ell m}\right)\right]m\gamma_{\ell m} \nonumber\\ 
    & \;+4\left[8-2{}_{s}B_{\ell m}-{}_{s}B_{\ell m}^2+{}_{s}B_{\ell m}^3
    +2\left(-2+{}_{s}B_{\ell m}\right)
    \left(4+3{}_{s}B_{\ell m}\right)m^2\right]\gamma_{\ell m}^2 \nonumber\\
    & \;-8m\left(8-12{}_{s}B_{\ell m}+3{}_{s}B_{\ell m}^2
    +4\left(-2+{}_{s}B_{\ell m}\right)m^2\right)
    \gamma_{\ell m}^3 \nonumber \\ 
    & \;+2\left(42-22{}_{s}B_{\ell m}+3{}_{s}B_{\ell m}^2
    +8\left(-11+3{}_{s}B_{\ell m}\right)m^2
    +8m^4\right)\gamma_{\ell m}^4 \nonumber\\ 
    & \;-8 m\left[3{}_{s}B_{\ell m}
    +4\left(-4+m^2\right)\right]\gamma_{\ell m}^5
    +4\left(-7+{}_{s}B_{\ell m}+6m^2\right)\gamma_{\ell m}^6
    -8m\gamma_{\ell m}^7+\gamma_{\ell m}^8\,,
\end{align}    
\end{widetext}
where $\gamma_{\ell m}=\chi M\omega_{\ell m}$, ${}_{s}B_{\ell m}={}_{s}A_{\ell m}+s$, $s$ is the spin weight, and ${}_{s}A_{\ell m}$ is the angular separation constant in the Teukolsky equations \cite{Teukolsky:1973ha}, with $\chi = a/M$ the dimensionless spin parameter of the BH background. In the slow-rotation framework, ${}_{s}A_{\ell m}$ can be expanded as
\begin{equation} \label{eq:A_lm}
    {}_{s}A_{\ell m}=\ell(\ell+1)-s(s +1)
    -\frac{2\chi mM\omega s^2}{\ell(\ell+1)}
    +\mathcal{O}(\chi^2)\,.
\end{equation}
We have used the explicit form of  $\mathfrak{C}$ in \cite{Cano:2023tmv, Cano:2023jbk}, which is consistent with the original expression in \cite{Teukolsky:1974yv, Starobinsky:1973aij, Starobinskil:1974nkd, Ori:2002uv}. For convenience, let us also define 
\begin{equation} \label{eq:Ddagger}
    \mathcal{D}^{\dagger}_{\ell m}
    \equiv-\frac{2}{\mathfrak{C}}\Delta^2(r)
    (D^\dagger_{\ell m})^4\Delta^2(r)\,,\;
    \mathcal{D}_{\ell m}
    \equiv\frac{32}{\mathfrak{C}}(D_{\ell m})^4\,.
\end{equation}
 
In \cite{Wagle:2023fwl}, we used the CCK procedures to evaluate all the source terms in Eqs.~\eqref{eq:EOM_scalar_11}--\eqref{eq:master_eqn_non_typeD_Psi4}. We found that the $(\ell,m)$ and $(\ell,-m)$ modes of ${}_{2}\hat{R}_{\ell m}(r)$ (or ${}_{-2}\hat{R}_{\ell m}(r)$) in the IRG (or the ORG) are coupled to each other in the source terms. Thus, to solve Eqs.~\eqref{eq:EOM_scalar_11}--\eqref{eq:master_eqn_non_typeD_Psi4} consistently, we need to solve the  $(\ell,m)$ and $(\ell,-m)$ modes of $\Psi_0$ (or $\Psi_4$) jointly by using the following ansatz
\begin{widetext}
\begin{subequations} \label{eq:solution_ansatz}
\begin{align}
    & {}_{s}\Psi^{(0,1)}
    ={}_{s}R_{\ell m}^{(0,1)}(r)
    {}_{s}\mathcal{Y}_{\ell m}(\theta,\phi)e^{-i\omega_{\ell m}t}
    +\eta_{\ell m}\;{}_{s}R_{\ell -m}^{(0,1)}(r)
    {}_{s}\mathcal{Y}_{\ell\,-m}(\theta,\phi)e^{i\bar{\omega}_{\ell m}t}\,, 
    \label{eq:solution_ansatz_01} \\
    & {}_{s}\Psi^{(1,1)}
    ={}_{s}R^{(1,1)}_{\ell m}(r)
    {}_{s}\mathcal{Y}_{\ell m}(\theta,\phi)e^{-i\omega_{\ell m}t}
    +\eta_{\ell m}\;{}_{s}R^{(1,1)}_{\ell -m}(r)
    {}_{s}\mathcal{Y}_{\ell\,-m}(\theta,\phi)e^{i\bar{\omega}_{\ell m}t}\,,
    \label{eq:solution_ansatz_11}
\end{align} 
\end{subequations}   
\end{widetext}
where ${}_{2}\Psi\equiv\Psi_{0}$ and ${}_{-2}\Psi\equiv\rho^{-4}\Psi_{4}$. Following \cite{Li:2023ulk, Wagle:2023fwl}, to solve the $(\ell, m)$ and $(\ell,-m)$ modes consistently, we have imposed in Eq.~\eqref{eq:solution_ansatz} that
\begin{equation} \label{eq:freq_symmetry}
    \omega_{\ell -m}=-\bar{\omega}_{\ell m}\,,
\end{equation}
which is the same symmetry satisfied by the QNM frequencies in GR. Notice that the coefficient $\eta_{\ell m}$ is well-defined once we fix the normalization of ${}_{s}R_{\ell m}^{(0,1)}(r)$ to be
\begin{equation} \label{eq:conjugate_R}
    {}_{s}R_{\ell -m}^{(0,1)}(r)=(-1)^m {}_{s}\bar{R}_{\ell m}^{(0,1)}(r)\,,
\end{equation}
and the same symmetry can be imposed on ${}_{s}\hat{R}_{\ell m}(r)$. As we will discuss in Sec.~\ref{sec:simplify_radial_eqns}, the same symmetry in Eq.~\eqref{eq:conjugate_R} can be also imposed on ${}_{s}R_{\ell m}^{(1,1)}(r)$. Since we have only regrouped the solution into pairs of $(\ell,m)$ and $(\ell,-m)$ modes in Eq.~\eqref{eq:solution_ansatz}, the number of modes within $\Psi_{0,4}^{(1,1)}$ remains unchanged after summing all $\ell$ and $m$. Nonetheless, we need to solve the coefficient $\eta_{\ell m}$ relating the two modes within each pair, as demonstrated in Sec.~\ref{sec:modified_teuk_eqn}. For this reason, we will always assume $m\geq0$ in the remaining work without losing generality. At $m=0$, there are still two modes with the frequencies $\omega_{\ell 0}$ and $-\bar{\omega}_{\ell 0}$, respectively, in Eq.~\eqref{eq:solution_ansatz}. Furthermore, we only consider prograde modes in this work, as the retrograde modes do not couple to the prograde modes within the modified Teukolsky equation and can be treated independently in a similar fashion. Similarly, since $\vartheta$ is a real (pseudo)scalar field, we expand its perturbation $\vartheta^{(1,1)}$ as
\begin{widetext}
\begin{align} \label{eq:scalar_separation}
    \vartheta^{(1,1)}
    =& \;\frac{\Theta_{\ell m}^{(1,1)}(r)}{r}{}_{0}
    \mathcal{Y}_{\ell m}(\theta,\phi)e^{-i\omega_{\ell m}t}
    +\frac{\Theta_{\ell -m}^{(1,1)}(r)}{r}
    {}_{0}\mathcal{Y}_{\ell -m}(\theta,\phi)
    e^{i\bar{\omega}_{\ell m}t} \nonumber\\
    \equiv& \;\frac{\Theta_{\ell m}^{(1,1)}(r)}{r}{}_{0}
    \mathcal{Y}_{\ell m}(\theta,\phi)e^{-i\omega_{\ell m}t}
    +\frac{\bar{\Theta}_{\ell m}^{(1,1)}(r)}{r}
    {}_{0}\bar{\mathcal{Y}}_{\ell m}(\theta,\phi)
    e^{i\bar{\omega}_{\ell m}t}\,.
\end{align}    
\end{widetext}
Notice that we include an extra factor of $1/r$ in Eq.~\eqref{eq:scalar_separation} following \cite{Cardoso:2009pk,Molina:2010fb, Wagle:2021tam}, so we will multiply the scalar field equation found in \cite{Wagle:2023fwl} by a factor of $r$, as discussed in Sec.~\ref{sec:scalar_eqn}. Since $\bar{\omega}_{\ell m}=\omega_{\ell -m}$  and ${}_{0}\bar{\mathcal{Y}}_{\ell m}(\theta,\phi)=(-1)^m{}_{0}\mathcal{Y}_{\ell -m}(\theta,\phi)$, we need
\begin{align} \label{eq:scalar_conjugate}
   \Theta_{\ell m}^{(1,1)}(r)=(-1)^{m}\bar{\Theta}_{\ell -m}^{(1,1)}(r)\,.
\end{align}
Using the ansatz in Eqs.~\eqref{eq:solution_ansatz} and \eqref{eq:scalar_separation} and projecting Eqs.~\eqref{eq:EOM_scalar_11}, \eqref{eq:master_eqn_non_typeD_Psi0}, and \eqref{eq:master_eqn_non_typeD_Psi4} to ${}_{0}\mathcal{Y}_{\ell m}(\theta,\phi)$, ${}_{2}\mathcal{Y}_{\ell m}(\theta,\phi)$, and ${}_{-2}\mathcal{Y}_{\ell m}(\theta,\phi)$, respectively, we obtain the purely-radial, ordinary differential equations governing $\left\{\vartheta^{(1,1)},\Psi_0^{(1,1)},\Psi_4^{(1,1)}\right\}$ in \cite{Wagle:2023fwl}. In Secs.~\ref{sec:scalar_eqn} and \ref{sec:modified_teuk_eqn}, we will closely examine and simplify these equations for the convenience of calculating QNM frequencies in this work. We will focus on the equations of $\vartheta^{(1,1)}$ and $\Psi_0^{(1,1)}$ in the IRG as a demonstration, and the results in the ORG will be presented at the end of each section.

\section{Scalar field equation}
\label{sec:scalar_eqn}

In this section, we review and simplify the master equation governing the scalar field perturbation $\vartheta^{(1,1)}$ in \cite{Wagle:2023fwl}. In \cite{Wagle:2023fwl}, we found that up to $\mathcal{O}(\zeta^1,\chi^1,\epsilon^1)$, the radial part $\Theta_{\ell m}^{(1,1)}(r)$ of $\vartheta^{(1,1)}$ in the IRG satisfies
\begin{widetext}
\begin{subequations} \label{eq:scalar_radial_IRG_1}
\begin{align}
    & \left[r(r-r_s)\partial_{r}^2 
    +r_s\partial_r+\frac{\omega^2r^3-4\chi m M^2 \omega}{r-r_s}
    -\frac{r_s}{r}-{}_{0}A_{\ell m}\right]\Theta_{\ell m}^{(1,1)}(r) \nonumber\\
    &=V^{R}_{\ell m}(r)+V^{\square}_{\ell m}(r)
    +\bar{\eta}_{\ell m}\left(V^{\dagger R}_{\ell-m}(r)
    +V^{\dagger \square}_{\ell-m}(r)\right)\,,
    \label{eq:scalar_radial_IRG_1_lm} \\
    & \left[r(r-r_s)\partial_{r}^2 
    +r_s\partial_r+\frac{\omega^2r^3+4\chi m M^2 \omega}{r-r_s}
    -\frac{r_s}{r}-{}_{0}A_{\ell-m}\right]\Theta_{\ell -m}^{(1,1)}(r) \nonumber\\
    &=\eta_{\ell m}\left(V^{R}_{\ell -m}(r)
    +V^{\square}_{\ell-m}(r)\right)
    +V^{\dagger R}_{\ell m}(r)
    +V^{\dagger\square}_{\ell m}(r)\,,
    \label{eq:scalar_radial_IRG_1_l-m}
\end{align}
\end{subequations}
where $r_s=2M$ is the Schwarzschild radius of the background BH, the angular separation constant ${}_{0}A_{\ell m}$ is given in Eq.~\eqref{eq:A_lm}, and
\begin{subequations} \label{eq:scalar_source_IRG_1}
\begin{align} 
    V^{R}_{\ell m}(r)
    =& \; i \left(g_1^{\ell m}(r)\,{}_{2}\hat{R}_{\ell m}(r)
    + g_2^{\ell m}(r)\,{}_{2}\hat{R}'_{\ell m}(r)\right)
    \Lambda^{\ell\ell m}_{00}
    +\chi\left(g_3^{\ell m}(r)\,{}_{2}\hat{R}_{\ell m}(r)
    +g_4^{\ell m}(r)\,{}_{2}\hat{R}_{\ell m}'(r)\right)
    \Lambda^{\ell\ell m}_{10s}\,, \label{eq:R*R_radial_IRG}\\
    V^{\square}_{\ell m}(r)
    =& \;\chi\left(h_1^{\ell m}(r)\,{}_{2}\hat{R}_{\ell m}(r)
    +h_2^{\ell m}(r)\,{}_{2}\hat{R}_{\ell m}'(r)\right)\Lambda^{\ell\ell m}_{10s}\,. \label{eq:box_radial_IRG} 
\end{align}
\end{subequations}
\end{widetext}
The functions $V^{\dagger R}_{\ell-m}$ and $V^{\dagger\square}_{\ell-m}$ are given by the complex conjugate of Eqs.~\eqref{eq:R*R_radial_IRG} and~\eqref{eq:box_radial_IRG}, respectively, but with the constants $\{\Lambda^{\ell_1\ell_2m}_{s_1s_2}, \Lambda^{\ell_1\ell_2m}_{s_1s_2c}, \Lambda^{\ell_1\ell_2m}_{s_1s_2s}\} \to \{\Lambda^{\dagger\ell_1\ell_2m}_{s_1s_2},\Lambda^{\dagger\ell_1\ell_2m}_{s_1s_2c},\Lambda^{\dagger\ell_1\ell_2m}_{s_1s_2s}\}$, where the definitions for the latter are given in Appendix~\ref{appendix:projection_coeff}. Here, $f'(r)$ denotes the partial derivative of $f(r)$ in $r$. Notice that Eq.~\eqref{eq:scalar_radial_IRG_1_l-m} comes from the terms generated by the $(\ell,-m)$ mode of the ansatz in Eq.~\eqref{eq:solution_ansatz_01}. We also redefine all the radial functions $g_{i}^{\ell m}(r)=g_i^{\ell m}(r,\omega,M)$ and $h_{i}^{\ell m}(r)=h_{i}^{\ell m}(r,\omega,M)$ in \cite{Wagle:2023fwl} as follows:
\begin{align} \label{eq:redefine_g_h}
    & g_1^{\ell m}(r)\rightarrow
    i\frac{M^2r^3}{16\pi^{\frac{1}{2}}}g_1^{\ell m}(r)\,,\;
    && g_2^{\ell m}(r)\rightarrow
    i\frac{M^2r^3}{16\pi^{\frac{1}{2}}}g_2^{\ell m}(r)\,,\nonumber\\
    & g_3^{\ell m}(r)\rightarrow
    -\frac{M^2r^3}{16\pi^{\frac{1}{2}}}g_3^{\ell m}(r)\,,\;
    && g_4^{\ell m}(r)\rightarrow
    -\frac{M^2r^3}{16\pi^{\frac{1}{2}}}g_4^{\ell m}(r)\,,\nonumber\\
    & h_1^{\ell m}(r)\rightarrow-r^3h_1^{\ell m}(r)\,,\;
    && h_2^{\ell m}(r)\rightarrow-r^3h_2^{\ell m}(r)\,.
\end{align}
In this case, $g_i^{\ell m}(r)$ and $h_i^{\ell m}(r)$ become real functions in $(r,-i\omega,im)$, where the imaginary unit $i$ only comes from the combination $-i\omega$ or $im$. Using $\omega_{\ell m}=-\bar{\omega}_{\ell -m}$ in Eq.~\eqref{eq:freq_symmetry}, we then have that
\begin{equation} \label{eq:parity_g_h}
    \bar{g}_{i}^{\ell -m}(r)=g_{i}^{\ell m}(r)\,,\quad
    \bar{h}_{i}^{\ell -m}(r)=h_{i}^{\ell m}(r)\,,
\end{equation}
where the expression of $g_{i}^{\ell m}(r)$ and $h_{i}^{\ell m}(r)$ before the redefinition can be found in \cite{Wagle:2023fwl} and the supplementary Mathematica notebook \cite{Pratikmodteuk}. In addition, $g_{1}^{\ell m}(r)$ and $g_{2}^{\ell m}(r)$ contain terms up to $\mathcal{O}(\chi^1)$, and $\chi$ only shows up in the combination $\chi m$. The other radial functions $g_i^{\ell m}(r)$ and $h_i^{\ell m}(r)$ in Eq.~\eqref{eq:scalar_source_IRG_1} do not contain any factor of $\chi$ or $m$. The coefficients $\{\Lambda^{\ell_1\ell_2m}_{s_1s_2}, \Lambda^{\ell_1\ell_2m}_{s_1s_2c}, \Lambda^{\ell_1\ell_2m}_{s_1s_2s}\}$ and $\{\Lambda^{\dagger\ell_1\ell_2m}_{s_1s_2},
\Lambda^{\dagger\ell_1\ell_2m}_{s_1s_2c},
\Lambda^{\dagger\ell_1\ell_2m}_{s_1s_2s}\}$ come from projecting the angular functions in the source terms to ${}_{0}\mathcal{Y}_{\ell m}(\theta,\phi)$, where their definition can be found in \cite{Wagle:2023fwl}. Using Eqs.~\eqref{eq:relation_12}, \eqref{eq:relation_34}, and \eqref{eq:relation_56} in Appendix~\ref{appendix:projection_coeff}, we can show that
\begin{equation} 
    \Lambda^{\ell\ell m}_{10s}=m\Lambda^{\ell\ell 1}_{10s}\,,\quad
    \Lambda^{\dagger\ell\ell m}_{10s}
    =(-1)^{m+1}m\Lambda^{\ell\ell 1}_{10s}\,.
\end{equation}
Since both the radial Teukolsky function ${}_{s}R_{\ell m}^{(0,1)}(r)$ and the radial part ${}_{s}\hat{R}_{\ell m}(r)$ of the Hertz potential satisfy
Eq.~\eqref{eq:conjugate_R}, Eq.~\eqref{eq:scalar_source_IRG_1} reduces to
\begin{widetext}
\begin{subequations} \label{eq:scalar_source_IRG_2}
\begin{align} 
    & V^{R}_{\ell m}(r)
    =i\left(g_1^{\ell m}(r)\,{}_{2}\hat{R}_{\ell m}(r)
    +g_2^{\ell m}(r)\,{}_{2}\hat{R}'_{\ell m}(r)\right)
    +\chi m\left(g_3^{\ell m}(r)\,{}_{2}\hat{R}_{\ell m}(r)
    +g_4^{\ell m}(r)\,{}_{2}\hat{R}_{\ell m}'(r)\right)
    \Lambda^{\ell\ell 1}_{10s}\,, \\
    & V^{\square}_{\ell m}(r)
    =\chi m\left(h_1^{\ell m}(r)\,{}_{2}\hat{R}_{\ell m}(r)
    +h_2^{\ell m}(r)\,{}_{2}\hat{R}_{\ell m}'(r)\right)\Lambda^{\ell\ell 1}_{10s}\,, \\
    & V^{\dagger R}_{\ell -m}(r)=-V^{R}_{\ell m}(r)\,,\quad
    V^{\dagger\square}_{\ell -m}(r)=-V^{\square}_{\ell m}(r)\,.
\end{align}
\end{subequations}
Therefore, Eq.~\eqref{eq:scalar_radial_IRG_1} becomes
\begin{subequations} \label{eq:scalar_radial_IRG_2}
\begin{align} 
    \left[r(r-r_s)\partial_{r}^2
    +r_s\partial_r+\frac{\omega^2r^3-4\chi m M^2 \omega}{r-r_s}
    -\frac{r_s}{r}-{}_{0}A_{\ell m}\right]\Theta_{\ell m}^{(1,1)}(r)
    &=(1-\bar{\eta}_{\ell m})\left(V^{R}_{\ell m}(r)
    +V^{\square}_{\ell m}(r)\right)\,, 
    \label{eq:scalar_radial_IRG_2_lm} \\
    \left[r(r-r_s)\partial_{r}^2
    +r_s\partial_r+\frac{\omega^2r^3+4\chi m M^2 \omega}{r-r_s}
    -\frac{r_s}{r}-{}_{0}A_{\ell-m}\right]\Theta_{\ell-m}^{(1,1)}(r)
    &=(\eta_{\ell m}-1)\left(V^{R}_{\ell -m}(r)
    +V^{\square}_{\ell -m}(r)\right)\,.
    \label{eq:scalar_radial_IRG_2_l-m}
\end{align}
\end{subequations}
\end{widetext}
As shown in \cite{Li:2023ulk}, the metric perturbation generated by the modes with $\bar{\eta}_{\ell m}=1$ is of even parity, so the even-parity metric perturbation is not coupled to the scalar field up to $\mathcal{O}(\zeta^1,\chi^1,\epsilon^1)$ in dCS gravity, a result that is consistent with \cite{Wagle:2021tam}. Furthermore, using Eqs.~\eqref{eq:conjugate_R}, \eqref{eq:parity_g_h}, and \eqref{eq:scalar_radial_IRG_2}, one can show that
\begin{widetext}
\begin{align} \label{eq:scalar_radial_IRG_3}
    \left[r(r-r_s)\partial_{r}^2
    +r_s\partial_r+\frac{\omega^2r^3-4\chi m M^2 \omega}{r-r_s}
    -\frac{r_s}{r}-{}_{0}A_{\ell m}\right]\bar{\Theta}_{\ell -m}^{(1,1)}(r)
    =(-1)^{m}(1-\bar{\eta}_{\ell m})\left(
    V^{R}_{\ell m}(r)+V^{\square}_{\ell m}(r)\right)\,,
\end{align}
which is consistent with our assumption in Eq.~\eqref{eq:scalar_conjugate}. 

Similarly, in the ORG, we find
\begin{subequations} \label{eq:scalar_radial_ORG}
\begin{align} 
    \left[r(r-r_s)\partial_{r}^2
    +r_s\partial_r+\frac{\omega^2r^3-4\chi m M^2 \omega}{r-r_s}
    -\frac{r_s}{r}-{}_{0}A_{\ell m}\right]\Theta_{\ell m}^{(1,1)}(r) 
    & =(1-\bar{\eta}_{\ell m})\left(U^{R}_{\ell m}(r)
    +U^{\square}_{\ell m}(r)\right)\,, 
    \label{eq:scalar_radial_ORG_lm} \\
    \left[r(r-r_s)\partial_{r}^2
    +r_s\partial_r+\frac{\omega^2r^3+4\chi m M^2 \omega}{r-r_s}
    -\frac{r_s}{r}-{}_{0}A_{\ell-m}\right]\Theta_{\ell-m}^{(1,1)}(r) 
    & =(\eta_{\ell m}-1)\left(U^{R}_{\ell -m}(r)
    +U^{\square}_{\ell -m}(r)\right)\,,
    \label{eq:scalar_radial_ORG_l-m}
\end{align}
\end{subequations}
where
\begin{subequations} \label{eq:scalar_source_ORG}
\begin{align} 
    & U^{R}_{\ell m}(r)
    =i\left(\textgoth{g}_1^{\ell m}(r)\,{}_{-2}\hat{R}_{\ell m}(r)
    +\textgoth{g}_2^{\ell m}(r)\,{}_{-2}\hat{R}'_{\ell m}(r)\right)
    +\chi m\left(\textgoth{g}_3^{\ell m}(r)\,{}_{-2}\hat{R}_{\ell m}(r)
    +\textgoth{g}_4^{\ell m}(r)\,{}_{-2}\hat{R}_{\ell m}'(r)\right)
    \Lambda^{\ell\ell 1}_{10s}\,, \\
    & U^{\square}_{\ell m}(r)
    =\chi m\left(\textgoth{h}_1^{\ell m}(r)\,{}_{-2}\hat{R}_{\ell m}(r)
    +\textgoth{h}_2^{\ell m}(r)\,{}_{-2}\hat{R}_{\ell m}'(r)\right)\Lambda^{\ell\ell 1}_{10s}\,, \\
    & U^{\dagger R}_{\ell -m}(r)=-U^{R}_{\ell m}(r)\,,\quad
    U^{\dagger\square}_{\ell -m}(r)=-U^{\square}_{\ell m}(r)\,,
\end{align}
\end{subequations}
\end{widetext}
and we have used Eq.~\eqref{eq:relation_34} to replace $\Lambda^{\ell\ell 1}_{-10s}$ by $\Lambda^{\ell\ell 1}_{10s}$. Similar to the IRG case, we have redefined $\textgoth{g}_{i}^{\ell m}(r)$ and $\textgoth{h}_{i}^{\ell m}(r)$ in \cite{Wagle:2023fwl} such that
\begin{equation} \label{eq:parity_gothg_gothh}
    \bar{\textgoth{g}}_{i}^{\ell -m}(r)
    =\textgoth{g}_{i}^{\ell m}(r)\,,\quad
    \bar{\textgoth{h}}_{i}^{\ell -m}(r)
    =\textgoth{h}_{i}^{\ell m}(r)\,.
\end{equation}
The redefinition of $\textgoth{g}_{i}^{\ell m}(r)$ and $\textgoth{h}_{i}^{\ell m}(r)$ can be obtained by replacing $g$ and $h$ with $\textgoth{g}$ and $\textgoth{h}$ in Eq.~\eqref{eq:redefine_g_h}, respectively. In addition, $\textgoth{g}_{1}^{\ell m}(r)$ and $\textgoth{g}_{2}^{\ell m}(r)$ contain terms up to $\mathcal{O}(\chi^1)$, and $\chi$ only shows up in the combination $\chi m$. The other radial functions $\textgoth{g}_i^{\ell m}(r)$ and $\textgoth{h}_i^{\ell m}(r)$ in Eq.~\eqref{eq:scalar_source_ORG} do not contain any factor of $\chi$ or $m$. The original $\textgoth{g}_i^{\ell m}(r)$ and $\textgoth{h}_i^{\ell m}(r)$ of \cite{Wagle:2023fwl} are listed in the Mathematica notebook \cite{Pratikmodteuk}.

\section{The modified Teukolsky equation}
\label{sec:modified_teuk_eqn}

In this section, we review and simplify the modified Teukolsky equation for $\Psi_0$ in dCS gravity up to $\mathcal{O}(\zeta^1,\chi^1,\epsilon^1)$, which was found in \cite{Wagle:2023fwl}. In that work, we showed that $\Psi_0^{(1,1)}$ satisfies\footnote{For a better representation of the physical meaning of each source term, we have renamed $\{\mathcal{S}_A,\tilde{\mathcal{S}}_A,
\mathcal{S}_B,\tilde{\mathcal{S}}_B\}$ given in~\cite{Wagle:2023fwl} to $\{\mathcal{S}_{T_{\vartheta}}, \tilde{\mathcal{S}}_{T_{\vartheta}}\mathcal{S}_{T_{\Psi}},\tilde{\mathcal{S}}_{T_{\Psi}}\}$, respectively.}
\begin{align} \label{eq:master_eqn_dCS_Psi0}
    H_{0}^{(0,0)}\Psi_0^{(1,1)}
    =2r^2\left(\mathcal{S}_{\geo}^{(1,1)}
    \right. &+ \left. \mathcal{S}_{T_{\vartheta}}^{(1,1)} + \tilde{\mathcal{S}}_{T_{\vartheta}}^{(1,1)} \right.
    \\ \nonumber & \left.+\mathcal{S}_{T_{\Psi}}^{(1,1)}+\tilde{\mathcal{S}}_{T_{\Psi}}^{(1,1)}\right)\,,
\end{align}
where $H_{0}^{(0,0)}$ is the Teukolsky operator for $\Psi_0$ in GR and expanded to $\mathcal{O}(\chi)$,
\begin{subequations} \label{eq:H0_00_TK}
\begin{align}
    & H_{0}^{(0,0)}
    =H_{0}^{(0,0,0)}+\chi H_{0}^{(0,1,0)}\,, \\
    \begin{split} \label{eq:H0_000_TK} 
        & H_{0}^{(0,0,0)}
        =-r(r-r_s)\partial^2_{r}-6(r-M)\partial_r
        -\frac{C(r)}{r-r_s} \\ 
        & \;-\partial^2_{\theta}-\cot{\theta}\partial_{\theta}
        +\left(4+m^{2}+4m\cos\theta\right)\csc^2{\theta}-6\,,
    \end{split} \\
    \begin{split} \label{eq:H0_010_TK}
        & H_{0}^{(0,1,0)}
        =-4M\left[\frac{m(i(r-M)-M\omega r)}{r(r-r_s)}
        -\omega\cos{\theta}\right]\,, \\
    \end{split} \\
    & \label{eq:Cfunc} C(r)
    =4i\omega r(r-3M)+\omega^{2}r^{3}\,.\nonumber
\end{align}    
\end{subequations}
In Eq.~\eqref{eq:master_eqn_dCS_Psi0}, $\mathcal{S}_{\geo}^{(1,1)}$ is the source term due to the dCS correction to the geometry of rotating BHs. The terms $\mathcal{S}_{T_{\vartheta}}^{(1,1)}$ and $\tilde{\mathcal{S}}_{T_{\vartheta}}^{(1,1)}$ are sources associated with the effective stress-energy tensor and driven by the scalar field perturbation $\vartheta^{(1,1)}$. The source terms $\mathcal{S}_{T_{\Psi}}^{(1,1)}$ and $\tilde{\mathcal{S}}_{T_{\Psi}}^{(1,1)}$ are also associated with the effective stress-energy tensor but driven by the Weyl scalar perturbation $\Psi_0^{(0,1)}$ in GR. 

As discussed earlier in Sec.~\ref{sec:review_master_eqn}, we need to solve the $(\ell, m)$ and $(\ell,-m)$ modes of $\Psi_0$ jointly since they are coupled in the source terms. In \cite{Li:2023ulk}, it was found that the Teukolsky operator in GR satisfies the symmetry:
\begin{equation}
    \hat{\mathcal{P}}H_0^{(0,0)}=H_0^{(0,0)}\,,
\end{equation}
where $\hat{\mathcal{P}}$ is an operator combining complex conjugation with a parity transformation, i.e.,
\begin{equation} \label{eq:def_P}
    \hat{\mathcal{P}}f
    =\hat{\mathcal{C}}\hat{P}f
    =\hat{\mathcal{C}}f(\pi-\theta,\phi+\pi)
    =\bar{f}(\pi-\theta,\phi+\pi)\,.
\end{equation}
For this reason, Ref.~\cite{Li:2023ulk} showed that one could solve Eq.~\eqref{eq:master_eqn_dCS_Psi0} jointly with its $\hat{\mathcal{P}}$ transformation such that the latter will generate a consistency relation for the $(\ell,m)$ mode from the equation of the $(\ell,-m)$ mode. We can then reduce the modified Teukolsky equation of $\Psi_0^{(1,1)}$ to a two-dimensional eigenvalue problem and compute the QNM frequencies \cite{Li:2023ulk}, as shown in more detail at the end of this subsection. In the following subsections, we will sequentially simplify these three groups of source terms $\mathcal{S}_{\geo}^{(1,1)}$, $\left\{\mathcal{S}_{T_{\vartheta}}^{(1,1)},\tilde{\mathcal{S}}_{T_{\vartheta}}^{(1,1)}\right\}$, and $\left\{\mathcal{S}_{T_{\Psi}}^{(1,1)},\tilde{\mathcal{S}}_{T_{\Psi}}^{(1,1)}\right\}$. We will derive their transformation under $\hat{\mathcal{P}}$ and show that the solutions to the modified Teukolsky equation are of definite parity with $\eta_{\ell m}=\pm1$.

\subsection{The $\mathcal{S}_{\geo}^{(1,1)}$ source term}
\label{sec:sGeo_simplify}

\begin{widetext}
First, we found in \cite{Wagle:2023fwl} that the ``geometrical'' source term $\mathcal{S}_{\geo}^{(1,1)}$ driven by $\Psi_0^{(0,1)}$ in Eq.~\eqref{eq:solution_ansatz_01} in dCS gravity is 
\begin{align} \label{eq:source_psi0_geo}
    \mathcal{S}_{\geo}^{(1,1)}
    =-e^{-i\omega_{\ell m}t}H_0^{\ell m(1,0)}
    \left[{}_{2}R_{\ell m}^{(0,1)}(r){}_{2}Y_{\ell m}(\theta,\phi)\right]
    -\eta_{\ell m}\times(m\rightarrow-m)\,,
\end{align}
where ${}_{s}Y_{\ell m}(\theta,\phi)$ are spin-weighted spherical harmonics due to the expansion of ${}_{s}\mathcal{Y}_{\ell m}(\theta,\phi)$ over $\chi$, and
\begin{align} \label{eq:H0_110}
    H_0^{\ell m(1,0)}
    =& \;\frac{i\chi mM^4}{448r^9(r-r_s)}
    \left(C_1(r)+4i\omega_{\ell m}r^2C_2(r)\right)
    -\frac{i\chi M^4}{16r^9}\cos{\theta}
    \left(C_3(r)-\frac{i\omega_{\ell m}r^2C_4(r)}{2}\right) \nonumber\\
    & \;+\frac{i\chi M^4}{32r^8}\left[(r-r_s)C_4(r)\cos{\theta}\partial_r
    -\frac{C_5(r)}{2r}\sin{\theta}\partial_{\theta}\right]\,.
\end{align}
Notice that when taking $m\rightarrow-m$ in Eq.~\eqref{eq:source_psi0_geo} for $m=0$, we obtain the mode with frequency $-\bar{\omega}_{\ell0}$. The radial functions $C_i(r)$ are real in $r$ and listed in \cite{Wagle:2023fwl}. Using that \cite{Wagle:2023fwl}
\begin{equation}
    \partial_{\theta}\left({}_{2}Y_{\ell m}(\theta,\phi)\right)
    =\frac{1}{2}\left(\sqrt{(\ell+2)(\ell-1)}
    {}_{1}Y_{\ell m}(\theta,\phi)
    -\sqrt{(\ell+3)(\ell-2)}
    {}_{3}Y_{\ell m}(\theta,\phi)\right)\,,
\end{equation}
we can rewrite $\mathcal{S}_{\geo}^{(1,1)}$ as
\begin{equation} \label{eq:S_geo_simplify}
    \mathcal{S}_{\geo}^{(1,1)}
    =e^{-i\omega_{\ell m}t}
    \mathcal{O}_{\geo}^{\ell m}\,{}_{2}R_{\ell m}^{(0,1)}
    +\eta_{\ell m}\times(m\rightarrow-m)\,,
\end{equation}
with the operator $\mathcal{O}_{\geo}^{\ell m}$ defined to be
\begin{equation} \label{eq:S_geo_radial}
\begin{aligned}
    \mathcal{O}_{\geo}^{\ell m}
    =& \;-\frac{i\chi mM^4}{448r^9(r-r_s)}\,{}_{2}Y_{\ell m}(\theta,\phi)
    \left(C_1(r)+4i\omega_{\ell m} r^2C_2(r)\right) \\
    & \;+\frac{i\chi M^4}{16r^9}\cos{\theta}\,{}_{2}Y_{\ell m}(\theta,\phi) 
    \left[C_3(r)-C_4(r)\left(\frac{i\omega_{\ell m}r^2}{2}
    +\frac{r(r-r_s)}{2}\partial_r\right)\right]
    \\ 
    & \;+\frac{i\chi M^4}{128r^9}C_5(r)
    \left(\sqrt{(\ell+2)(\ell-1)}\sin{\theta}\,{}_{1}Y_{\ell m}(\theta,\phi)
    -\sqrt{(\ell+3)(\ell-2)}\sin{\theta}\,{}_{3}Y_{\ell m}(\theta,\phi)\right)\,.
\end{aligned}
\end{equation}
\end{widetext}

To perform the $\hat{\mathcal{P}}$ transformation [i.e., Eq.~\eqref{eq:def_P}] on $\mathcal{S}_{\geo}^{(1,1)}$, we also need the following properties of spin-weighted spherical harmonics ${}_{s}Y_{\ell m}(\theta,\phi)$ \cite{Nichols:2012jn}:
\begin{equation} \label{eq:angular_transformation}
\begin{aligned}
    & {}_{s}Y_{\ell m}(\pi-\theta,\phi+\pi)
    =(-1)^{\ell}{}_{-s}Y_{\ell m}(\theta,\phi)\,, \\
    & {}_{s}\bar{Y}_{\ell m}(\theta,\phi)
    =(-1)^{m+s}{}_{-s}Y_{\ell-m}(\theta,\phi)\,,
\end{aligned}
\end{equation}
so we obtain
\begin{align} \label{eq:parity_Teuk_func}
    & \hat{\mathcal{P}}\left[{}_{\pm 2}Y_{\ell m}(\theta,\phi)\right]
    =(-1)^{\ell+m}{}_{\pm 2}Y_{\ell -m}(\theta,\phi)\,, \nonumber\\
    & \hat{\mathcal{P}}\left[\sin{\theta}{}_{\pm 1}Y_{\ell m}(\theta,\phi)\right]
    =(-1)^{\ell+m+1}\sin{\theta}
    {}_{\pm 1}Y_{\ell -m}(\theta,\phi)\,, \nonumber\\
    & \hat{\mathcal{P}}\left[\cos{\theta}{}_{\pm 2}Y_{\ell m}(\theta,\phi)\right]
    =(-1)^{\ell+m+1}\cos{\theta}
    {}_{\pm 2}Y_{\ell -m}(\theta,\phi)\,, \nonumber\\
    & \hat{\mathcal{P}}\left[\sin{\theta}{}_{\pm 3}Y_{\ell m}(\theta,\phi)\right]
    =(-1)^{\ell+m+1}\sin{\theta}
    {}_{\pm 3}Y_{\ell -m}(\theta,\phi)\,.
\end{align}
Using Eq.~\eqref{eq:parity_Teuk_func} and that $C_i(r)$ are real functions in $r$, we obtain
\begin{equation} \label{eq:O_geo_P_transform}
    \hat{\mathcal{P}}\left[\mathcal{O}_{\geo}^{\ell m}\,
    {}_{2}R^{(0,1)}_{\ell m}(r)\right]
    =(-1)^{\ell}\mathcal{O}_{\geo}^{\ell -m}\,{}_{2}R^{(0,1)}_{\ell -m}(r)\,.
\end{equation}

\subsection{The $\mathcal{S}_{T_{\vartheta}}^{(1,1)}$ and $\tilde{\mathcal{S}}_{T_{\vartheta}}^{(1,1)}$ source terms}
\label{sec:sA_simplify}

Next, we know from \cite{Wagle:2023fwl} that the source terms $\mathcal{S}_{T_{\vartheta}}^{(1,1)}$ and $\tilde{\mathcal{S}}_{T_{\vartheta}}^{(1,1)}$ driven by $\vartheta^{(1,1)}$ in Eq.~\eqref{eq:scalar_separation} take the form
\begin{widetext}
\begin{subequations} \label{eq:SAset}
\begin{align} 
    \mathcal{S}_{T_{\vartheta}}^{(1,1)} 
    =& \;e^{-i\omega_{\ell m} t}
    \left[\mathcal{A}_{1}^{\ell m}(r){}_2Y_{\ell m}(\theta,\phi)
    +i\chi \mathcal{A}_{2}^{\ell m}(r)\sin\theta\,{}_1Y_{\ell m}(\theta,\phi)
    +i\chi \mathcal{A}_{3}^{\ell m}(r)\cos\theta\,{}_2Y_{\ell m}(\theta,\phi)\right]\,,
    \label{eq:sA_coord} \\
    \tilde{\mathcal{S}}_{T_{\vartheta}}^{(1,1)} 
    =& \;e^{+i\bar{\omega}_{\ell m} t}
    \left[-\bar{\mathcal{A}}_{1}^{\ell m}(r){}_{-2}\bar{Y}_{\ell m}(\theta,\phi)
    +i\chi\bar{\mathcal{A}}_{2}^{\ell m}(r)\sin\theta\,{}_{-1}
    \bar{Y}_{\ell m}(\theta,\phi)
    -i\chi\bar{\mathcal{A}}_{3}^{\ell m}(r)\cos\theta\,{}_{-2}
    \bar{Y}_{\ell m}(\theta,\phi)\right]\,,
    \label{eq:sAc_coord}
\end{align}
\end{subequations}
\end{widetext}
where $\mathcal{S}_{T_{\vartheta}}^{(1,1)}$ and $\tilde{\mathcal{S}}_{T_{\vartheta}}^{(1,1)}$ come from the first and the second term of Eq.~\eqref{eq:scalar_separation}, respectively. The spin-weighted spherical harmonics ${}_{s}Y_{\ell m}(\theta,\phi)$ comes from expanding ${}_{s}\mathcal{Y}_{\ell m}(\theta,\phi)$ over $\chi$. We have also dropped the terms proportional to ${}_{0}b^m_{\ell,\ell\pm 1}$ in \cite{Wagle:2023fwl} since they contribute at $\mathcal{O}(\chi^2)$ after the angular projection, as discussed in \cite{Wagle:2023fwl}. The radial function $\mathcal{A}^{\ell m}_{i}(r)$ takes the form
\begin{equation} \label{eq:A_i_m}
    \mathcal{A}_i^{\ell m}(r)
    =i\mathcal{O}^{\ell m}_i\Theta_{\ell m}(r)
    +\alpha_{\ell m}\mathcal{Q}^{\ell m}_i{}_{2}\hat{R}_{\ell m}(r)\,,
\end{equation}
where $\mathcal{O}^{\ell m}_i$ and $\mathcal{Q}^{\ell m}_i$ are differential operators in $r$ and contain up to first derivatives in $r$. The coefficient $\alpha_{\ell m}=1-\bar{\eta}_{\ell m}$ for $m>0$ and $\alpha_{\ell m}=\eta_{\ell -m}-1$ for $m<0$. When $m=0$, $\alpha_{\ell m}$ is $1-\bar{\eta}_{\ell 0}$ and $\eta_{\ell 0}-1$ for the modes with the frequency $\omega_{\ell 0}$ and $-\bar{\omega}_{\ell 0}$, respectively. Furthermore, $\mathcal{O}^{\ell m}_i$ and $\mathcal{Q}^{\ell m}_i$ satisfy
\begin{equation} \label{eq:O_Q_conjugate}
    \bar{\mathcal{O}}^{\ell -m}_i=\mathcal{O}^{\ell m}_i\,,\quad
    \bar{\mathcal{Q}}^{\ell -m}_i=\mathcal{Q}^{\ell m}_i\,.
\end{equation}
Notice that the term $\mathcal{Q}^{\ell m}_i\,{}_{2}\hat{R}_{\ell m}(r)$ comes from replacing $\partial_r^2\Theta_{\ell m}^{(1,1)}(r)$ with $\partial_r\Theta_{\ell m}^{(1,1)}(r)$ and $\Theta_{\ell m}(r)$ using Eq.~\eqref{eq:scalar_radial_IRG_2}, though $\mathcal{S}_{T_{\vartheta}}^{(1,1)}$ and $\tilde{\mathcal{S}}_{T_{\vartheta}}^{(1,1)}$ are driven by the scalar field perturbation $\vartheta^{(1,1)}$. Similarly, using Eqs.~\eqref{eq:scalar_conjugate} and \eqref{eq:O_Q_conjugate}, one can obtain
\begin{equation} \label{eq:A_i_-m}
    \bar{\mathcal{A}}_i^{\ell -m}(r)
    =(-1)^{m+1}\left[i\mathcal{O}^{\ell m}_i\Theta_{\ell m}^{(1,1)}(r)
    +\alpha_{\ell m}\mathcal{Q}^{\ell m}_i{}_{2}\hat{R}_{\ell m}(r)\right]\,,
\end{equation}
where $\mathcal{A}_i^{\ell -m}(r)$ comes from the terms generated by the $(\ell,-m)$ mode of the solution ansatz in Eq.~\eqref{eq:solution_ansatz_01}.

To analyze the structure of isospectrality breaking, we need to rewrite all the terms in Eq.~\eqref{eq:A_i_m} in terms of ${}_{2}\hat{R}_{\ell m}(r)$ such that we can focus on the vector space of ${}_{2}\hat{R}_{\ell m}(r)$ following \cite{Li:2023ulk}. Denoting the operator acting on $\Theta_{\ell m}^{(1,1)}(r)$ on the left-hand side of Eq.~\eqref{eq:scalar_radial_IRG_2} as $\mathcal{H}_{\vartheta}$, so
\begin{align}
    \Theta_{\ell m}^{(1,1)}(r)
    =& \;\alpha_{\ell m}\mathcal{H}_{\mathcal{\vartheta}}^{-1}
    \left[V^R_{\ell m}(r)+V_{\ell m}^{\square}(r)\right] \nonumber\\
    \equiv& \;i\alpha_{\ell m}\mathcal{H}_{\mathcal{\vartheta}}^{-1}
    \mathcal{V}^{\ell m}{}_{2}\hat{R}_{\ell m}(r)\,,
\end{align}
where we have defined an operator $\mathcal{V}^{\ell m}$ according to that $V_{\ell m}^R(r)$ and $\mathcal{V}_{\ell m}^\square(r)$ are driven by ${}_{2}\hat{R}_{\ell m}(r)$ in Eq.~\eqref{eq:scalar_source_IRG_2}. From Eq.~\eqref{eq:scalar_source_IRG_2}, we observe that $\mathcal{V}^{\ell m}$ contains up to the first derivative in $r$. We have also extracted a factor of $i$ such that 
\begin{equation} \label{eq:calV_conjugate}
    \bar{\mathcal{V}}^{\ell -m}=\mathcal{V}^{\ell m}\,.
\end{equation}
The operator $\mathcal{H}_{\vartheta}$ also satisfies the same symmetry in Eq.~\eqref{eq:calV_conjugate}. Thus, we can rewrite Eqs.~\eqref{eq:A_i_m} and \eqref{eq:A_i_-m} as
\begin{subequations} \label{eq:A_i_m_simplified}
\begin{align} 
    \mathcal{A}_i^{\ell m}(r)
    =& \;\alpha_{\ell m}
    \hat{A}_{i}^{\ell m}{}_{2}\hat{R}_{\ell m}(r)\,, \\
    \bar{\mathcal{A}}_i^{\ell -m}(r)
    =& \;(-1)^{m+1}\alpha_{\ell m}
    \hat{A}_{i}^{\ell m}{}_{2}\hat{R}_{\ell m}(r)\,,
\end{align}
\end{subequations}
with
\begin{equation} \label{eq:hat_A_i}
    \hat{A}_{i}^{\ell m}
    \equiv\mathcal{Q}^{\ell m}_i-\mathcal{O}^{\ell m}_i
    \mathcal{H}^{-1}_{\vartheta}\mathcal{V}^{\ell m}\,.
\end{equation}
Using Eqs.~\eqref{eq:O_Q_conjugate}, \eqref{eq:calV_conjugate}, and \eqref{eq:hat_A_i}, we obtain
\begin{equation} \label{eq:A_i_P_transform}
    \hat{\mathcal{P}}\hat{A}_{i}^{\ell m}
    =\bar{\hat{A}}_{i}^{\ell m}
    =\hat{A}_{i}^{\ell -m}\,.
\end{equation}

Leveraging Eqs.~\eqref{eq:freq_symmetry}, \eqref{eq:angular_transformation}, and \eqref{eq:A_i_m_simplified}, we can rewrite $\mathcal{S}_{T_{\vartheta}}^{(1,1)}$ and $\tilde{\mathcal{S}}_{T_{\vartheta}}^{(1,1)}$ in Eq.~\eqref{eq:SAset} as
\begin{align} \label{eq:S_A_simplify}
    \mathcal{S}_{T_{\vartheta}}^{(1,1)}
    =& \;\tilde{\mathcal{S}}_{T_{\vartheta}}^{(1,1)}(m\rightarrow-m) \nonumber\\
    =& \;(1-\bar{\eta}_{\ell m})e^{-i\omega_{\ell m}t}
    \mathcal{O}_{T_{\vartheta}}^{\ell m}{}_{2}R_{\ell m}^{(0,1)}(r)\,,
\end{align}
such that
\begin{align}
    \mathcal{O}_{T_{\vartheta}}^{\ell m}
    =& \;\left[{}_2Y_{\ell m}(\theta,\phi)\hat{A}_1^{\ell m}
    +i\chi\sin\theta\,{}_1Y_{\ell m}(\theta,\phi)
    \hat{A}_2^{\ell m}\right. \nonumber\\
    & \;\left.+i\chi\cos\theta\,{}_2Y_{\ell m}(\theta,\phi)\hat{A}_3^{\ell m}\right]\mathcal{D}^{\dagger}_{\ell m}\,.
\end{align}
It is not surprising that $\mathcal{S}_{T_{\vartheta}}^{(1,1)}=\tilde{\mathcal{S}}_{T_{\vartheta}}^{(1,1)}(m\rightarrow-m)$ since $\vartheta$ is real, and $\bar{\Theta}_{\ell -m}^{(1,1)}(r)$ is related to $\Theta_{\ell m}^{(1,1)}(r)$ via Eq.~\eqref{eq:scalar_conjugate}. Since the pairs $\left\{\Theta_{\ell m}^{(1,1)}(r),\mathcal{S}_{T_{\vartheta}}^{(1,1)}\right\}$ and $\left\{\Theta_{\ell -m}^{(1,1)}(r),\tilde{\mathcal{S}}_{T_{\vartheta}}^{(1,1)}\right\}$ contain the same information, we can just solve the radial modified Teukolsky equation driven by $\mathcal{S}_{T_{\vartheta}}^{(1,1)}$ jointly with the equation of $\Theta_{\ell m}^{(1,1)}(r)$. Using Eqs.~\eqref{eq:conjugate_R}, \eqref{eq:parity_Teuk_func}, and \eqref{eq:A_i_P_transform}, we can show that
\begin{equation} \label{eq:O_A_P_transform}
    \hat{\mathcal{P}}\left[\mathcal{O}_{T_{\vartheta}}^{\ell m}
    {}_{2}R^{(0,1)}_{\ell m}(r)\right]
    =(-1)^{\ell}\mathcal{O}_{T_{\vartheta}}^{\ell-m}
    {}_{2}R^{(0,1)}_{\ell -m}(r)\,.
\end{equation}

\subsection{The $\mathcal{S}_{T_{\Psi}}^{(1,1)}$ and $\tilde{\mathcal{S}}_{T_{\Psi}}^{(1,1)}$ source terms}
\label{sec:sB_simplify}

Lastly, in \cite{Wagle:2023fwl}, we find that the source terms $\mathcal{S}_{T_{\Psi}}^{(1,1)}$ and $\tilde{\mathcal{S}}_{T_{\Psi}}^{(1,1)}$ driven by $\Psi_0^{(0,1)}$ in Eq.~\eqref{eq:solution_ansatz_01} can be written as
\begin{widetext}
\begin{subequations}\label{eq:SBset}
\begin{align} 
    \mathcal{S}_{T_{\Psi}}^{(1,1)} 
    =& \;i\chi e^{-i\omega_{\ell m} t}
    \left[\mathcal{B}_{1}^{\ell m}(r)\sin\theta\,{}_1Y_{\ell m}(\theta,\phi)
    +\mathcal{B}_{2}^{\ell m}(r)\cos\theta\,{}_2Y_{\ell m}(\theta,\phi)
    +\mathcal{B}_{3}^{\ell m}(r)\sin\theta\,{}_3Y_{\ell m}(\theta,\phi)\right]+\eta_{\ell m}\times(m\rightarrow-m)\,, 
    \label{eq:sB_coord} \\ 
    \tilde{\mathcal{S}}_{T_{\Psi}}^{(1,1)} 
    =& \;i\chi e^{+i\bar{\omega}_{\ell m}t}
    \left[\tilde{\mathcal{B}}_{1}^{\ell m}(r)
    \sin\theta\,{}_{-1}\bar{Y}_{\ell m}(\theta,\phi)
    +\tilde{\mathcal{B}}_{2}^{\ell m}(r)
    \cos\theta\,{}_{-2}\bar{Y}_{\ell m}(\theta,\phi)\right] +\bar{\eta}_{\ell m}\times(m\rightarrow-m)\,, 
    \label{eq:sBc_coord}
\end{align}
\end{subequations}
\end{widetext}
where we have extracted an overall factor of $i$ from all the radial functions $\mathcal{B}_i^{\ell m}(r)$ and $\tilde{\mathcal{B}}_i^{\ell m}(r)$, such that
\begin{equation} \label{eq:B_i_conjugate}
    \bar{\mathcal{B}}_i^{\ell -m}(r)
    =\mathcal{B}_i^{\ell m}(r)\,,\quad
    \bar{\tilde{\mathcal{B}}}_i^{\ell -m}(r)=\tilde{\mathcal{B}}_{i}^{\ell m}(r)\,.
\end{equation}
The functions $\mathcal{B}_i^{\ell m}(r)$ and $\tilde{\mathcal{B}}_i^{\ell m}(r)$ contain up to the first derivative in $r$ of ${}_{2}\hat{R}_{\ell m}(r)$ and ${}_{2}\bar{\hat{R}}_{\ell m}(r)$, respectively. Following Sec.~\ref{sec:sA_simplify}, let us define
\begin{equation}
    \mathcal{B}_i^{\ell m}(r)
    \equiv \hat{B}_i^{\ell m}{}_{2}\hat{R}_{\ell m}(r)\,,\quad
    \tilde{\mathcal{B}}_i^{\ell m}
    \equiv \hat{\tilde{B}}_i^{\ell m}{}_{2}\bar{\hat{R}}_{\ell m}(r)\,,
\end{equation}
where $\hat{B}_i^{\ell m}$ and $\hat{\tilde{B}}_i^{\ell m}$ contain up to the first derivative in $r$ and satisfy the same symmetry as Eq.~\eqref{eq:B_i_conjugate}, so
\begin{equation} \label{eq:B_i_P_transform}
    \hat{\mathcal{P}}\hat{B}_{i}^{\ell m}
    =\hat{B}_{i}^{\ell -m}\,,\quad
    \hat{\mathcal{P}}\hat{\tilde{B}}_{i}^{\ell m}
    =\hat{\tilde{B}}_{i}^{\ell -m}\,.
\end{equation}
Thus, we can rewrite Eq.~\eqref{eq:SBset} as
\begin{subequations} \label{eq:S_B_simplify}
\begin{align} 
    & \mathcal{S}_{T_{\Psi}}^{(1,1)}
    =e^{-i\omega_{\ell m}t}\mathcal{O}_{T_{\Psi}}^{\ell m}
    {}_{2}R^{(0,1)}_{\ell m}(r)
    +\eta_{\ell m}\times(m\rightarrow-m)\,, \\
    & \tilde{\mathcal{S}}_{T_{\Psi}}^{(1,1)}
    =\bar{\eta}_{\ell m}e^{-i\omega_{\ell m}t}
    \tilde{\mathcal{O}}_{T_{\Psi}}^{\ell m}
    {}_{2}R^{(0,1)}_{\ell m}(r)+(m\rightarrow-m)\,,
\end{align}    
\end{subequations}
with
\begin{subequations}
\begin{align}
    \mathcal{O}_{T_{\Psi}}^{\ell m}
    =& \;i\chi\left[\sin{\theta}{}_1Y_{\ell m}(\theta,\phi)\hat{B}_1^{\ell m}
    +\cos\theta\,{}_2Y_{\ell m}(\theta,\phi)
    \hat{B}_2^{\ell m} \right.\nonumber \\
    & \;\left.+\sin\theta\,{}_3Y_{\ell m}(\theta,\phi)
    \hat{B}_3^{\ell m}\right]\mathcal{D}^{\dagger}_{\ell m}\,, \\
    \tilde{\mathcal{O}}_{T_{\Psi}}^{\ell m}
    =& \;i\chi\left[-\sin{\theta}{}_1Y_{\ell m}(\theta,\phi)
    \hat{\tilde{B}}_1^{\ell -m}
    +\cos\theta\,{}_2Y_{\ell m}(\theta,\phi)
    \hat{\tilde{B}}_2^{\ell -m}\right] \nonumber\\
    & \;\mathcal{D}^{\dagger}_{\ell m}\,,
\end{align}
\end{subequations}
where we have used Eqs.~\eqref{eq:conjugate_R} and \eqref{eq:angular_transformation} to replace $\left\{{}_{s}\bar{\hat{R}}_{\ell -m}(r),{}_{-s}\bar{Y}_{\ell -m}(\theta,\phi)\right\}$ with $\left\{{}_{s}\hat{R}_{\ell m}(r),{}_{s}Y_{\ell m}(\theta,\phi)\right\}$ in Eq.~\eqref{eq:S_B_simplify} and that $\omega_{\ell m}=-\bar{\omega}_{\ell -m}$. With Eqs.~\eqref{eq:parity_Teuk_func} and \eqref{eq:B_i_P_transform}, we can show that
\begin{subequations} \label{eq:O_B_P_transform}
\begin{align}
    & \hat{\mathcal{P}}\left[\mathcal{O}_{T_{\Psi}}^{\ell m}
    {}_{2}R^{(0,1)}_{\ell m}(r)\right]
    =(-1)^{\ell}\mathcal{O}_{T_{\Psi}}^{\ell -m}
    {}_{2}R^{(0,1)}_{\ell -m}(r)\,,\\
    & \hat{\mathcal{P}}\left[\tilde{\mathcal{O}}_{T_{\Psi}}^{\ell m}
    {}_{2}R^{(0,1)}_{\ell m}(r)\right]
    =(-1)^{\ell}\tilde{\mathcal{O}}_{T_{\Psi}}^{\ell -m}
    {}_{2}R^{(0,1)}_{\ell -m}(r)\,.
\end{align}
\end{subequations}

\subsection{Simplification of the radial master equations}
\label{sec:simplify_radial_eqns}

In this subsection, we will use the results in Secs.~\ref{sec:sGeo_simplify}--\ref{sec:sB_simplify} to simplify the modified Teukolsky equation of $\Psi_0^{(1,1)}$ in Eq.~\eqref{eq:master_eqn_dCS_Psi0} and extract its radial part. Combining Eqs.~\eqref{eq:S_geo_simplify}, \eqref{eq:S_A_simplify}, and \eqref{eq:S_B_simplify}, we can reduce Eq.~\eqref{eq:master_eqn_dCS_Psi0} to
\begin{widetext}
\begin{subequations} \label{eq:master_eqn_Psi0_simplify}
\begin{align}
    & H_{0}^{\ell m(0,0)}\left[{}_{2}R_{\ell m}^{(1,1)}(r)
    {}_{2}Y_{\ell m}(\theta,\phi)\right]
    =2r^2\left[\left(\mathcal{O}_{\geo}^{\ell m}
    +\mathcal{O}_{T_{\vartheta}}^{\ell m}+\mathcal{O}_{T_{\Psi}}^{\ell m}\right)
    -\bar{\eta}_{\ell m}\left(\mathcal{O}_{T_{\vartheta}}^{\ell m}
    -\tilde{\mathcal{O}}_{T_{\Psi}}^{\ell m}\right)\right]
    {}_{2}R_{\ell m}^{(0,1)}(r)\,,
    \label{eq:master_eqn_Psi0_lm_simplify} \\
    & \eta_{\ell m}H_{0}^{\ell -m(0,0)}\left[{}_{2}R_{\ell -m}^{(1,1)}(r)
    {}_{2}Y_{\ell -m}(\theta,\phi)\right]
    =2r^2\left[\eta_{\ell m}\left(\mathcal{O}_{\geo}^{\ell -m}
    +\mathcal{O}_{T_{\vartheta}}^{\ell -m}+\mathcal{O}_{T_{\Psi}}^{\ell -m}\right)
    -\left(\mathcal{O}_{T_{\vartheta}}^{\ell -m}
    -\tilde{\mathcal{O}}_{T_{\Psi}}^{\ell -m}\right)\right]
    {}_{2}R_{\ell -m}^{(0,1)}(r)\,.
    \label{eq:master_eqn_Psi0_l-m_simplify}
\end{align}
\end{subequations}
One can perform a further $\hat{\mathcal{P}}$ transformation on Eq.~\eqref{eq:master_eqn_Psi0_l-m_simplify} and use Eqs.~\eqref{eq:O_geo_P_transform}, \eqref{eq:O_A_P_transform}, and \eqref{eq:O_B_P_transform}, so Eq.~\eqref{eq:master_eqn_Psi0_l-m_simplify} becomes
\begin{equation} \label{eq:master_eqn_Psi0_l-m_P_simplify}
    \bar{\eta}_{\ell m}H_{0}^{\ell m(0,0)}\left[{}_{2}R_{\ell m}^{(1,1)}(r)
    {}_{2}Y_{\ell m}(\theta,\phi)\right]
    =2r^2\left[\bar{\eta}_{\ell m}\left(\mathcal{O}_{\geo}^{\ell m}
    +\mathcal{O}_{T_{\vartheta}}^{\ell m}+\mathcal{O}_{T_{\Psi}}^{\ell m}\right)
    -\left(\mathcal{O}_{T_{\vartheta}}^{\ell m}
    -\tilde{\mathcal{O}}_{T_{\Psi}}^{\ell m}\right)\right]
    {}_{2}R_{\ell m}^{(0,1)}(r)\,,
\end{equation}
\end{widetext}
where we have chosen the normalization that ${}_{2}\bar{R}_{\ell -m}^{(1,1)}(r)=(-1)^m{}_{2}R_{\ell m}^{(1,1)}(r)$. Comparing Eqs.~\eqref{eq:master_eqn_Psi0_lm_simplify} and \eqref{eq:master_eqn_Psi0_l-m_P_simplify}, we notice that only $\eta_{\ell m}=\pm 1$ can make these two equations consistent. As found in \cite{Li:2023ulk}, $\eta_{\ell m}=\pm 1$ correspond to even- and odd-parity metric perturbations, respectively. Furthermore, for even-parity modes ($\eta_{\ell m}=1$), the terms driven by the scalar field perturbation $\vartheta^{(1,1)}$ (i.e., terms with $\mathcal{O}_{T_{\vartheta}}^{\ell m}$), cancel out exactly. This is due to the fact that the scalar field equation is not driven by the even-parity metric perturbation, as shown in Sec.~\ref{sec:scalar_eqn}. In total, our modified Teukolsky equation has the same structure of isospectrality breaking as the modified RW and ZM equations in \cite{Wagle:2021tam}, where the even- and odd-parity modes decouple, and the scalar field couples to the odd-parity modes only. For this reason, the $(\ell,m)$ and $(\ell,-m)$ modes of the modified Teukolsky equation contain redundant information, as shown in Eqs.~\eqref{eq:master_eqn_Psi0_lm_simplify} and \eqref{eq:master_eqn_Psi0_l-m_P_simplify} for dCS gravity and in \cite{Li:2023ulk} more generally. Thus, we only need to solve the $(\ell,m)$ mode of the modified Teukolsky equation [Eq.~\eqref{eq:master_eqn_Psi0_lm_simplify}] with the scalar field equation [Eq.~\eqref{eq:scalar_radial_IRG_2_lm}].

Now, let us follow the same procedures in \cite{Wagle:2023fwl} to reduce the simplified modified Teukolsky equation in Eq.~\eqref{eq:master_eqn_Psi0_lm_simplify} into a purely radial equation and express it in terms of the radial functions defined in \cite{Wagle:2023fwl}. Integrating Eq.~\eqref{eq:master_eqn_Psi0_lm_simplify} over ${}_{2}\mathcal{Y}_{\ell m}(\theta,\phi)$, we obtain
\begin{widetext}
\begin{equation} \label{eq:master_eqn_Psi0_radial_simplify} 
\begin{aligned}
    & \left[r(r-r_s)\partial^2_{r}+6(r-M)\partial_r
    +\frac{4i\omega_{\ell m} r(r-3M)+\omega_{\ell m}^{2}r^{3}}{r-r_s}
    +\frac{4i\chi mM((r-M)+iM\omega_{\ell m}r)}{r(r-r_s)}
    -{}_{2}A_{\ell m}\right]{}_{2}R_{\ell m}^{(1,1)}(r) \\
    & =-2r^2\left[\left(\mathscr{O}_{\geo}^{\ell m}
    +\mathscr{O}_{T_{\vartheta}}^{\ell m}+\mathscr{O}_{T_{\Psi}}^{\ell m}\right)
    \mp\left(\mathscr{O}_{T_{\vartheta}}^{\ell m}
    -\tilde{\mathscr{O}}_{T_{\Psi}}^{\ell m}\right)\right]
    {}_{2}R_{\ell m}^{(0,1)}(r)\,,
\end{aligned}
\end{equation}
where ${}_{2}A_{\ell m}$ is given in Eq.~\eqref{eq:A_lm}, the $\pm$ sign correspond to the solutions with $\eta_{\ell m}=\pm1$, respectively, and
\begin{subequations} \label{eq:source_simplified_IRG}
\begin{align}
    \mathscr{O}_{\geo}^{\ell m}
    =& \;-\frac{i\chi mM^4}{448r^9(r-r_s)}
    \left(C_1(r)+4i\omega_{\ell m} r^2C_2(r)\right)
    +\frac{i\chi mM^4}{16r^9}\left[C_3(r)-C_4(r)
    \left(\frac{i\omega_{\ell m} r^2}{2}+\frac{r(r-r_s)}{2}\partial_r\right)\right]
    \Lambda^{\ell\ell 1}_{22c} \nonumber\\ 
    & \;+\frac{i\chi mM^4}{128r^9}C_5(r)
    \left(\sqrt{(\ell+2)(\ell-1)}\Lambda^{\ell\ell 1}_{12s}
    -\sqrt{(\ell+3)(\ell-2)}\Lambda^{\ell\ell 1}_{32s}\right)\,, 
    \label{eq:O_geo} \\
    \mathscr{O}_{T_{\vartheta}}^{\ell m}
    =& \;\left[\hat{A}_1^{\ell m}
    +i\chi m\Lambda^{\ell\ell 1}_{12s}\hat{A}_2^{\ell m}
    +i\chi m\Lambda^{\ell\ell 1}_{22c}\hat{A}_3^{\ell m}\right]
    \mathcal{D}^{\dagger}_{\ell m}\,, \label{eq:O_A} \\
    \mathscr{O}_{T_{\Psi}}^{\ell m}
    =& \;i\chi m\left[\Lambda^{\ell\ell 1}_{12s}\hat{B}_1^{\ell m}
    +\Lambda^{\ell\ell 1}_{22c}\hat{B}_2^{\ell m}
    +\Lambda^{\ell\ell 1}_{32s}\hat{B}_3^{\ell m}\right]
    \mathcal{D}^{\dagger}_{\ell m}\,, \label{eq:O_B} \\
    \tilde{\mathscr{O}}_{T_{\Psi}}^{\ell m}
    =& \;-i\chi m\left[\Lambda^{\ell\ell 1}_{12s}\hat{\tilde{B}}_1^{\ell-m}
    -\Lambda^{\ell\ell 1}_{22c}\hat{\tilde{B}}_2^{\ell-m}\right]
    \mathcal{D}^{\dagger}_{\ell m}\,, \label{eq:O_B_tilde}
\end{align}
\end{subequations}
\end{widetext}
where, $\mathcal{D}^{\dagger}_{\ell m}$ is defined in Eq.~\eqref{eq:Ddagger} and we have used 
\begin{align}
    & \Lambda^{\ell\ell m}_{12s}
    =m\Lambda^{\ell\ell 1}_{12s}\,,\quad
    \Lambda^{\dagger\ell\ell m}_{-12s}
    =(-1)^{m+1}m\Lambda^{\ell\ell 1}_{12s}\,,\nonumber\\
    & \Lambda^{\ell\ell m}_{22c}
    =m\Lambda^{\ell\ell 1}_{22c}\,,\quad
    \Lambda^{\dagger\ell\ell\ell m}_{-22c}
    =(-1)^m m\Lambda^{\ell\ell 1}_{22c}\,,\nonumber\\
    & \Lambda^{\ell\ell m}_{32s}
    =m\Lambda^{\ell\ell 1}_{32s}\,,\quad
    \Lambda^{\dagger\ell\ell\ell m}_{-32s}
    =(-1)^{m+1}m\Lambda^{\ell\ell 1}_{32s}\,,
\end{align}
which can be derived from Eqs.~\eqref{eq:relation_12} and \eqref{eq:relation_56} in Appendix~\ref{appendix:projection_coeff}. The radial functions $C_i(r)$ are given in \cite{Wagle:2023fwl}, and the radial operators $\{\hat{A}_i^{\ell m},\hat{B}_i^{\ell m},\hat{\tilde{B}}_i^{\ell m}\}$ are given explicitly in Appendix~\ref{appendix:radial_operators}.

All the source terms in Eq.~\eqref{eq:source_simplified_IRG} are proportional to $\chi m$ except $\hat{A}_1^{\ell m}$, so they are evaluated on a Schwarzschild background. The operator $\hat{A}_1^{\ell m}$ contains terms at both $\mathcal{O}(\chi^0)$ and $\mathcal{O}(\chi^1)$, the latter of which only depends on the combination $\chi m$. Thus, the dCS correction $\omega_{\ell m}^{(1,0)}$ to the QNM frequency of a slowly-rotating BH should be expanded as
\begin{equation}
    \omega_{\ell m}^{(1,0)}
    =\omega_{\ell m}^{(1,0,0)}+\chi m\omega_{\ell 1}^{(1,1,0)}
    +\mathcal{O}(\chi^2)\,,
\end{equation}
consistent with the result in \cite{Wagle:2021tam}. Although the master equations in \cite{Wagle:2021tam} contain terms proportional to $\chi$ only, these terms do not contribute to the QNM frequencies since the equations are invariant under the simultaneous transformation of $\chi\rightarrow-\chi$ and $m\rightarrow-m$, as shown in more detail in \cite{Wagle:2021tam}. Furthermore, only after removing these terms do the even- and odd-parity decouple in \cite{Wagle:2021tam}. In contrast, all the source terms at $\mathcal{O}(\chi^1)$ here are proportional to $\chi m$, so we do not need to drop any term manually; they are naturally absent. The even- and odd-parity modes also naturally decouple in our case.

\begin{widetext}
Similarly, in the ORG, we obtain the following equation for $\Psi_4^{(1,1)}$:
\begin{equation} \label{eq:master_eqn_Psi4_radial_simplify} 
\begin{aligned}
    & \left[r(r-r_s)\partial^2_{r}-2(r-M)\partial_r
    -\frac{4i\omega_{\ell m} r(r-3M)-\omega_{\ell m}^{2}r^{3}}{r-r_s}
    -\frac{4i\chi mM((r-M)-iM\omega_{\ell m}r)}{r(r-r_s)}
    -{}_{-2}A_{\ell m}\right]{}_{-2}R_{\ell m}^{(1,1)}(r) \\
    & =-2r^6\left[\left(\mathscr{Q}_{\geo}^{\ell m}
    +\mathscr{Q}_{T_{\vartheta}}^{\ell m}+\mathscr{Q}_{T_{\Psi}}^{\ell m}\right)
    \mp\left(\mathscr{Q}_{T_{\vartheta}}^{\ell m}
    -\tilde{\mathscr{Q}}_{T_{\Psi}}^{\ell m}\right)\right]
    {}_{-2}R_{\ell m}^{(0,1)}(r)\,,
\end{aligned}
\end{equation}
where ${}_{-2}A_{\ell m}$ is given in Eq.~\eqref{eq:A_lm} and
\begin{subequations} \label{eq:source_simplified_ORG}
\begin{align}
    \mathscr{Q}_{\geo}^{\ell m}
    =& \;\frac{i\chi mM^4}{448r^{13}(r-r_s)}
    \left(D_1(r)-4i\omega_{\ell m} r^2D_2(r)\right)
    +\frac{i\chi mM^4}{16r^{13}}\left[D_3(r)-D_4(r)
    \left(\frac{i\omega_{\ell m} r^2}{2}-\frac{r(r-r_s)}{2}\partial_r\right)\right]
    \Lambda^{\ell\ell 1}_{22c} \nonumber\\ 
    & \;-\frac{i\chi mM^4}{128r^{13}}D_5(r)
    \left(\sqrt{(\ell+2)(\ell-1)}\Lambda^{\ell\ell 1}_{12s}
    -\sqrt{(\ell+3)(\ell-2)}\Lambda^{\ell\ell 1}_{32s}\right)\,, 
    \label{eq:Q_geo} \\
    \mathscr{Q}_{T_{\vartheta}}^{\ell m}
    =& \;\left[\hat{\mathscr{A}}_1^{\ell m}
    +i\chi m\Lambda^{\ell\ell 1}_{12s}\hat{\mathscr{A}}_2^{\ell m}
    -i\chi m\Lambda^{\ell\ell 1}_{22c}\hat{\mathscr{A}}_3^{\ell m}\right]
    \mathcal{D}_{\ell m}\,, \label{eq:Q_A} \\
    \mathscr{Q}_{T_{\Psi}}^{\ell m}
    =& \;i\chi m\left[\Lambda^{\ell\ell 1}_{12s}\hat{\mathscr{B}}_1^{\ell m}
    -\Lambda^{\ell\ell 1}_{22c}\hat{\mathscr{B}}_2^{\ell m}
    +\Lambda^{\ell\ell 1}_{32s}\hat{\mathscr{B}}_3^{\ell m}\right]
    \mathcal{D}_{\ell m}\,, \label{eq:Q_B} \\
    \tilde{\mathscr{Q}}_{T_{\Psi}}^{\ell m}
    =& \;-i\chi m\left[\Lambda^{\ell\ell 1}_{12s}\hat{\tilde{\mathscr{B}}}_1^{\ell-m}
    +\Lambda^{\ell\ell 1}_{22c}\hat{\tilde{\mathscr{B}}}_2^{\ell-m}\right]
    \mathcal{D}_{\ell m}\,, \label{eq:Q_B_tilde}
\end{align}
\end{subequations}
\end{widetext}
where $\mathcal{D}_{\ell m}$ is defined in Eq.~\eqref{eq:Ddagger}, and we have used 
\begin{align}
    & \Lambda^{\ell\ell m}_{-1-2s}
    =m\Lambda^{\ell\ell 1}_{12s}\,,\quad
    && \Lambda^{\dagger\ell\ell m}_{1-2s}
    =(-1)^{m+1}m\Lambda^{\ell\ell 1}_{12s}\,,\nonumber\\
    & \Lambda^{\ell\ell m}_{-2-2c}
    =-m\Lambda^{\ell\ell 1}_{22c}\,,\quad
    && \Lambda^{\dagger\ell\ell\ell m}_{2-2c}
    =(-1)^{m+1}m\Lambda^{\ell\ell 1}_{22c}\,,\nonumber\\
    & \Lambda^{\ell\ell m}_{-3-2s}
    =m\Lambda^{\ell\ell 1}_{32s}\,,\quad
    && \Lambda^{\dagger\ell\ell\ell m}_{3-2s}
    =(-1)^{m+1}m\Lambda^{\ell\ell 1}_{32s}\,,
\end{align}
which can be derived from Eqs.~\eqref{eq:relation_12}, \eqref{eq:relation_34}, and \eqref{eq:relation_56} in Appendix~\ref{appendix:projection_coeff}. The radial functions $D_i(r)$ are given in \cite{Wagle:2023fwl}, and the radial operators $\{\hat{\mathscr{A}}_i^{\ell m},\hat{\mathscr{B}}_i^{\ell m},\hat{\tilde{\mathscr{B}}}_i^{\ell m}\}$ are given explicitly in Appendix~\ref{appendix:radial_operators}. Since Eqs.~\eqref{eq:scalar_radial_ORG_lm} and \eqref{eq:master_eqn_Psi4_radial_simplify} are derived in a different gauge from the one used by Eqs.~\eqref{eq:scalar_radial_IRG_2_lm} and \eqref{eq:master_eqn_Psi0_radial_simplify}, we will compute the QNM frequency shifts from both pairs and use the results from Eqs.~\eqref{eq:scalar_radial_ORG_lm} and \eqref{eq:master_eqn_Psi4_radial_simplify} as a consistency check of the results from Eqs.~\eqref{eq:scalar_radial_IRG_2_lm} and \eqref{eq:master_eqn_Psi0_radial_simplify} in the next section.

\section{Calculation of the QNM frequency shifts}
\label{sec:QNM_freq}

In this section, we will calculate the QNM frequencies of a slowly-rotating BH in dCS gravity up to $\mathcal{O}(\zeta^1,\chi^1,\epsilon^1)$ using Eqs.~\eqref{eq:scalar_radial_IRG_2_lm}, \eqref{eq:scalar_radial_ORG_lm}, \eqref{eq:master_eqn_Psi0_radial_simplify} and \eqref{eq:master_eqn_Psi4_radial_simplify}. We will first review the EVP method in \cite{Zimmerman:2014aha, Mark:2014aja, Hussain:2022ins, Li:2023ulk} and show how to apply it to the case of dCS gravity. We will then present the QNM spectra of non-rotating and slowly-rotating BHs in dCS gravity.

\subsection{The EVP method}
\label{sec:EVP}

To compute the QNM frequency shifts $\omega_{\ell m}^{(1,0)}$, we choose to follow the EVP approach developed in \cite{Zimmerman:2014aha, Mark:2014aja, Hussain:2022ins, Li:2023ulk}. As one can notice in Eqs.~\eqref{eq:master_eqn_Psi0_radial_simplify} and \eqref{eq:master_eqn_Psi4_radial_simplify}, the solutions to the homogeneous part of the equation (i.e., ${}_{\pm2}R_{\ell m}^{(0,1)}(r)$), or the Teukolsky equation in GR, also drive the source terms, potentially leading to secularly-growing solutions. To avoid this issue, Refs.~\cite{Zimmerman:2014aha, Mark:2014aja} developed the EVP method, following the Poincar\'{e}-Lindstedt method of solving the secular perturbation problem, by introducing an additional expansion in the QNM frequency to cancel off secularly growing terms. More specifically, consider a system of the form
\begin{subequations} \label{eq:EVP_eqn}
\begin{align} 
    & {}_{s}\mathcal{H}_{\ell m}^{(0,0)}{}_{s}R_{\ell m}^{(0,1)}=0\,,
    \label{eq:EVP_eqn_01} \\
    & {}_{s}\mathcal{H}_{\ell m}^{(0,0)}{}_{s}R_{\ell m}^{(1,1)}
    ={}_{s}\mathscr{V}_{\ell m}^{(1,0)}{}_{s}R_{\ell m}^{(0,1)}\,,
    \label{eq:EVP_eqn_11}
\end{align}  
\end{subequations}
where ${}_{s}\mathcal{H}_{\ell m}^{(0,0)}$ is the radial Teukolsky operator for a spin-$s$ field in GR, and ${}_{s}\mathscr{V}_{\ell m}^{(1,0)}$ is some differential operator containing up to first derivative in $r$. One can expand the QNM frequency $\omega_{\ell m}$ associated with ${}_{s}R_{\ell m}^{(0,1)}$ and ${}_{s}R_{\ell m}^{(1,1)}$ as
\begin{align} \label{eq:QNM_slow_expansion}
    \omega_{\ell m}
    =& \;\omega_{\ell m}^{(0,0)}+\zeta\omega_{\ell m}^{(1,0)}
    +\mathcal{O}(\zeta^2) \nonumber\\
    =& \;\left(\omega_{\ell m}^{(0,0,0)}
    +\chi m\omega_{\ell 1}^{(0,1,0)}\right)
    +\zeta\left(\omega_{\ell m}^{(1,0,0)}
    +\chi m\omega_{\ell 1}^{(1,1,0)}\right) \nonumber\\
    & \;+\mathcal{O}(\zeta^2,\chi^2)\,,
\end{align}
where we include an additional expansion in $\chi$ in the second line, considering the slow-rotation approximation used in \cite{Wagle:2023fwl} and this work. As discussed in Sec.~\ref{sec:simplify_radial_eqns}, the $\mathcal{O}(\chi^1)$ terms in the modified Teukolsky equation depend linearly on $m$, so we extract a factor of $m$ from $\omega_{\ell m}^{(0,1,0)}$ and $\omega_{\ell m}^{(1,1,0)}$ by expressing $\omega_{\ell m}^{(0,1,0)}=m\omega_{\ell 1}^{(0,1,0)}$ and $\omega_{\ell m}^{(1,1,0)}=m\omega_{\ell 1}^{(1,1,0)}$. Since ${}_{s}\mathcal{H}_{\ell m}^{(0,0)}$ depends on $\omega_{\ell m}$, Eq.~\eqref{eq:EVP_eqn} expands to
\begin{equation} \label{eq:EVP_eqn_expand}
    {}_{s}\mathcal{H}_{\ell m}^{(0,0)}{}_{s}R_{\ell m}^{(1,1)}
    +\omega_{\ell m}^{(1,0)}\partial_\omega
    \left({}_{s}\mathcal{H}_{\ell m}^{(0,0)}\right)
    {}_{s}R_{\ell m}^{(0,1)}
    ={}_{s}\mathscr{V}_{\ell m}^{(1,0)}{}_{s}R_{\ell m}^{(0,1)}\,,
\end{equation}
where all the operators are evaluated at the GR QNM frequency $\omega_{\ell m}^{(0,0)}$. The second term on the left-hand side of Eq.~\eqref{eq:EVP_eqn_expand} comes from the expansion of $\omega_{\ell m}$ in Eq.~\eqref{eq:EVP_eqn_01}.

\begin{figure}[t]
    \centering
    \includegraphics[width=0.4\textwidth]{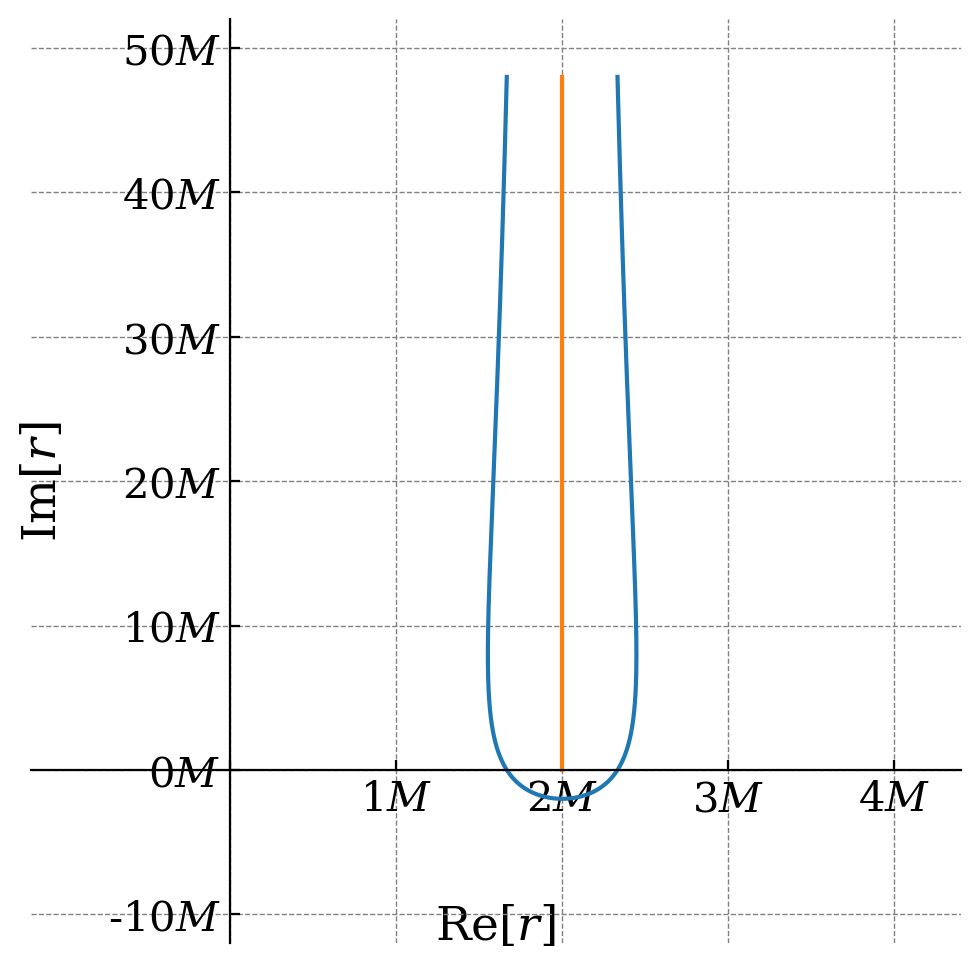}
    \caption{The contour $\mathscr{C}$ in Eq.~\eqref{eq:contour} with $\xi_{\max}=10M$. The blue line is the contour $\mathscr{C}$, while the orange line is the branch cut at $r=2M$. } 
    \label{fig:contour}
\end{figure}

The key step in the EVP method is to construct an inner product such that the Teukolsky operator in GR is self-adjoint, i.e.,
\begin{equation} \label{eq:self_adjoint}
    \left\langle\varphi_1(r)\Big|{}_{s}
    \mathcal{H}_{\ell m}^{(0,0)}\varphi_2(r)\right\rangle
    =\left\langle {}_{s}\mathcal{H}_{\ell m}^{(0,0)}
    \varphi_1(r)\Big|\varphi_2(r)\right\rangle\,,
\end{equation}
where $\varphi_{1,2}(r)$ are radial functions with similar asymptotic behavior as the radial Teukolsky function of spin weight $s$ in GR. This inner product can be defined as a contour integral over complex $r$ \cite{Zimmerman:2014aha, Mark:2014aja},
\begin{equation} \label{eq:inner_product}
    \langle\varphi_{1}(r)|\varphi_{2}(r)\rangle
    =\int_{\mathscr{C}}\Delta^{s}(r)
    \varphi_{1}(r)\varphi_{2}(r) dr\,,
\end{equation}
where the contour $\mathscr{C}$ is around the branch cut along the positive imaginary axis at the outer horizon $r_{+}$. In our case, since $r_{+}=2M$ up to $\mathcal{O}(\chi)$, we parametrize the contour as
\begin{align} \label{eq:contour}
    & r_{\mathscr{C}}(\xi)
    =2M+\frac{4M^2\xi}{20M^2+\xi^2}
    +i\left(\frac{\xi^2}{2M}-2M\right)\,,\nonumber\\
    & \xi\in[-\xi_{\max},\xi_{\max}]
\end{align}
such that $r(\pm\xi_{\max})$ are the right and left ends of the contour, respectively. In Fig.~\ref{fig:contour}, we plot the contour for $\xi_{\max}=10M$. Now taking the inner product of ${}_{s}R_{\ell m}^{(0,1)}$ with Eq.~\eqref{eq:EVP_eqn_expand}, and using Eqs.~\eqref{eq:EVP_eqn_01} and \eqref{eq:self_adjoint}, we find that the first term in Eq.~\eqref{eq:EVP_eqn_expand} vanishes, i.e.,
\begin{equation}
    \left\langle {}_{s}R_{\ell m}^{(0,1)}\Big|
    {}_{s}\mathcal{H}_{\ell m}^{(0,0)}
    {}_{s}R_{\ell m}^{(1,1)}\right\rangle
    =\left\langle {}_{s}\mathcal{H}_{\ell m}^{(0,0)}
    {}_{s}R_{\ell m}^{(0,1)}
    \Big|{}_{s}R_{\ell m}^{(1,1)}\right\rangle
    =0\,.
\end{equation}

Thus, the QNM frequency shift $\omega_{\ell m}^{(1,0)}$ satisfies
\begin{equation} \label{eq:freq_shift_EVP}
    \omega_{\ell m}^{(1,0)}
    =\left\langle{}_{s}\mathscr{V}_{\ell m}^{(1,0)}\right\rangle\Big/
    \left\langle\partial_{\omega}
    \left({}_{s}\mathcal{H}_{\ell m}^{(0,0)}\right)
    \right\rangle\,,
\end{equation}
where we use the simplified notation
\begin{equation} \label{eq:inner_product_simplify}
    \left\langle\hat{\mathcal{O}}\right\rangle
    =\left\langle{}_{s}R_{\ell m}^{(0,1)}\Big|\hat{\mathcal{O}}\,
    {}_{s}R_{\ell m}^{(0,1)}\right\rangle\,.
\end{equation}
Note that all the terms within the inner product are evaluated at the GR frequencies $\omega_{\ell m}^{(0,0)}$.

\begin{figure*}[t]
    \centering
    \includegraphics[width=\linewidth]{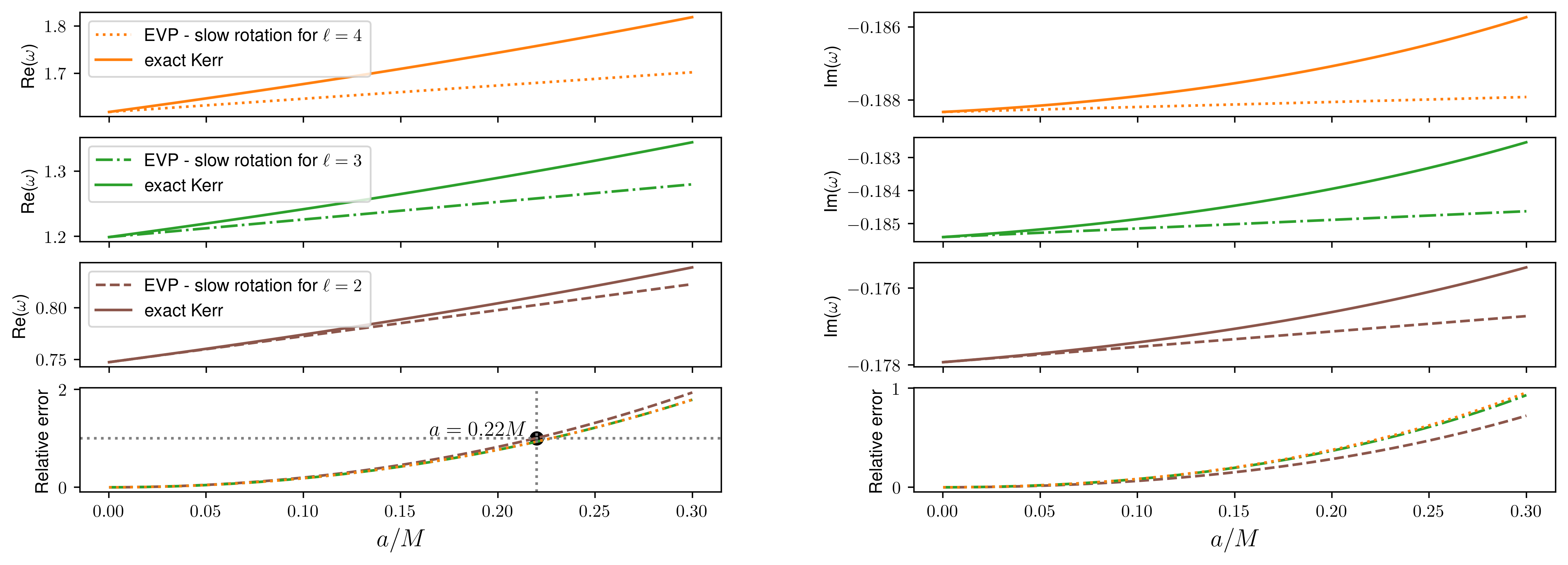}
    \caption{Comparison between the real and the imaginary parts of the QNM frequencies for the modes at $\ell=m=2,3,4$ calculated using the exact Kerr metric with Leaver's method and its expansion to leading order in $\chi$ with the EVP method. The dependence of the real (imaginary) part of the QNM frequency on the spin $a$ is shown in the left (right) top panel. The left (right) bottom panel shows the relative error defined by Eq.~\eqref{eq:relerror} between these two calculations. The legends shown in the left panels apply to the corresponding right panels. We observe that the relative error defined by Eq.~\eqref{eq:relerror} is $1\%$ at $a\approx 0.22M$ for $\textrm{Re}[\omega_{22}]$. We have set $M=1/2$ in this plot.}
    \label{fig:slowroterror}
\end{figure*}

One application of the EVP method is to compute the slow-rotation correction to the Schwarzschild QNM spectrum for a slowly-rotating BH in GR. Taking $\Psi_0$ as an example, we have the following set of equations up to $\mathcal{O}(\chi^1)$:
\begin{subequations}
\begin{align}
    & \mathcal{H}_{0}^{\ell m(0,0,0)}{}_{2}R_{\ell m}^{(0,0,1)}=0\,, \\
    & \mathcal{H}_{0}^{\ell m(0,0,0)}{}_{2}R_{\ell m}^{(0,1,1)}
    =-\mathcal{H}_{0}^{\ell m(0,1,0)}{}_{2}R_{\ell m}^{(0,0,1)}\,,
\end{align}
\end{subequations}
where from Eq.~\eqref{eq:master_eqn_Psi0_radial_simplify}, we obtain
\begin{widetext}
\begin{subequations}
\begin{align}
    \mathcal{H}_{0}^{\ell m(0,0,0)}
    =& \;r(r-r_s)\partial^2_{r}+6(r-M)\partial_r
    +\frac{4i\omega_{\ell m} r(r-3M)+\omega_{\ell m}^{2}r^{3}}{r-r_s}
    -\ell(\ell+1)+6\,, \\
    \mathcal{H}_{0}^{\ell m(0,1,0)}
    =& \;\frac{4imM((r-M)+iM\omega_{\ell m}r)}{r(r-r_s)}
    +\frac{8mM\omega_{\ell m}}{\ell(\ell+1)}\,,
\end{align}    
\end{subequations}
by setting $\zeta = 0$.
Then, following the procedures above, we obtain
\begin{equation} \label{eq:GR_QNM_slow_rot}
    \omega_{0,\ell 1}^{(0,1,0)}
    =\frac{\left\langle 4iM\left((r-M)+iM\omega^{(0,0,0)}_{\ell m}r\right)/(r-r_s)
    +8M\omega_{\ell m}^{(0,0,0)}/(\ell(\ell+1))\right\rangle}
    {\left\langle\left[4ir(r-3M)+2\omega^{(0,0,0)}_{\ell m}r^3\right]
    /(r-r_s)\right\rangle}\,,
\end{equation}
\end{widetext}
where the inner product is defined in Eqs.~\eqref{eq:inner_product} and \eqref{eq:inner_product_simplify} with ${}_{s}R_{\ell m}^{(0,1)}$ replaced by ${}_{2}R_{\ell m}^{(0,0,1)}$.

To test the accuracy of the slow-rotation approximation, we calculated $\omega_{0,\ell 1}^{(0,1,0)}$ from Eq.~\eqref{eq:GR_QNM_slow_rot} and compared it to previously obtained results for the exact Kerr metric (i.e., without performing any expansion in the spin parameter) using Leaver's method in \cite{Leaver:1985ax} and tabulated in \cite{BertiRingdown, Berti:2005ys, Berti:2009kk, Cook:2014cta}. These results are presented in Fig.~\ref{fig:slowroterror} for the fundamental modes at $\ell=m=2,3,4$. We also list the values of $\omega_{0,\ell m}^{(0,0,0)}$ and $\omega_{0,\ell 1}^{(0,1,0)}$ in Appendix~\ref{appendix:GR_values} for completeness. For the convenience of comparison, we define the relative difference between any $\omega_2$ and any $\omega_1$ as
\begin{subequations} \label{eq:relerror}
\begin{align}
    \delta\left(\textrm{Re}[\omega_1],\textrm{Re}[\omega_2]\right) 
    & =\left|\textrm{Re}[\omega_2]/\textrm{Re}[\omega_1]-1 \right|\,, \\
    \delta\left(\textrm{Im}[\omega_1],\textrm{Im}[\omega_2]\right) 
    & =\left|\textrm{Im}[\omega_2]/\textrm{Im}[\omega_1]-1\right|\,.
\end{align}
\end{subequations}
In the bottom panels of Fig.~\ref{fig:slowroterror}, we plot $\delta\left(\textrm{Re}[\omega_{\Kerr}],\textrm{Re}[\omega_{\slow}]\right)$ and $\delta\left(\textrm{Im}[\omega_{\Kerr}],\textrm{Im}[\omega_{\slow}]\right)$, where $\omega_{\Kerr}$ are the QNM frequencies calculated from the exact Kerr metric~\cite{Leaver:1985ax, Berti:2009kk, Cook:2014cta}, and $\omega_{\slow}$ are the QNM frequencies we calculated using the EVP method after expanding the Kerr metric to $\mathcal{O}(\chi^1)$. We find that a relative error of $1\%$ is reached at a spin parameter of $a\approx0.22M$ for $\textrm{Re}\left[\omega_{22}^{(0,0)}\right]$, while the relative error is lower for the $\ell=m=3,4$ modes. Meanwhile, the imaginary part of the QNM frequencies has a smaller relative error. From this analysis, we infer that the slow-rotation QNM spectra in dCS gravity obtained in this work ought to remain valid for BHs with a spin $a\lesssim 0.22M$ with a slow-rotation approximation error of $\lesssim 1\%$. Nevertheless, this estimate only uses our knowledge in GR and should be re-examined after calculating the QNM spectra of BHs with general spins in dCS gravity in our follow-up work.

\subsection{Application of the EVP method to dCS gravity}
\label{sec:EVP_dCS}

For beyond-GR theories, although one generally has a coupled system of $(\ell,\pm m)$ modes or even- and odd-parity modes, Refs.~\cite{Cano:2021myl, Hussain:2022ins, Cano:2023tmv, Li:2023ulk} showed how to reduce the system to a two-dimensional eigenvalue problem, so one can still apply the EVP method in \cite{Zimmerman:2014aha, Mark:2014aja}. Nonetheless, up to $\mathcal{O}(\zeta^1,\chi^1,\epsilon^1)$ in dCS gravity, the modified Teukolsky equation is invariant under a $\hat{\mathcal{P}}$ transformation, as shown in Sec.~\ref{sec:simplify_radial_eqns}, so the equations for the $(\ell,m)$ and $(\ell,-m)$ modes contain the same information. Thus, as shown below, after using Eqs.~\eqref{eq:scalar_radial_IRG_2_lm} and \eqref{eq:scalar_radial_ORG_lm} to solve for the scalar field's radial part $\Theta_{\ell m}^{(1,1)}(r)$ in terms of ${}_{s}R^{(0,1)}_{\ell m}(r)$, we can simply apply the one-dimensional EVP method to Eqs.~\eqref{eq:master_eqn_Psi0_radial_simplify} and ~\eqref{eq:master_eqn_Psi4_radial_simplify} and obtain\footnote{Solving for $\Theta_{\ell m}^{(1,1)}(r)$ along the contour in Eq.~\eqref{eq:contour} requires correctly mapping the boundary conditions of $\Theta_{\ell m}^{(1,1)}(r)$ and ${}_{s}R^{(0,1)}_{\ell m}(r)$ from the real line to the contour, the validity of which will be shown in detail in \cite{Dong:2025abc}.}
\begin{widetext}
\begin{subequations} \label{eq:QNM_shift}
\begin{align} 
    & \omega_{0,\ell m}^{\pm(1,0)}
    =\frac{\left\langle -2r^2\left[\left(\mathscr{O}_{\geo}^{\ell m}
    +\mathscr{O}_{T_{\vartheta}}^{\ell m}+\mathscr{O}_{T_{\Psi}}^{\ell m}\right)
    \mp\left(\mathscr{O}_{T_{\vartheta}}^{\ell m}
    -\tilde{\mathscr{O}}_{T_{\Psi}}^{\ell m}\right)\right]\right\rangle}
    {\left\langle\partial_{\omega}\mathcal{H}_0^{\ell m}\right\rangle}\,, 
    \label{eq:QNM_shift_Psi0} \\
    & \omega_{4,\ell m}^{\pm(1,0)}
    =\frac{\left\langle -2r^6\left[\left(\mathscr{Q}_{\geo}^{\ell m}
    +\mathscr{Q}_{T_{\vartheta}}^{\ell m}+\mathscr{Q}_{T_{\Psi}}^{\ell m}\right)
    \mp\left(\mathscr{Q}_{T_{\vartheta}}^{\ell m}
    -\tilde{\mathscr{Q}}_{T_{\Psi}}^{\ell m}\right)\right]\right\rangle}
    {\left\langle\partial_{\omega}\mathcal{H}_4^{\ell m}\right\rangle}\,,
    \label{eq:QNM_shift_Psi4}
\end{align}    
\end{subequations}
where $\omega_{0,\ell m}^{\pm(1,0)}$ and $\omega_{4,\ell m}^{\pm(1,0)}$ refer to the QNM frequency shifts computed from the equation of $\Psi_0^{(1,1)}$ in the IRG [i.e., Eq.~\eqref{eq:master_eqn_Psi0_radial_simplify}] and $\Psi_4^{(1,1)}$ in the ORG [i.e., Eq.~\eqref{eq:master_eqn_Psi4_radial_simplify}], respectively. The superscripts $\pm$ label the QNM frequencies of even-parity [i.e., set $\eta_{\ell m}=1$ in Eq.~\eqref{eq:solution_ansatz}] and odd-parity [i.e., set $\eta_{\ell m}=-1$ in Eq.~\eqref{eq:solution_ansatz}] modes, respectively. The operators $\mathcal{H}_0^{\ell m}$ and $\mathcal{H}_4^{\ell m}$ are the radial Teukolsky operators acting on ${}_{2}R_{\ell m}^{(1,1)}$ and ${}_{-2}R_{\ell m}^{(1,1)}$ in Eqs.~\eqref{eq:master_eqn_Psi0_radial_simplify} and \eqref{eq:master_eqn_Psi4_radial_simplify}, respectively. We have dropped the order-counting superscript $(0,0)$ of $\mathcal{H}_0^{\ell m}$ and $\mathcal{H}_4^{\ell m}$ for simplicity. Their derivatives in $\omega$ are
\begin{subequations}
\begin{align}
    & \partial_{\omega}\mathcal{H}_0^{\ell m}
    =\frac{4ir(r-3M)+2\omega_{\ell m}r^3}{r-r_s}
    +\chi m\left[\frac{8M}{\ell(\ell+1)}-\frac{4M^2}{r-r_s}\right]\,, \\
    & \partial_{\omega}\mathcal{H}_4^{\ell m}
    =\frac{-4ir(r-3M)+2\omega_{\ell m}r^3}{r-r_s}
    +\chi m\left[\frac{8M}{\ell(\ell+1)}-\frac{4M^2}{r-r_s}\right]\,. 
\end{align}     
\end{subequations} 
\end{widetext}
Since the operators in Eq.~\eqref{eq:QNM_shift} contain both $\mathcal{O}(\chi^0)$ and $\mathcal{O}(\chi^1)$ contributions, one can expand $\omega_{0,\ell m}^{\pm(1,0)}$ and $\omega_{4,\ell m}^{\pm(1,0)}$ following Eq.~\eqref{eq:QNM_slow_expansion}, i.e.,
\begin{subequations} \label{eq:QNM_slow_expansion_dCS}
\begin{align}
    & \omega_{i,\ell m}^{+(1,0)}
    =\chi m\omega_{i,\ell 1}^{+(1,1,0)}+\mathcal{O}(\chi^2)\,,
    \label{eq:QNM_slow_expansion_dCS_even} \\
    & \omega_{i,\ell m}^{-(1,0)}
    =\omega_{i,\ell m}^{-(1,0,0)}+\chi m\omega_{i,\ell 1}^{-(1,1,0)}
    +\mathcal{O}(\chi^2)\,,
    \label{eq:QNM_slow_expansion_dCS_odd}
\end{align}
\end{subequations}
where $i=0,4$. Notice that $\omega_{i,\ell m}^{+(1,0)}$ starts from the $\mathcal{O}(\chi^1)$ term since at $\mathcal{O}(\chi^0)$, only $\mathscr{O}_{T_{\vartheta}}^{\ell m}$ and $\mathscr{Q}_{T_{\vartheta}}^{\ell m}$ are nonzero in Eq.~\eqref{eq:QNM_shift}. For even-parity modes, $\mathscr{O}_{T_{\vartheta}}^{\ell m}$ and $\mathscr{Q}_{T_{\vartheta}}^{\ell m}$ cancel out since the scalar field perturbation only couple to the odd-parity modes, as discussed in Sec.~\ref{sec:scalar_eqn} and \ref{sec:simplify_radial_eqns}. In Secs.~\ref{sec:QNM_nonrotating} and \ref{sec:QNM_slow_rotation}, we will evaluate each individual piece in Eq.~\eqref{eq:QNM_slow_expansion_dCS}.

\begin{table*}[]
    \centering
    \begin{tabular}{cccccc}
        \hline\hline
        $\ell$ & Overtones & MTF (IRG) & MTF (ORG) 
        & RW \\
        \hline
        & $n=0$ &  $-0.0307678-0.0156999i$ & $-0.0307678-0.0156999i$ 
        &  $-0.0307678-0.0156999i$ \\
        $\ell=2$ & $n=1$ & $-0.0637401-0.0511483i$ & $-0.0637400-0.0511481i$ 
        & $-0.0637400-0.0511481i$  \\
        & $n=2$ & $-0.1464548-0.1014012i$ & $-0.1464536-0.1013966i$ 
        & $-0.1464567-0.1013978i$  \\
        \hline
        & $n=0$ & $-0.1142723-0.0206452i$ & $-0.1142724-0.0206454i$ 
        & $-0.1142723-0.0206454i$ \\
        $\ell=3$ & $n=1$ & $-0.1446872-0.0639291i$ & $-0.1446866-0.0639276i$ 
        & $-0.1446866-0.0639278i$ \\
        & $n=2$ & $-0.2084786-0.1140437i$ & $-0.2084679-0.1140426i$ 
        & $-0.2084838-0.1140453i$ \\
        \hline
        & $n=0$ & $-0.2728671-0.0259840i$ & $-0.2728667-0.0259848i$ 
        & $-0.2728665-0.0259846i$ \\
        $\ell=4$ & $n=1$ & $-0.3037111-0.0800387i$ & $-0.3037152-0.0800351i$ 
        & $-0.3037114-0.0800317i$ \\
        & $n=2$ & $-0.3667305-0.1407570i$ & $-0.3667488-0.1407812i$ 
        & $-0.3667476-0.1407778i$ \\
        \hline \hline
    \end{tabular}
    \caption{The QNM frequency shifts for the odd-parity modes $\omega_{\ell m}^{-(1,0,0)}$ of a non-rotating BH in dCS gravity for $\ell=2,3,4$ and the overtones $n=0,1,2$. Due to spherical symmetry, $\omega_{\ell m}^{-(1,0,0)}$ with the same $\ell$ but different $m$ are the same. The word ``MTF'' is an acronym for the modified Teukolsky formalism. The columns ``MTF (IRG)'' and ``MTF (ORG)'' are the results in the IRG ($\omega_{0,\ell m}^{+(1,0,0)}$) and the ORG ($\omega_{0,\ell m}^{-(1,0,0)}$) of this work, respectively. The results in the column ``RW'' use the EVP method to solve for $\omega_{\ell m}^{-(1,0,0)}$ from the RW equation and the scalar field equation of dCS gravity in \cite{Cardoso:2009pk, Molina:2010fb, Pani:2011xj, Wagle:2021tam, Srivastava:2021imr}. We have set $M=1/2$ in this table.}
    \label{tab:QNM_results_nonrotating}
\end{table*}

\subsection{The QNMs of non-rotating BHs in dCS gravity}
\label{sec:QNM_nonrotating}

In this subsection, we evaluate the QNM frequency shifts for a non-rotating BH in dCS gravity using the EVP approach discussed in Secs.~\ref{sec:EVP} and \ref{sec:EVP_dCS}. When $\chi=0$, Eq.~\eqref{eq:QNM_shift} reduces to
\begin{subequations} \label{eq:QNM_shift_nonrotating}
\begin{align}
    & \omega_{0,\ell m}^{-(1,0,0)}
    =\frac{\left\langle -2r^2\left(\hat{A}_{1}^{\ell m}
    +\hat{A}_{1}^{\ell m}\right)
    \mathcal{D}^{\dagger}_{\ell m}\right\rangle}
    {\left\langle\left[4ir(r-3M)+2\omega^{(0,0,0)}_{\ell m}r^3\right]
    /(r-r_s)\right\rangle}\,, 
    \label{eq:QNM_shift_Psi0_nonrotating} \\
    & \omega_{4,\ell m}^{-(1,0,0)}
    =\frac{\left\langle -2r^6\left(\hat{\mathscr{A}}_{1}^{\ell m}
    +\hat{\mathscr{A}}_{1}^{\ell m}\right)
    \mathcal{D}_{\ell m}\right\rangle}
    {\left\langle\left[-4ir(r-3M)+2\omega^{(0,0,0)}_{\ell m}r^3\right]
    /(r-r_s)\right\rangle}\,, 
    \label{eq:QNM_shift_Psi4_nonrotating}
\end{align}
\end{subequations}
while $\omega_{0,\ell m}^{+(1,0,0)}=\omega_{4,\ell m}^{+(1,0,0)}=0$. As discussed in Sec.~\ref{sec:sA_simplify}, to evaluate the source terms associated with $\hat{A}_{1}^{\ell m}$ or $\hat{\mathscr{A}}_{1}^{\ell m}$ in Eq.~\eqref{eq:QNM_shift_nonrotating}, one needs to first solve the scalar field equation in Eq.~\eqref{eq:scalar_radial_IRG_2_lm} or \eqref{eq:scalar_radial_ORG_lm}. For non-rotating BHs in the IRG, Eq.~\eqref{eq:scalar_radial_IRG_2_lm} becomes
\begin{widetext}
\begin{equation} \label{eq:scalar_IRG_Schw}
    \left[r(r-r_s)\partial_{r}^2 
    +r_s\partial_r+\frac{\omega_{\ell m}^2r^3}{r-r_s}-\frac{r_s}{r}
    -{}_{0}A_{\ell m}\right]\Theta_{\ell m}^{(1,1)}(r) 
    =-\frac{2i}{\mathcal{C}_2}(1-\eta_{\ell m})
    \left(g_1^{\ell m}(r)+g_2^{\ell m}(r)\partial_r\right)
    {}_{-2}R^{(0,1)}_{\ell m}(r)\,,
\end{equation}
where the radial functions $g_1^{\ell m}(r)$ and $g_2^{\ell m}(r)$ are
\begin{align}
   & g_1^{\ell m}(r)
   =-3\sqrt{\Lambda_{\ell}}M^3 
   \frac{\left(2\omega_{\ell m}^2r^2
   -8i\omega_{\ell m}r-\ell^2-\ell-4\right)r^2
   +2M\left(9i\omega_{\ell m}r+\ell^2+\ell+10\right)r-24M^2}
   {4\sqrt{\pi}r^4(r-r_s)^2}\,, \nonumber\\
   & g_2^{\ell m}(r)
   =-3\sqrt{\Lambda_{\ell}}M^3\frac{i\omega_{\ell m}r^2+r-3M}{2\sqrt{\pi}r^3(r-r_s)}\,,
   \quad
   \Lambda_{\ell}=(\ell+2)(\ell+1)\ell(\ell-1)\,,
\end{align}
\end{widetext}
and $\mathcal{C}_2$ is a complex constant in the Teukolsky-Starobinsky identity \cite{Teukolsky:1974yv, Starobinsky:1973aij, Starobinskil:1974nkd}, 
\begin{subequations} \label{eq:TS_identity}
\begin{align}
    & (D_{m\omega})^4{}_{-2}R_{\ell m}^{(0,1)}(r)
    =\mathcal{C}_2\,{}_{2}\hat{R}_{\ell m}(r)\,, \\
    & \Delta^2(D^\dagger_{m\omega})^4\left[\Delta^2\,{}_{2}R_{\ell m}^{(0,1)}(r)\right]
    =\mathcal{C}_{-2}\,{}_{-2}\hat{R}_{\ell m}(r)\,.
\end{align}
\end{subequations}
The complex constants $\mathcal{C}_2$ and $\mathcal{C}_{-2}$ satisfy $\mathcal{C}_2\mathcal{C}_{-2}=\mathfrak{C}$, with $\mathfrak{C}$ being the Teukolsky-Starobinsky constant in Eq.~\eqref{eq:TS_constant}. Here, we have used Eq.~\eqref{eq:TS_identity} to reduce Eq.~\eqref{eq:hertztoteukall} to
\begin{subequations}
\begin{align}
    & {}_{2}\hat{R}_{\ell m}(r)=-\frac{2}{\mathcal{C}_2}{}_{-2}R_{\ell m}^{(0,1)}(r)\,,\\
    & {}_{-2}\hat{R}_{\ell m}(r)=\frac{32}{\mathcal{C}_{-2}}
    {}_{2}R_{\ell m}^{(0,1)}(r)\,,
\end{align}
\end{subequations}
so we can replace the radial functions ${}_{\pm2}\hat{R}_{\ell m}(r)$ of the Hertz potential with the radial functions ${}_{\mp 2}R_{\ell m}^{(0,1)}(r)$ of $\Psi_{4}$ or $\Psi_{0}$, respectively. Fixing the normalization of ${}_{\pm 2}R_{\ell m}^{(0,1)}(r)$, one can also set $\mathcal{C}_{-2}=\bar{\mathcal{C}}_2$. In contrast, the Teukolsky-Starobinsky constant $\mathfrak{C}$ is normalization independent.

In this work, we compute the radial Teukolsly functions ${}_{\pm 2}R_{\ell m}^{(0,1)}(r)$ using Leaver's method in \cite{Leaver:1985ax} and evaluate the coefficients $\mathcal{C}_{\pm2}$ directly. Specifically, we use the radial Teukolsky equations to reduce the maximum number of derivatives in Eq.~\eqref{eq:TS_identity} to one, apply the simplified operators on ${}_{\pm 2}R_{\ell m}^{(0,1)}(r)$, and evaluate the ratio between the resulting wave functions and ${}_{\mp 2}R_{\ell m}^{(0,1)}$ along the contour $\mathscr{C}$ in Eq.~\eqref{eq:contour} (or Fig.~\ref{fig:contour}). We find that this ratio is rather stable until very large $|\xi|$, corresponding to $r$ with a large imaginary part. In this case, we simply use the ratio at $r(\xi=0)=2(1-i)M$ as the value of $\mathcal{C}_{\pm2}$.

Using the Chandrasekhar transformation in \cite{Chandrasekhar_1983}, we can also transform the radial Teukolsky functions ${}_{\pm 2}R_{\ell m}^{(0,1)}(r)$ to the RW function $\Psi_{\ell m}^{\RW(0,1)}(r)$, i.e.,
\begin{widetext}
\begin{align} \label{eq:chandra_transform}
    & {}_{\pm 2}R_{\ell m}^{(0,1)}
    =f_{\pm2}(r)\left[V_{Z}(r)+\left(\frac{2}{r^2}(r-3M)\mp
    2i\omega_{\ell m}\right)\Lambda_{\pm}\right]
    \Psi_{\ell m}^{\RW(0,1)}(r)\,, \nonumber\\
    & V_Z(r)=\left(1-\frac{r_s}{r}\right)
    \left(\frac{\ell(\ell+1)}{r^2}-\frac{6M}{r^3}\right)\,,\quad
    f_{2}(r)=r^{3}\Delta^{-2}(r)\,,\quad
    f_{-2}(r)=r^{3}\,,\quad
    \Lambda_{\pm}=\frac{d}{dr_{*}}\mp i\omega_{\ell m}\,,
\end{align}    
\end{widetext}
where $r_{*}$ is the tortoise coordinate, i.e., $dr_*/dr=(r^2+a^2)/\Delta(r)$, and $V_{Z}(r)$ is determined by the potential in the RW equation. We can then directly compare Eq.~\eqref{eq:scalar_IRG_Schw} to the results in \cite{Cardoso:2009pk, Molina:2010fb, Wagle:2021tam}. Under the transformation in Eq.~\eqref{eq:chandra_transform}, Eq.~\eqref{eq:scalar_IRG_Schw} becomes
\begin{align} \label{eq:Teuk_to_RW}
    & \left[r(r-r_s)\partial_{r}^2 
    +r_s\partial_r+\frac{\omega_{\ell m}^2r^3}{r-r_s}-\frac{r_s}{r}
    -{}_{0}A_{\ell m}\right]\Theta_{\ell m}^{(1,1)}(r) \nonumber\\
    & =-\frac{3i\sqrt{\Lambda_{\ell}}M^3}{\sqrt{\pi}r^3}
     \Psi_{\ell m}^{\RW(0,1)}(r)\,,
\end{align}
where we assume $\mathcal{C}_{-2}=\bar{\mathcal{C}}_2$. Equation~\eqref{eq:Teuk_to_RW} is consistent with the result in \cite{Wagle:2019mdq} up to an overall constant, which can be compensated when evaluating the modified Teukolsky equation. We have also tried applying the same Chandrasekhar transformation in Eq.~\eqref{eq:chandra_transform} to the modified Teukolsky equation, but the results do not match the one in \cite{Cardoso:2009pk, Molina:2010fb, Wagle:2021tam}. This is not surprising since ${}_{\pm2}R^{(1,1)}_{\ell m}(r)$ satisfy the modified Teukolsky equation in dCS gravity but not the Teukolsky equation in GR. We first need to extend the Chandrasekhar transformation in Eq.~\eqref{eq:chandra_transform} to the dCS case by including more terms, which we will work out in the future. On the other hand, the source terms in Eq.~\eqref{eq:scalar_IRG_Schw} are driven by the GR Teukolsky functions ${}_{\pm2}R^{(0,1)}_{\ell m}(r)$, so we can still apply the Chandrasekhar transformation in GR.

\begin{table*}[]
    \centering
        \begin{tabular}{cccccc}
        \hline\hline
        $\ell$ & Overtones & MTF (IRG) & MTF (ORG) 
        & ZM \\
        \hline
        & $n=0$ &  $-0.0019472+0.0010443i$ & $-0.0019472+0.0010443i$ 
        & $-0.0019472+0.0010443i$ \\
        $\ell=2$ & $n=1$ & $+0.0103192-0.0066778i$ & $+0.0103192-0.0066778i$ 
        & $+0.0103192-0.0066778i$ \\
        & $n=2$ & $+0.0279807-0.0434918i$ & $+0.0279806 -0.0434910i$ 
        & $+0.0279807-0.0434918i$ \\
        \hline
        & $n=0$ & $-0.0037626-0.0000358i$ & $-0.0037626-0.0000358i$ 
        & $-0.0037626-0.0000358i$ \\
        $\ell=3$ & $n=1$ & $-0.0003195-0.0018110i$ & $-0.0003195-0.0018110i$ 
        & $-0.0003195-0.0018110i$ \\
        & $n=2$ & $+0.0063750-0.0087907i$ & $+0.0063750-0.0087907i$ 
        & $+0.0063750-0.0087907i$ \\
        \hline
        & $n=0$ & $-0.0038431-0.0005217i$ & $-0.0038431-0.0005217i$ 
        & $-0.0038431-0.0005217i$ \\
        $\ell=4$ & $n=1$ & $-0.0027150-0.0020804i$ & $-0.0027150-0.0020804i$ 
        & $-0.0027150-0.0020804i$ \\
        & $n=2$ & $-0.0004310-0.0052115i$ & $-0.0004310-0.0052115i$ 
        & $-0.0004310-0.0052115i$ \\
        \hline \hline
    \end{tabular}
    \caption{The QNM frequency shifts $\omega_{\ell 1}^{+(1,1,0)}$ for the even-parity modes of a slowly-rotating BH in dCS gravity for $\ell=2,3,4$ and the overtones $n=0,1,2$. Same as Table~\ref{tab:QNM_results_nonrotating}, the columns ``MTF (IRG)'' and ``MTF (ORG)'' present the results in the IRG ($\omega_{0,\ell 1}^{+(1,1,0)}$) and the ORG ($\omega_{4,\ell 1}^{+(1,1,0)}$) of this work, respectively. The results in the column ``ZM'' use the EVP method to solve for $\omega_{\ell 1}^{+(1,1,0)}$ from the ZM equation of dCS gravity in \cite{Wagle:2021tam, Srivastava:2021imr}. We have set $M=1/2$ in this table.}
    \label{tab:QNM_results_rotating_even}
\end{table*}

To solve the scalar field from Eq.~\eqref{eq:scalar_IRG_Schw}, we can use the Green function method. Rewriting the left-hand side of Eq.~\eqref{eq:scalar_IRG_Schw} in the tortoise coordinate $r_*$, we obtain
\begin{widetext}
\begin{equation} \label{eq:scalar_IRG_Schw_r*}
    \left[\partial_{r_*}^2+\frac{r-r_s}{r^3}
    \left(\frac{\omega_{\ell m}^2r^3}{r-r_s}-\frac{r_s}{r}
    -{}_{0}A_{\ell m}\right)\right]\Theta_{\ell m}^{(1,1)}(r)
    =-\frac{4i}{\mathcal{C}_2}\frac{r-r_s}{r^3}
    \left(g_1^{\ell m}(r)+g_2^{\ell m}(r)\partial_r\right)
    {}_{-2}R^{(0,1)}_{\ell m}(r)\,,
\end{equation}
where we have set $\eta_{\ell m}=-1$ since the even-parity modes with $\eta_{\ell m}=1$ do not couple to the scalar field. In this case, the solution to Eq.~\eqref{eq:scalar_IRG_Schw_r*} along the contour $\mathscr{C}$ is
\begin{equation} \label{eq:scalar_sol}
    \Theta_{\ell m}^{(1,1)}(\xi)
    =\frac{\Theta_{\ell m}^R(\xi)\int_{-\infty}^{\xi}
    \Theta_{\ell m}^L(\xi')\mathcal{S}^{\ell m}_{\vartheta}(\xi')
    \partial_{\xi'}r_{*}\,d\xi'
    +\Theta_{\ell m}^L(\xi)\int_{\xi}^{\infty}
    \Theta_{\ell m}^R(\xi')\mathcal{S}^{\ell m}_{\vartheta}(\xi')
    \partial_{\xi'}r_{*}\,d\xi'}
    {\left(\Theta_{\ell m}^L(\xi)\partial_{\xi}\Theta_{\ell m}^R(\xi)
    -\Theta_{\ell m}^R(\xi)\partial_{\xi}\Theta_{\ell m}^L(\xi)\right)
    \partial_{r_*}\xi}\,,
\end{equation}
\end{widetext}
where $f(\xi)$ and $f(\xi')$ mean evaluating $f(r)$ at $r=r_{\mathscr{C}}(\xi)$ and $r=r_{\mathscr{C}}(\xi')$, respectively. Numerically, we evaluate $\xi=\pm\infty$ at $\xi=\pm\xi_{\max}$, respectively. The function $\mathcal{S}^{\ell m}_{\vartheta}(\xi)$ is the source term on the right-hand side of Eq.~\eqref{eq:scalar_IRG_Schw_r*}. The functions $\Theta_{\ell m}^R(\xi)$ and $\Theta_{\ell m}^L(\xi)$ are solutions to the left-hand side of Eq.~\eqref{eq:scalar_IRG_Schw_r*} with asymptotic behaviors $\Theta_{\ell m}^R(\xi\rightarrow\infty)\propto e^{i\omega_{\ell m}r_{*}(\xi)}$ and $\Theta_{\ell m}^L(\xi\rightarrow-\infty)\propto e^{-i\omega_{\ell m}r_{*}(\xi)}$, respectively. We compute $\Theta_{\ell m}^R(\xi)$ and $\Theta_{\ell m}^L(\xi)$ by numerically integrating their asymptotic expansions from $\xi=\xi_{\max}$ and $\xi=-\xi_{\max}$ along the contour to $\xi=-\xi_{\max}$ and $\xi=\xi_{\max}$, respectively. After obtaining $\Theta_{\ell m}^{(1,1)}(r)$ along the contour $\mathscr{C}$, we then plug it back into Eq.~\eqref{eq:QNM_shift_Psi0_nonrotating}. Notice that by solving $\Theta_{\ell m}^{(1,1)}(r)$, we effectively compute the piece $\mathcal{H}_{\vartheta}^{-1}\mathcal{V}^{\ell m}{}_{2}\mathcal{D}^{\dagger}_{\ell m}\,{}_{2}R_{\ell m}^{(0,1)}$ of $\hat{A}_1^{\ell m}\mathcal{D}^{\dagger}_{\ell m}\,{}_{2}R_{\ell m}^{(0,1)}$ [see Eq.~\eqref{eq:hat_A_i}]. We can then compute the inner product in Eq.~\eqref{eq:QNM_shift_Psi0_nonrotating} using the solution of $\Theta_{\ell m}^{(1,1)}(r)$ and ${}_{2}R_{\ell m}^{(0,1)}(r)$. A similar calculation can be also done in the ORG for $\Psi_4^{(1,1)}$ using Eq.~\eqref{eq:QNM_shift_Psi4_nonrotating}.

In Table~\ref{tab:QNM_results_nonrotating} and Fig.~\ref{fig:odd_scatter}, we present the results of $\omega_{\ell m}^{-(1,0,0)}$ for $\ell=2,3,4$ and the overtones $n=0,1,2$. To obtain these results, we have kept the first $150$ terms in the continuous fraction of Leaver's method in \cite{Leaver:1985ax} to obtain ${}_{\pm 2}R_{\ell m}^{(0,1)}(r)$. When numerically integrating $\Theta^{R,L}_{\ell m}(\zeta)$ from their asymptotic expansion, we have considered the first eight terms in the expansion. We notice that $\omega_{0,\ell m}^{-(1,0,0)}$ computed from $\Psi_0$ in the IRG are consistent with $\omega_{4,\ell m}^{-(1,0,0)}$ computed from $\Psi_4$ in the ORG. The relative differences [i.e., Eq.~\eqref{eq:relerror}] between $\omega_{0,\ell m}^{-(1,0,0)}$ and $\omega_{4,\ell m}^{-(1,0,0)}$ are $\lesssim 10^{-4}$ for all the modes, as shown in Fig.~\ref{fig:odd_scatter}. This indicates that these two independent calculations within the modified Teukolsky formalism are self-consistent, and the QNM frequencies do not depend on the gauge chosen. We also observe that the relative differences between $\omega_{0,\ell m}^{-(1,0,0)}$ and $\omega_{4,\ell m}^{-(1,0,0)}$ generally increase with the overtone number $n$. This is possibly due to the fact that $\textrm{Im}\left[\omega_{\ell m}^{(0,0)}\right]$ decreases with $n$, which is always negative, so the source terms, which are driven by the GR QNMs, are more oscillatory along the contour $\mathscr{C}$ for a larger $n$, resulting in larger numerical inaccuracies. Since this is the first calculation of overtones in dCS gravity, we also observe from Table~\ref{tab:QNM_results_nonrotating} and Fig.~\ref{fig:odd_scatter} that both the real and imaginary parts of $\omega_{\ell m}^{-(1,0,0)}$ decrease with the overtone number $n$, which follows the same trend of QNM frequencies in GR, as one can see in Table~\ref{tab:QNM_results_GR}. On the other hand, unlike in GR, the real part of $\omega_{\ell m}^{-(1,0,0)}$ is negative for all the overtones $n$ we have considered.

\subsection{The QNMs of slowly-rotating BHs in dCS gravity}
\label{sec:QNM_slow_rotation}

In this subsection, we apply the EVP method in Secs.~\ref{sec:EVP} and \ref{sec:EVP_dCS} to compute the dCS correction to the QNM spectrum at $\mathcal{O}(\zeta^1,\chi^1,\epsilon^1)$. As discussed in \cite{Wagle:2023fwl} and shown in Eq.~\eqref{eq:QNM_shift}, dCS corrections to the QNM spectrum of both even- and odd-parity modes are nonzero at $\mathcal{O}(\chi)$ since the BH geometry is deformed at this order. For this case, we separately compute the corrections to the QNMs of different parity in the remaining subsection. 

\subsubsection{Even-parity modes}
\label{sec:QNM_slow_rotation_even}

\begin{figure}[t]
    \centering
    \includegraphics[width=0.95\linewidth]{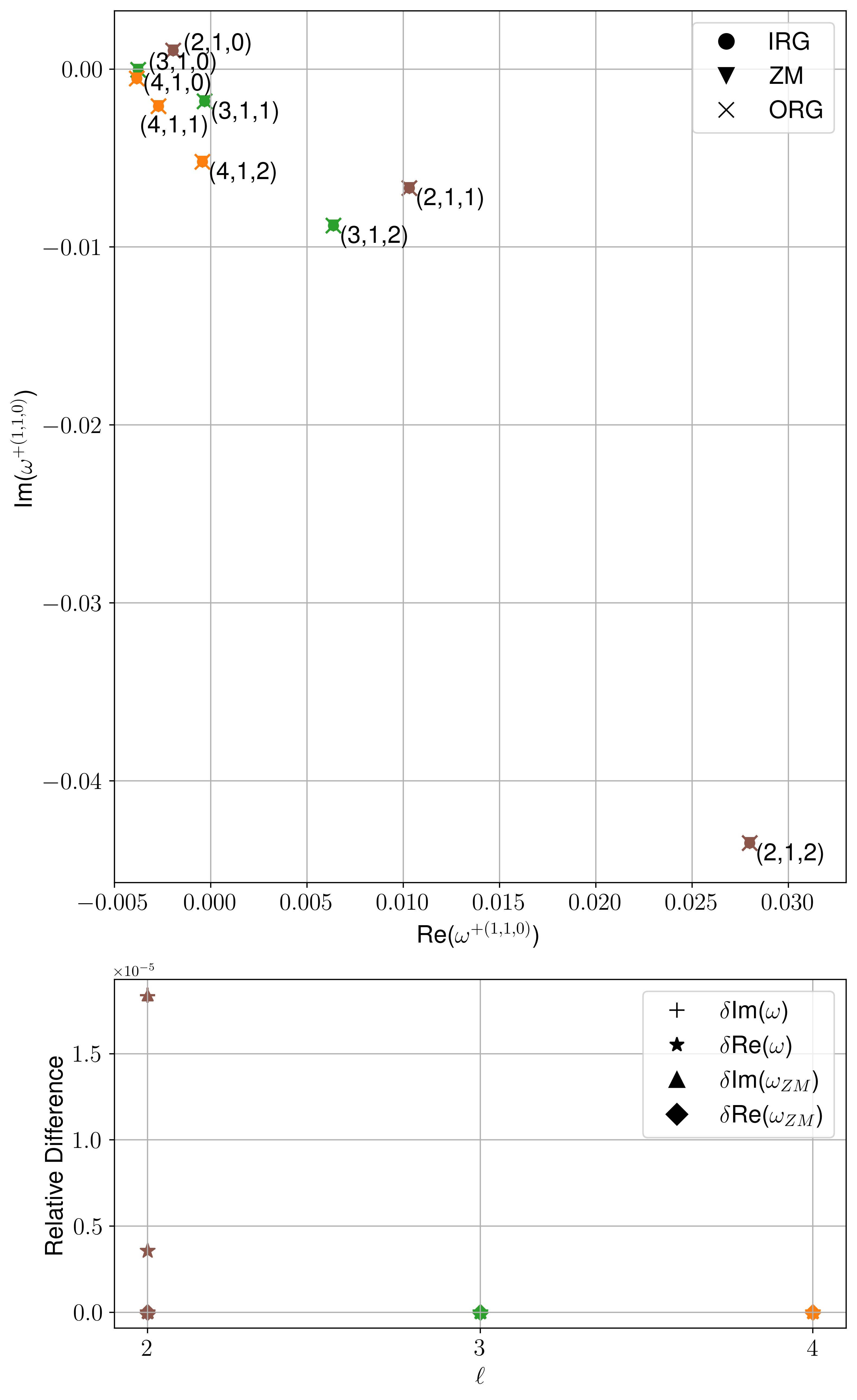}
    \caption{QNM frequency shifts for the even-parity modes of a slowly-rotating BH in dCS gravity up to $\mathcal{O}(\chi^1)$ for $\ell=2,3,4$ and the overtones $n=0,1,2$. The index of each point labels the mode $(\ell,m,n)$. The top panel plots $\omega_{\ell 1}^{+(1,1,0)}$ computed in Sec.~\ref{sec:QNM_slow_rotation_even} with their values listed in Table~\ref{tab:QNM_results_rotating_even}. The marker ``ZM'' labels the results computed from the ZM equation [Eq.~\eqref{eq:ZM_expand}] in the RW gauge found by \cite{Wagle:2021tam,Srivastava:2021imr}. The bottom panel shows the relative differences of the QNM frequency shifts obtained in the ORG [marked as ``$\delta\rm{Im}(\omega)$" and ``$\delta\rm{Re}(\omega)$''] and the RW gauge [marked as ``$\delta\rm{Im}(\omega_{ZM})$" and ``$\delta\rm{Re}(\omega_{ZM})$''] from those obtained in the IRG, respectively. For simplicity, we drop the mode labels in the bottom panels. All the other labels have the same meaning as Fig.~\ref{fig:summary_scatter}. We have set $M=1/2$ in this plot.}
    \label{fig:even_scatter}
\end{figure}

For even-parity modes, the scalar field perturbation $\vartheta^{(1,1)}$ decouple from the Weyl scalar perturbations $\Psi_{0,4}^{(1,1)}$, as shown in Eq.~\eqref{eq:QNM_shift} (i.e., $\mathscr{O}_{T_{\vartheta}}^{\ell m}$ and  $\mathscr{Q}_{T_{\vartheta}}^{\ell m}$ are canceled out). In this case, from Eq.~\eqref{eq:QNM_shift}, one obtains
\begin{subequations} \label{eq:QNM_shift_even}
\begin{align}
    & \omega_{0,\ell m}^{+(1,0)}
    =\frac{\left\langle -2r^2\left(\mathscr{O}_{\geo}^{\ell m}
    +\mathscr{O}_{T_{\Psi}}^{\ell m}
    +\tilde{\mathscr{O}}_{T_{\Psi}}^{\ell m}\right)\right\rangle}
    {\left\langle\partial_{\omega}\mathcal{H}_0^{\ell m}\right\rangle}\,, 
    \label{eq:QNM_shift_Psi0_even} \\
    & \omega_{4,\ell m}^{+(1,0)}
    =\frac{\left\langle -2r^6\left(\mathscr{Q}_{\geo}^{\ell m}
    +\mathscr{Q}_{T_{\Psi}}^{\ell m}
    +\tilde{\mathscr{Q}}_{T_{\Psi}}^{\ell m}\right)\right\rangle}
    {\left\langle\partial_{\omega}\mathcal{H}_4^{\ell m}\right\rangle}\,,
    \label{eq:QNM_shift_Psi4_even}
\end{align}    
\end{subequations}
where $\{\mathscr{O}_{\geo}^{\ell m},\mathscr{O}_{T_{\Psi}}^{\ell m},\tilde{\mathscr{O}}_{T_{\Psi}}^{\ell m},\mathscr{Q}_{\geo}^{\ell m},\mathscr{Q}_{T_{\Psi}}^{\ell m},\tilde{\mathscr{Q}}_{T_{\Psi}}^{\ell m}\}$ are all independent of $\vartheta^{(1,1)}$ or its radial part $\Theta_{\ell m}^{(1,1)}(r)$. Instead of solving a coupled set of equations like in Sec.~\ref{sec:QNM_nonrotating}, one can now carry out the contour integral in Eq.~\eqref{eq:inner_product} of the source terms in Eq.~\eqref{eq:QNM_shift_even} along the contour in Eq.~\eqref{eq:contour}. Since these source terms depend on ${}_{\pm2}R_{\ell m}^{(0,1)}(r)$, we use Leaver's method in \cite{Leaver:1985ax} with $150$ terms included in the continuous fraction to compute ${}_{\pm2}R_{\ell m}^{(0,1)}(r)$. Furthermore, all the operators in the nominator of Eq.~\eqref{eq:QNM_shift_even} have an overall factor of $\chi m$, so all the radial functions within these operators are evaluated at $\mathcal{O}(\chi^0)$. Since the functions $\{\hat{B}_i^{\ell m}, \hat{\tilde{B}}_i^{\ell m}, \hat{\mathscr{B}}_i^{\ell m}, \hat{\tilde{\mathscr{B}}}_i^{\ell m}\}$ in Eqs.~\eqref{eq:O_B}, \eqref{eq:O_B_tilde}, \eqref{eq:Q_B}, and \eqref{eq:Q_B_tilde} do not depend on $m$, one can extract the factor $\chi m$ out from $\omega_{0,\ell m}^{+(1,0)}$ and $\omega_{4,\ell m}^{+(1,0)}$ following Eq.~\eqref{eq:QNM_slow_expansion_dCS}. In Table \ref{tab:QNM_results_rotating_even} and Fig.~\ref{fig:even_scatter}, we present the results of $\omega_{0,\ell 1}^{+(1,1,0)}$ and $\omega_{4,\ell 1}^{+(1,1,0)}$ for $\ell=2,3,4$ and the overtones $n=0,1,2$.

\begin{figure*}[t]
    \centering
    \includegraphics[width=\linewidth]{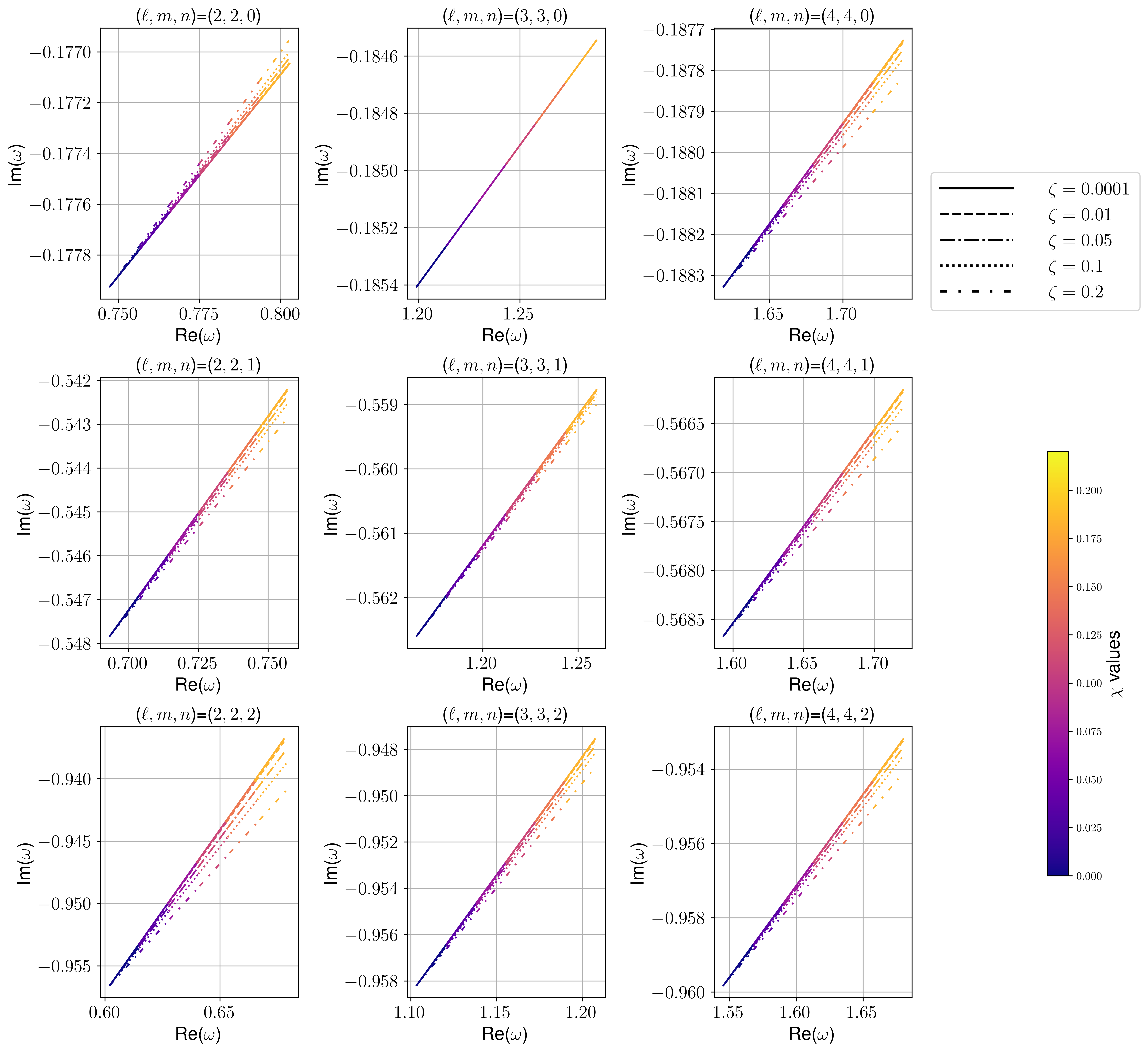}
    \caption{Total QNM frequencies $\omega_{\ell m}^{+}=\omega_{\ell m}^{(0,0)}+\zeta\omega_{\ell m}^{+(1,0)}$ for even-parity modes at $\ell=m=2,3,4$ and the overtones $n=0,1,2$ up to $\mathcal{O}(\chi^1)$ in dCS gravity. The values of $\omega_{\ell m}^{(0,0)}$ are retrieved from Table~\ref{tab:QNM_results_GR}. The values of $\omega_{\ell m}^{+(1,0)}$ are found using Eq.~\eqref{eq:QNM_slow_expansion_dCS_even} and the ``MTF (IRG)'' results in Table~\ref{tab:QNM_results_rotating_even}. The line color represents the magnitude of the spin $a=M\chi$, and different line styles represent different values of $\zeta$. We have set $a_{\max}=0.22M$ since it is the value below which the slow-rotation approximation is still relatively accurate in GR, as discussed in Sec.~\ref{sec:EVP}. The $(3,3,0)$ mode does exhibit a shift in the QNM frequencies as a function of $\zeta$. However, the shift in the imaginary part of the $(3,3,0)$ mode is two orders of magnitude smaller than all other modes presented here and thus in the chosen scale, appears as independent of $\zeta$. We have set $M=1/2$ in this plot. }
    \label{fig:even_overtone}
\end{figure*}

From Table~\ref{tab:QNM_results_rotating_even} and Fig.~\ref{fig:even_scatter}, we observe that $\omega_{0,\ell 1}^{+(1,1,0)}$ computed from $\Psi_0$ in the IRG are consistent with $\omega_{4,\ell 1}^{+(1,1,0)}$ computed from $\Psi_4$ in the ORG. For most modes, $\omega_{0,\ell 1}^{+(1,1,0)}$ and $\omega_{4,\ell 1}^{+(1,1,0)}$ are identical for the first seven digits after the decimal point displayed here. For all the modes except $\ell=2,n=2$, we found the relative difference between $\omega_{0,\ell 1}^{+(1,1,0)}$ and $\omega_{4,\ell 1}^{+(1,1,0)}$ [i.e., Eq.~\eqref{eq:relerror}] is $\lesssim10^{-6}$ for both the real and imaginary parts of the frequency. At $\ell=2,n=2$, the relative difference between $\omega_{0,\ell 1}^{+(1,1,0)}$ and $\omega_{0,\ell 1}^{+(1,1,0)}$ is $\sim 10^{-6}$ and $\sim10^{-5}$ for the real and imaginary part, respectively. Similar to the case of non-rotating BHs, the relative differences generally increase with the overtone number $n$. We want to emphasize again that the consistency across $\omega_{0,\ell 1}^{+(1,1,0)}$ and $\omega_{0,\ell 1}^{+(1,1,0)}$ is nontrivial since they are computed in two different gauges for two different NP quantities. Furthermore, if one computes the contribution of each piece within Eq.~\eqref{eq:QNM_shift_even} to the QNM frequency, such as $\mathscr{O}_{\geo}^{\ell m}$ and $\mathscr{Q}_{\geo}^{\ell m}$, they in general do not agree across the two gauges. Although we have associated physical meaning with each term in Eq.~\eqref{eq:QNM_shift_even}, it is only their sum a gauge invariant quantity in the end. In Sec.~\ref{sec:comparison_to_previous_results}, we will further show that the results using the modified Teukolsky equation are also highly consistent with the results using the ZM equation in \cite{Wagle:2021tam, Srivastava:2021imr}.

Furthermore, this is also the first time that the complete QNM frequencies of overtones are computed for slowly-rotating BHs in dCS gravity.\footnote{Reference~\cite{Srivastava:2021imr} also used the EVP method to calculate the overtone spectrum of slowly rotating BHs in dCS gravity but only for even-parity modes at $\ell=2$.} We observe from Table~\ref{tab:QNM_results_rotating_even} and Fig.~\ref{fig:even_scatter} that the real part of $\omega_{\ell 1}^{+(1,1,0)}$ increases with the overtone number $n$, while the imaginary part decreases with $n$ for all the $(\ell,m)$ modes we have studied. This is different from the GR case, where both the real and imaginary parts of $\omega_{\ell 1}^{(0,1,0)}$ increase with $n$ and stay positive, as one can observe in Table~\ref{tab:QNM_results_GR}. In Fig.~\ref{fig:even_overtone}, we also display the total, even-parity QNM frequencies $\omega_{\ell m}^{+}=\omega_{\ell m}^{(0,0)}+\zeta\omega_{\ell m}^{+(1,0)}$ as a function of $\chi$ and $\zeta$ after adding the dCS gravity corrections $\omega_{\ell m}^{(1,0)+}$, computed from the modified Teukolsky equation of $\Psi_0$ in the IRG, to the GR values $\omega_{\ell m}^{(0,0)}$ listed in Table~\ref{tab:QNM_results_GR}. Since $\omega_{\ell m}^{+(1,0,0)}=0$, all lines coincide at the point where $\chi=0$, i.e.~at zero spin. The imaginary part of $\omega_{\ell 1}^{+(1,1,0)}$ is negative for most of the modes, so the lines in most of the subplots shift downward when increasing $\zeta$. One exception is when $\ell=m=2$ and $n=0$, where $\omega_{\ell 1}^{+(1,1,0)}$ is still positive, so the lines shift upward when increasing $\zeta$. Moreover, since the damping time is inversely proportional to the magnitude of the imaginary part of the QNM frequency, it is evident from Fig.~\ref{fig:even_overtone} that for even-parity modes, the damping time of the $\ell=m=2,n=0$ mode increases (i.e.~the mode is longer lived), but that of all other modes decreases (i.e.~the mode is shorter lived) with respect to the GR case as the strength of dCS modification increases.

\subsubsection{Odd-parity modes}
\label{sec:QNM_slow_rotation_odd}

\begin{table*}[]
    \centering
    \begin{tabular}{cccccc}
        \hline\hline
        $\ell$ & Overtones & MTF (IRG) & MTF (ORG) 
        & RW \\
        \hline
        & $n=0$ &  $-0.0225978-0.0114406i$ & $-0.0226008-0.0114420i$ 
        &  $-0.0226009-0.0114401i$  \\
        $\ell=2$ & $n=1$ & $-0.0642208-0.0173969i$ & $-0.0641208-0.0173940i$ 
        & $-0.0642122-0.0174046i$  \\
        & $n=2$ & $-0.1624404+0.0523510i$ & $-0.1614131+0.0504675i$ 
        & $-0.1624372+0.0522496i$  \\
        \hline
        & $n=0$ & $-0.0459245-0.0039159i$ & $-0.0459350-0.0038795i$ 
        & $-0.0459340-0.0039073i$  \\
        $\ell=3$ & $n=1$ & $-0.0623013-0.0053067i$ & $-0.0624115-0.0054797i$ 
        & $-0.0623500-0.0053136i$ \\
        & $n=2$ & $-0.0944224+0.0143491i$ & $-0.0915884+0.0162253i$ 
        & $-0.0941648+0.0143046i$ \\
        \hline
        & $n=0$ & $-0.0785056+0.0005121i$ & $-0.0784914+0.0005517i$ 
        & $-0.0785223+0.0005359i$  \\
        $\ell=4$ & $n=1$ & $-0.0873724+0.0050579i$ & $-0.0871041+0.0051952i$ 
        & $-0.0873964+0.0050251i$  \\
        & $n=2$ & $-0.1049963+0.0202876i$ & $-0.1041546+0.0227117i$ 
        & $-0.1044711+0.0202965i$ \\
        \hline \hline
    \end{tabular}
    \caption{The QNM frequency shifts $\omega_{\ell 1}^{-(1,1,0)}$ for the odd-parity modes of a slowly-rotating BH in dCS gravity for $\ell=2,3,4$ and the overtones $n=0,1,2$. Similarly, the columns ``MTF (IRG)'' and ``MTF (ORG)'' contain the results in the IRG ($\omega_{0,\ell 1}^{-(1,1,0)}$) and the ORG ($\omega_{4,\ell 1}^{-(1,1,0)}$) of this work, respectively. The results in the column ``RW'' use the EVP method to solve for $\omega_{\ell 1}^{-(1,1,0)}$ from the RW equation and the scalar field equation of dCS gravity in \cite{Wagle:2021tam, Srivastava:2021imr}. We have set $M=1/2$ in this table.}
    \label{tab:QNM_results_rotating_odd}
\end{table*}

For odd-parity modes, similar to the case of non-rotating BHs in dCS gravity, one needs to solve the scalar field equation jointly with the modified Teukolsky equation. The general procedures are similar to the ones presented in Sec.~\ref{sec:QNM_nonrotating}. We first numerically integrate the asymptotic expansion of the radial part $\Theta_{\ell m}^{(1,1)}(r)$ of the scalar field perturbation $\vartheta^{(1,1)}$ from $\xi=\xi_{\min}$ to $\xi=\xi_{\max}$ and vice versa to obtain two independent solutions. From these two solutions, we use Green functions to construct the particular solution to Eqs.~\eqref{eq:scalar_radial_IRG_2_lm} and \eqref{eq:scalar_radial_ORG_lm} [i.e., Eq.~\eqref{eq:scalar_sol}], while $S_{\vartheta}^{\ell m}(r)$ now has additional $\mathcal{O}(\chi^1)$ terms, i.e.,
\begin{equation} \label{eq:scalar_slow_rotation_source_IRG}
    S_{\vartheta}^{\ell m}(r)=-\frac{4}{\mathcal{C}_2}\frac{r-r_s}{r^3}
    \left(V^R_{\ell m}(r)+V^{\square}_{\ell m}(r)\right)\,,
\end{equation}
where $V^R_{\ell m}(r)$ and $V^{\square}_{\ell m}(r)$ are radial functions in Eq.~\eqref{eq:scalar_source_IRG_2} involving ${}_{-2}R^{(0,1)}_{\ell m}(r)$, and similarly for the source term in the ORG. 

\begin{figure*}[t]
    \centering
    \includegraphics[width=0.95\linewidth]{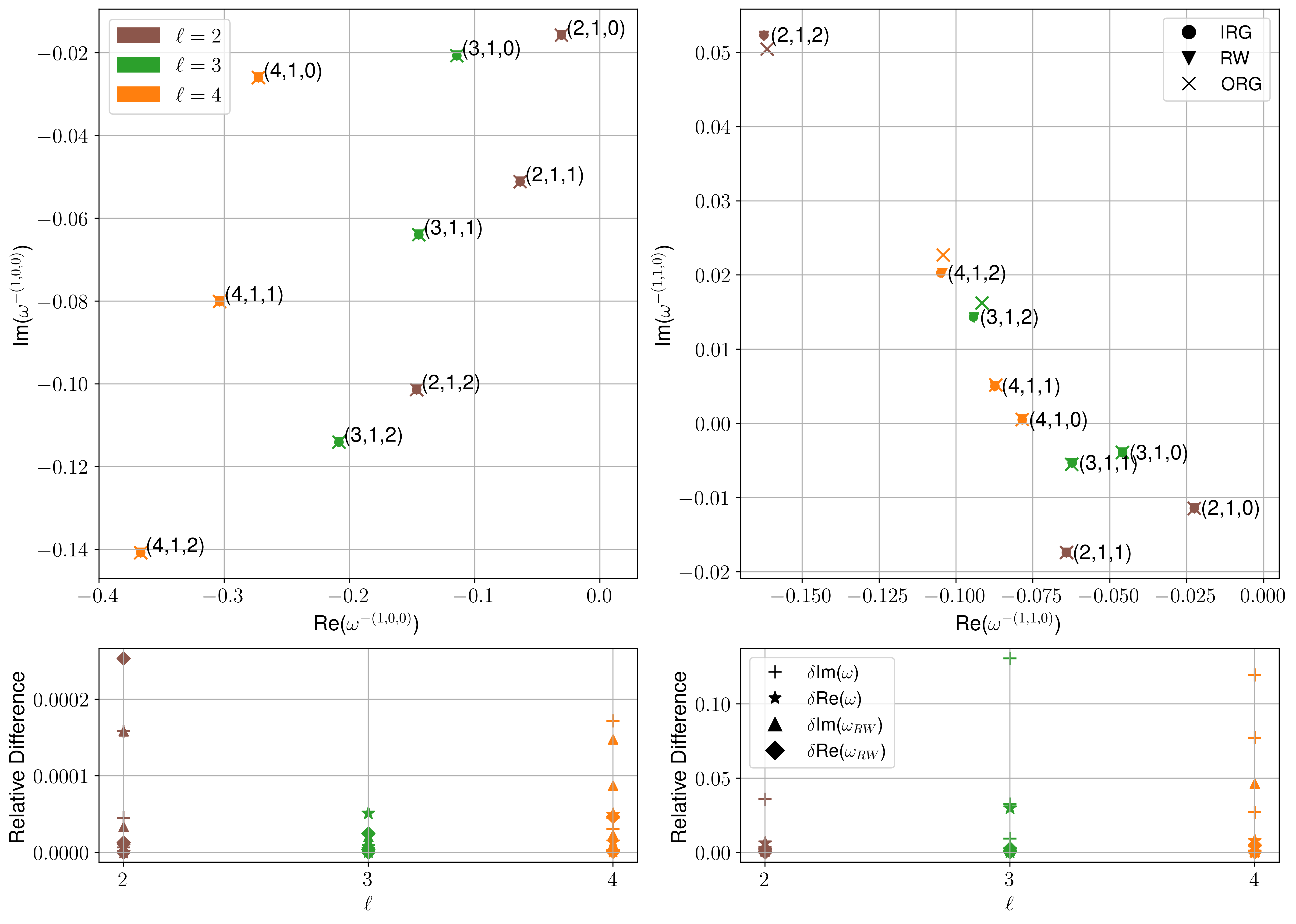}
    \caption{QNM frequency shifts for the odd-parity modes of a slowly rotating BH in dCS gravity up to $\mathcal{O}(\chi^1)$ for $\ell=2,3,4$ and overtones $n=0,1,2$. The index at each point labels the mode $(\ell,m,n)$. The top panels plot $\omega_{\ell 1}^{-(1,0,0)}$ (top left) and $\omega_{\ell 1}^{-(1,1,0)}$ (top right) computed in Secs.~\ref{sec:QNM_nonrotating} and \ref{sec:QNM_slow_rotation_odd} with their values listed in Tables~\ref{tab:QNM_results_nonrotating} and \ref{tab:QNM_results_rotating_odd}, respectively. The bottom panels show the relative differences [defined in Eq.~\eqref{eq:relerror}] of the QNM frequency shifts obtained in the ORG [marked as ``$\delta\rm{Im}(\omega)$" and ``$\delta\rm{Re}(\omega)$''] and the RW gauge [marked as ``$\delta\rm{Im}(\omega_{RW})$" and ``$\delta\rm{Re}(\omega_{RW})$''] from the ones in the IRG, respectively, for $\omega_{\ell 1}^{-(1,0,0)}$ (bottom left) and $\omega_{\ell 1}^{-(1,1,0)}$ (bottom right). In this plot, the legends in the left panels apply to the corresponding right panels, and vice versa. All the other labels share the same meaning as Fig.~\ref{fig:summary_scatter}. We have set $M=1/2$ in this plot.}
    \label{fig:odd_scatter}
\end{figure*}

\begin{figure*}[t]
    \centering
    \includegraphics[width=\linewidth]{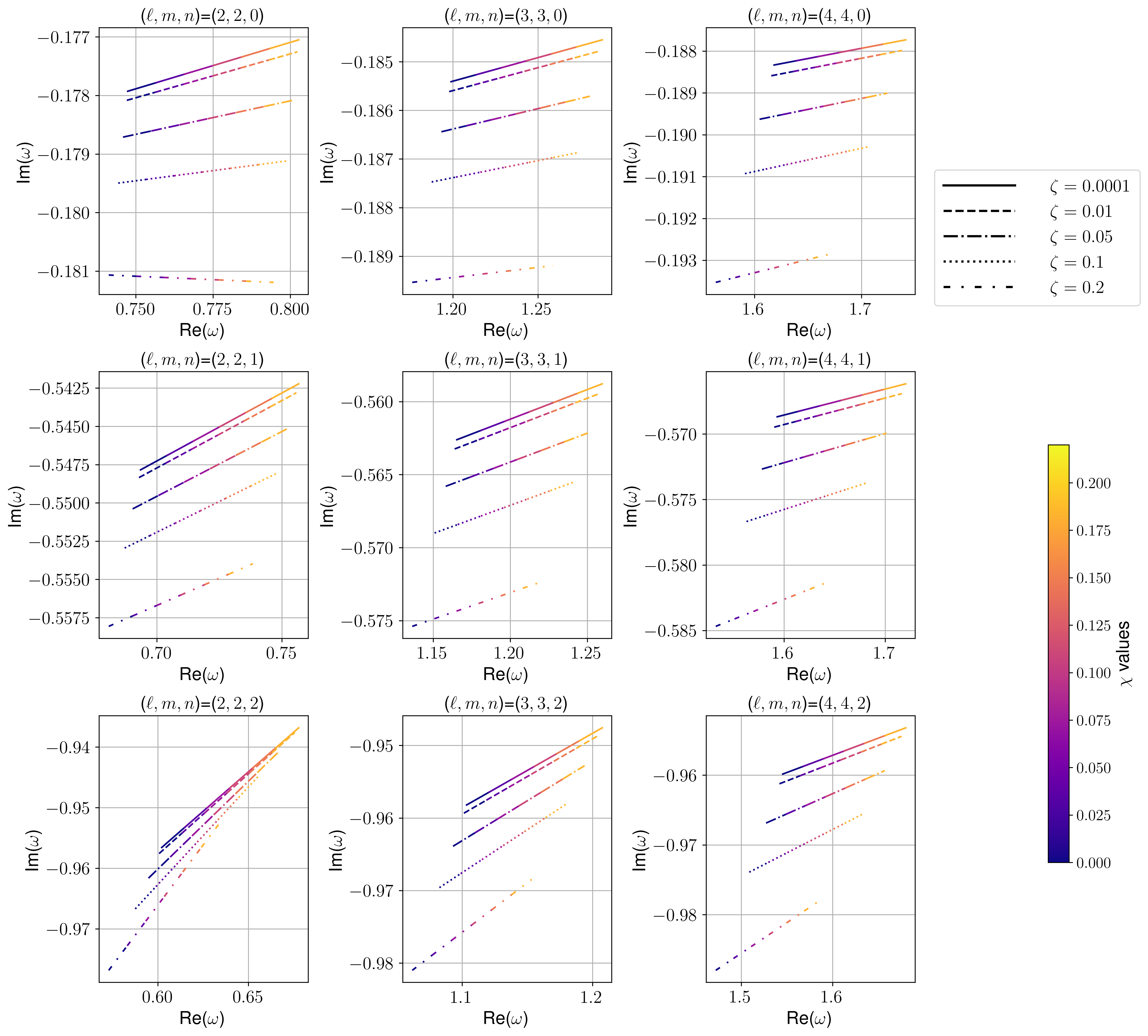}
    \caption{Total QNM frequencies $\omega_{\ell m}^{-}=\omega_{\ell m}^{(0,0)}+\zeta\omega_{\ell m}^{-(1,0)}$ for odd-parity modes at $\ell=m=2,3,4$ and overtones $n=0,1,2$ up to $\mathcal{O}(\chi^1)$ in dCS gravity. The values of $\omega_{\ell m}^{(0,0)}$ are retrieved from Table~\ref{tab:QNM_results_GR}. The values of $\omega_{\ell m}^{-(1,0)}$ are found using Eq.~\eqref{eq:QNM_slow_expansion_dCS_odd} and the ``MTF (IRG)'' results in Tables~\ref{tab:QNM_results_nonrotating} and \ref{tab:QNM_results_rotating_odd}. The line color and style have the same meaning as the ones in Fig.~\ref{fig:even_overtone} with the same set of $\zeta$ and $a$ being used. We have set $M=1/2$ in this plot.}
    \label{fig:odd_overtone}
\end{figure*}

To solve Eqs.~\eqref{eq:scalar_radial_IRG_2_lm} and \eqref{eq:scalar_radial_ORG_lm}, we need to first compute ${}_{\pm 2}R^{(0,1)}_{\ell m}(r)$ up to $\mathcal{O}(\chi^1)$ on the contour $r_{\mathscr{C}}(\xi)$ in Eq.~\eqref{eq:contour}. Similar to Sec.~\ref{sec:QNM_nonrotating}, one can use Leaver's method to calculate ${}_{\pm 2}R^{(0,1)}_{\ell m}(r)$ on the real line and analytically continue it to $r_{\mathscr{C}}(\xi)$. In this work, we instead choose to use the shooting method to numerically solve ${}_{\pm 2}R^{(0,1)}_{\ell m}(r)$ along $r_{\mathscr{C}}(\xi)$ for a fair comparison with the results using the modified RW/ZM equations, as we will discuss in Sec.~\ref{sec:perturb_eqn_metric}. This is largely due to the fact that Leaver's method for an RW/ZM equation in the slow-rotation expansion is not known to the authors' best knowledge. More specifically, we first make an initial guess of the QNM frequency and construct asymptotic expansions of ${}_{\pm 2}R^{(0,1)}_{\ell m}(r)$ at $\xi=\xi_{\min}$ and $\xi=\xi_{\max}$, respectively. We then numerically evolve these two expansions to $\xi=0$ using the left-hand side of Eqs.~\eqref{eq:master_eqn_Psi0_radial_simplify} and \eqref{eq:master_eqn_Psi4_radial_simplify}, which are Teukolsky equations in GR expanded to $\mathcal{O}(\chi^1)$. At the correct QNM frequencies, these two solutions become linearly dependent, so their Wronskian must vanish. By searching for frequencies that are zeros of the Wronskian, we obtain the GR QNM spectra and the associated radial functions ${}_{\pm 2}R^{(0,1)}_{\ell m}(r)$ along $r_{\mathscr{C}}(\xi)$. 

To test the accuracy of our shooting method, we have checked our GR QNM spectra against known results computed using Leaver's method in \cite{BertiRingdown, Berti:2005ys, Berti:2009kk}. We also cross-check the QNM spectrum obtained either using the Teukolsky equation for $\Psi_0$ or $\Psi_4$. At $\chi=0$, our GR QNM frequencies for $\ell=2,3,4$ and $n=0,1,2$ match all the digits of the corresponding results using $\Psi_0$ or $\Psi_4$ in \cite{BertiRingdown, Berti:2005ys, Berti:2009kk}. The relative differences [i.e., Eq.~\eqref{eq:relerror}] between the results using $\Psi_0$ or $\Psi_4$ are $\lesssim 10^{-19}$. At $\chi=10^{-4}$, where we extract $\mathcal{O}(\zeta^1,\chi^1)$ dCS corrections to the QNM spectra, the relative differences between our GR QNM frequencies and the results in \cite{BertiRingdown, Berti:2005ys, Berti:2009kk} are $\lesssim 10^{-9}$ for all the modes considered in this work. Since we use the slow-rotation expansion of Teukolsky equations, the relative differences between the results using $\Psi_0$ or $\Psi_4$ are $\lesssim 10^{-10}$ in this case.

Using ${}_{\pm 2}R^{(0,1)}_{\ell m}(r)$ to obtain $S_{\vartheta}^{\ell m}(r)$, we can now compute $\Theta_{\ell m}^{(1,1)}(r)$ up to $\mathcal{O}(\chi)$. After obtaining $\Theta_{\ell m}^{(1,1)}(r)$ and ${}_{\pm 2}R_{\ell m}(r)$, we use Eq.~\eqref{eq:QNM_shift} to compute the QNM frequency shifts in dCS gravity. For odd-parity modes, Eq.~\eqref{eq:QNM_shift} reduce to
\begin{subequations} \label{eq:QNM_shift_odd_slow}
\begin{align} 
    & \omega_{0,\ell m}^{\pm(1,0)}
    =\frac{\left\langle -2r^2\left(\mathscr{O}_{\geo}^{\ell m}
    +2\mathscr{O}_{T_{\vartheta}}^{\ell m}+\mathscr{O}_{T_{\Psi}}^{\ell m}
    -\tilde{\mathscr{O}}_{T_{\Psi}}^{\ell m}\right)\right\rangle}
    {\left\langle\partial_{\omega}\mathcal{H}_0^{\ell m}\right\rangle}\,, 
    \label{eq:QNM_shift_Psi0_odd_slow} \\
    & \omega_{4,\ell m}^{\pm(1,0)}
    =\frac{\left\langle -2r^6\left(\mathscr{Q}_{\geo}^{\ell m}
    +2\mathscr{Q}_{T_{\vartheta}}^{\ell m}+\mathscr{Q}_{T_{\Psi}}^{\ell m}
    -\tilde{\mathscr{Q}}_{T_{\Psi}}^{\ell m}\right)\right\rangle}
    {\left\langle\partial_{\omega}\mathcal{H}_4^{\ell m}\right\rangle}\,.
    \label{eq:QNM_shift_Psi4_odd_slow}
\end{align}    
\end{subequations}
The contributions of $\{\mathscr{O}_{\geo}^{\ell m},\mathscr{O}_{T_{\Psi}}^{\ell m},\tilde{\mathscr{O}}_{T_{\Psi}}^{\ell m},\mathscr{Q}_{\geo}^{\ell m},\mathscr{Q}_{T_{\Psi}}^{\ell m},\tilde{\mathscr{Q}}_{T_{\Psi}}^{\ell m}\}$ were computed in Sec.~\ref{sec:QNM_slow_rotation_even} for even-parity modes with ${}_{\pm2}R_{\ell m}^{(0,1)}(r)$ generated by Leaver's method. We recomputed these terms using ${}_{\pm2}R_{\ell m}^{(0,1)}(r)$ found by the shooting method and confirmed that they agree well with the results using Leaver's method. The only new contributions are from $\mathscr{O}_{T_{\vartheta}}^{\ell m}$ and $\mathscr{Q}_{T_{\vartheta}}^{\ell m}$, which are the terms coupled to the scalar field perturbations $\vartheta^{(1,1)}$. Since $\omega_{0,\ell m}^{-(1,0,0)}$ and $\omega_{4,\ell m}^{-(1,0,0)}$ were computed in Sec.~\ref{sec:QNM_nonrotating}, our goal here is to determine $\omega_{i,\ell 1}^{-(1,1,0)}$ ($i=0,4$) in Eq.~\eqref{eq:QNM_slow_expansion_dCS}. This can be simply done by numerically computing $\partial_\chi\omega_{i,\ell 1}^{-(1,0)}$ using the values of $\omega_{i,\ell 1}^{-(1,0)}$ at two different spins, i.e.,
\begin{equation}
    \omega_{i,\ell 1}^{-(1,1,0)}\approx
    \frac{\omega_{i,\ell 1}^{-(1,0)}(\chi_2)
    -\omega_{i,\ell 1}^{-(1,0)}(\chi_1)}{\chi_2-\chi_1}\,.
\end{equation}
For sufficient small $\delta\chi=\chi_2-\chi_1$, one can obtain $\omega_{i,\ell 1}^{-(1,1,0)}$ very accurately. Since the operators in Eq.~\eqref{eq:QNM_shift_odd_slow} only include terms up to $\mathcal{O}(\chi^1)$, they are only valid for small spin. Thus, we choose $\chi_1=0$ and a $\chi_2\ll1$. Furthermore, when solving $\Theta_{\ell m}^{(1,1)}(r)$ from Eqs.~\eqref{eq:scalar_radial_IRG_2_lm} and \eqref{eq:scalar_radial_ORG_lm} and evaluating the inner products of Eq.~\eqref{eq:QNM_shift_odd_slow}, we choose not to expand ${}_{\pm 2}R^{(0,1)}_{\ell m}(r)$ about $\chi$ explicitly as it is not straightforward to directly compute the $\mathcal{O}(\chi^1)$ contribution ${}_{\pm 2}R^{(0,1,1)}_{\ell m}(r)$. For this reason, Eq.~\eqref{eq:QNM_shift_odd_slow} implicitly contains $\mathcal{O}(\chi^2)$ contributions. However, Eqs.~\eqref{eq:scalar_radial_IRG_2_lm}, \eqref{eq:scalar_radial_ORG_lm}, \eqref{eq:master_eqn_Psi0_radial_simplify}, and \eqref{eq:master_eqn_Psi4_radial_simplify} are only valid up to $\mathcal{O}(\chi^1)$, so we only expect $\omega_{0,\ell m}^{\pm(1,0)}$ and $\omega_{4,\ell m}^{\pm(1,0)}$ to be correct and the calculation to be self-consistent up to $\mathcal{O}(\chi^1)$ in this work. Thus, those implicit $\mathcal{O}(\chi^2)$ contributions could result in small discrepencies between $\omega_{0,\ell m}^{-(1,0)}$ and $\omega_{4,\ell m}^{-(1,0)}$, which we also would like to suppress by choosing a small $\chi_2$.

Due to the coupling between the scalar field and $\Psi_{0,4}$, the computational cost for the odd-parity modes is much higher than the even-parity modes, so we have only extracted $\omega_{i,\ell m}^{\pm(1,1,0)}$ ($i=0,4$) at $\chi_2=10^{-3}$ and $\chi_2=10^{-4}$. A more careful calculation can be done by evaluating $\omega_{i,\ell m}^{\pm(1,0)}$ at more small spins, from which one then fits $\omega_{i,\ell m}^{\pm(1,1,0)}$. Despite this, we observe that the relative difference between $\omega_{0,\ell m}^{-(1,1,0)}$ and $\omega_{4,\ell m}^{-(1,1,0)}$ generally decreases with $\chi_2$ as expected. For example, for all the fundamental modes ($n=0$), the relative difference between $\omega_{0,\ell m}^{-(1,1,0)}$ and $\omega_{4,\ell m}^{-(1,1,0)}$ decreases by one order of magnitude from $\chi_2=10^{-3}$ to $\chi_2=10^{-4}$. For this reason, we use the values of $\omega_{0,\ell m}^{-(1,1,0)}$ and $\omega_{4,\ell m}^{-(1,1,0)}$ extracted at $\chi_2=10^{-4}$ as our results in Table~\ref{tab:QNM_results_rotating_odd}.

From Table~\ref{tab:QNM_results_rotating_odd} and Fig.~\ref{fig:odd_scatter}, we notice that $\omega_{0,\ell m}^{-(1,1,0)}$ and $\omega_{4,\ell m}^{-(1,1,0)}$ generally agree well with each other, validating the expectation that the QNM spectra computed from the modified Teukolsky equation are independent of the gauge chosen when reconstructing $h_{\mu\nu}^{(0,1)}$. Nonetheless, due to the coupling to the scalar field, the mismatch between $\omega_{0,\ell m}^{-(1,1,0)}$ and $\omega_{4,\ell m}^{-(1,1,0)}$ is generally larger than the mismatch between $\omega_{0,\ell m}^{+(1,1,0)}$ and $\omega_{4,\ell m}^{+(1,1,0)}$. The relative differences between the real and imaginary parts of $\omega_{0,\ell m}^{-(1,1,0)}$ and the ones of $\omega_{4,\ell m}^{-(1,1,0)}$ [i.e., Eq.~\eqref{eq:relerror}] are $\sim 10^{-4}$ at $\ell=2$, $n=0$ and $\lesssim 3\%$ for most of the modes. At $\ell=3,n=2$ and $\ell=4,n=2$, the relative difference in the real part is still $\lesssim 3\%$, while the relative difference in the imaginary part is $\sim 10\%$, largely due to the fact that ${}_{\pm 2}R^{(0,1)}_{\ell m}(r)$ becomes more oscillatory for larger $n$. Another exception is when $\ell=4,n=0$, where the relative differences between $\omega_{0,\ell m}^{-(1,1,0)}$ and $\omega_{4,\ell m}^{-(1,1,0)}$ in their real and imaginary parts are $\sim 10^{-4}$ and $\sim 8\%$, respectively. This mismatch in the imaginary part is comparable to the case when $\ell=4,n=2$ since $\left|\textrm{Im}\left[\omega_{\ell m}^{-(1,1,0)}\right]\right|\ll\left|\textrm{Re}\left[\omega_{\ell m}^{-(1,1,0)}\right]\right|$ when $\ell=4,n=0$, making the extraction of $\left|\textrm{Im}\left[\omega_{\ell m}^{-(1,1,0)}\right]\right|$ more challenging numerically. For the same reason, the mismatch in $\textrm{Im}\left[\omega_{\ell m}^{-(1,1,0)}\right]$ is much larger than $\textrm{Re}\left[\omega_{\ell m}^{-(1,1,0)}\right]$ for several other modes, where $\left|\textrm{Im}\left[\omega_{\ell m}^{-(1,1,0)}\right]\right|$ is one magnitude less than $\left|\textrm{Re}\left[\omega_{\ell m}^{-(1,1,0)}\right]\right|$.

In terms of how $\omega_{\ell 1}^{-(1,1,0)}$ changes with respect to the overtone number $n$, we observe from Table~\ref{tab:QNM_results_rotating_odd} and Fig.~\ref{fig:odd_scatter} that the real part of $\omega_{\ell 1}^{-(1,1,0)}$ always decreases with $n$ and stays negative, different from the behavior of $\omega_{\ell 1}^{(0,1,0)}$ in GR. The imaginary part of $\omega_{\ell 1}^{-(1,1,0)}$ generally increases with $n$, but with exceptions at $\ell=2,n=1$ and $\ell=3,n=1$. Moreover, comparing $\omega_{\ell 1}^{+(1,1,0)}$ in Table~\ref{tab:QNM_results_rotating_even} to $\omega_{\ell 1}^{-(1,0,0)}$ and $\omega_{\ell 1}^{-(1,1,0)}$ in Tables~\ref{tab:QNM_results_nonrotating} and \ref{tab:QNM_results_rotating_odd}, respectively, we observe that $\left|\textrm{Re}\left[\omega_{\ell m}^{+(1,1,0)}\right]\right|\ll\left|\textrm{Re}\left[\omega_{\ell m}^{-(1,0,0)}\right]\right|,\left|\textrm{Re}\left[\omega_{\ell m}^{-(1,1,0)}\right]\right|$ and similarly for the imaginary part of most of the modes. This feature is possibly due to the scalar field perturbation only driving odd-parity modes. Together with $\omega_{\ell m}^{+(1,1,0)}=0$, we expect to observe QNM frequency shifts mostly from odd-parity modes for slowly-rotating BHs in dCS gravity, implying we may need highly asymmetric systems to excite more odd-parity perturbations to detect these frequency shifts. Nevertheless, this asymmetry between even- and odd-parity modes requires further investigation for fast rotating BHs in our follow-up work.

In Fig.~\ref{fig:odd_overtone}, we also display the total odd-parity QNM frequencies $\omega_{\ell m}^{-}=\omega_{\ell m}^{(0,0)}+\zeta\omega_{\ell m}^{-(1,0)}$, where the dCS corrections $\omega_{\ell m}^{(1,0)-}$ come from the modified Teukolsky equation of $\Psi_0$ in the IRG, and the GR values $\omega_{\ell m}^{(0,0)}$ are retrieved from Table~\ref{tab:QNM_results_GR}. We first observe that the left endpoints of all lines shift leftward and downward when increasing $\zeta$, consistent with our finding in Sec.~\ref{sec:QNM_nonrotating} that $\omega_{\ell m}^{-(1,0,0)}$ is negative for all the modes we have studied. We also see that for certain modes, such as the $\ell=2,n=0$ one, the line flips direction at high $\zeta$. This is due to the overall sign difference between $\omega_{\ell m}^{(0,1,0)}$ and $\omega_{\ell m}^{-(1,1,0)}$, which together determine the slope of these lines. Since our calculation is only expected to work for small $\zeta$ (within the cut-off scale of the dCS EFT), this feature needs further investigation after including nonlinear contributions of $\zeta$ in future work.

In summary, adapting the EVP method to the modified Teukolsky equation, we computed $\omega_{0,\ell m}^{\pm (1,0)}$ and $\omega_{4,\ell m}^{\pm (1,0)}$ up to $\mathcal{O}(\chi)$ for $\ell=2,3,4$ and $n=0,1,2$. This is the first time that the QNM frequencies of overtones in dCS gravity are studied. For all the corrections to the QNM spectra up to $\mathcal{O}(\zeta^1,\chi^1)$, we found great agreement between the results using the IRG ($\omega_{0,\ell m}^{\pm (1,0)}$) and the ORG ($\omega_{4,\ell m}^{\pm (1,0)}$) for metric reconstruction for both even- and odd-parity modes. This agreement across two gauges serves as a good self-consistency check of our calculations. We also noticed that for higher overtones, due to the more oscillatory behavior of the wave functions along the contour, the mismatch between $\omega_{0,\ell m}^{\pm (1,0)}$ and $\omega_{4,\ell m}^{\pm (1,0)}$ becomes larger, so better numerical methods need to be developed for more accurately extracting the QNM spectra of those modes. In the next section, we will compare these results to the ones using the metric perturbation approach in \cite{Cardoso:2009pk, Molina:2010fb, Pani:2011xj, Wagle:2021tam, Srivastava:2021imr}.

\section{Comparison to previous results}
\label{sec:comparison_to_previous_results}

Though this work presents the first calculation of QNM frequencies in dCS gravity using the modified Teukolsky formalism, there were attempts in the past to calculate the frequencies of QNMs in dCS gravity for non-rotating and slowly-rotating BHs using metric perturbations \cite{Cardoso:2009pk, Molina:2010fb, Pani:2011xj, Wagle:2021tam, Srivastava:2021imr}. Instead of studying curvature perturbation equations, these previous works relied on the metric perturbation approach \cite{Regge:1957td}. In this section, we will review the results in these previous works and compare the QNM frequencies obtained in the past to the QNM frequencies obtained through the modified Teukolsky formalism.

\subsection{Perturbation equations from a metric perturbation approach}
\label{sec:perturb_eqn_metric}

In this subsection, we review the evolution equations obtained in \cite{Wagle:2021tam, Srivastava:2021imr} for studying perturbations of slowly-rotating BHs in dCS gravity. In \cite{Wagle:2021tam}, the radial differential equations for the scalar, axial, and polar perturbations of a slowly-rotating BH in dCS gravity, respectively, were found to be
\begin{subequations}
\label{eqs:perturbed_final_Wagle}
\begin{align}
    & \left[f(r)^2\partial_{r}^2+\frac{2M}{r^2}f(r)\partial_{r}
    +\omega^2-V^S_{\rm eff}(r,\chi,\zeta)\right]
    \Theta_{\ell m}(r) \nonumber \\ 
    & =\zeta^{\frac{1}{2}}M^2\kappa^{\frac{1}{2}}f(r)
    \left[g(r)+\chi Mm\left(h(r)+j(r)\partial_{r}\right)\right]
    \Psi^{\RW}_{\ell m}(r)\,, \label{eq:SF} \\
    & \left[f(r)^2\partial_{r}^2+\frac{2M}{r^2}f(r)\partial_{r}
    +\omega^2-V^A_{\rm eff}(r,\chi,\zeta)\right]
    \Psi^{\RW}_{\ell m}(r)\nonumber \\ 
    & =\zeta^{\frac{1}{2}}M^2\kappa^{\frac{1}{2}}f(r)
    \left[v(r)+\chi Mm\left(n(r)+p(r)\partial_{r}\right)\right]
    \Theta_{\ell m}(r)\,, \label{eq:RW} \\
    & \left[f(r)^2\partial_{r}^2+\frac{2M}{r^2}f(r)\partial_{r}
    +\omega^2-V^P_{\rm eff}(r,\chi,\zeta)\right]
    \Psi^{\ZM}_{\ell m}(r)=0 \,,\label{eq:ZM}	
\end{align}
\end{subequations}
where $\Theta_{\ell m}(r)$, $\Psi^{\RW}_{\ell m}(r)$, and $\Psi^{\ZM}_{\ell m}(r)$ are the radial part of the scalar field, the RW function, and the ZM function, respectively. The dimensionless coupling constant $\zeta$ is defined in Eq.~\eqref{eq:zeta}\footnote{Though~\cite{Wagle:2021tam} presents these equations using the dCS coupling constant $\alpha$, we chose to introduce the master equations using the dimensionless coupling constant $\zeta$ to maintain consistent notation throughout this work.} and $\kappa = (16\pi)^{-1}$. The radial function $f(r)=1-2M/r$, while $g(r)$, $h(r)$, $j(r)$, $v(r)$, $n(r)$, and $p(r)$ are presented in the Appendix of \cite{Wagle:2021tam}. The effective potentials $V^i_{\rm eff}(r,\chi,\zeta)$, where $i\in\left\{S,A,P \right\}$, are listed in \cite{Wagle:2021tam} and can be further expanded as 
\begin{align}
    V^S_{\rm eff}(r,\chi,\zeta) 
    &=V^{S(0,0)}_{\rm eff}(r,\chi)+\zeta V^{S(1,0)}_{\rm eff}(r,\chi)\,, \nonumber \\
    V^A_{\rm eff}(r,\chi,\zeta) 
    &=V^{A(0,0)}_{\rm eff}(r,\chi)+\zeta V^{A(1,0)}_{\rm eff}(r,\chi)\,, \nonumber \\
    V^P_{\rm eff}(r,\chi,\zeta) 
    &=V^{P(0,0)}_{\rm eff}(r,\chi)+\zeta V^{P(1,0)}_{\rm eff}(r,\chi)\,.
\end{align}
We then apply the two-parameter expansion defined in Eqs.~\eqref{eq:expansion_scalar} and \eqref{eq:expansion_NP} to Eq.~\eqref{eqs:perturbed_final_Wagle} and keep the terms up to $\mathcal{O}(\zeta^1,\epsilon^1)$ to obtain 
\begin{widetext}
\begin{subequations}
\label{eqs:perturbed_final_Wagle_expanded}
\begin{align} 
    \mathcal{H}_{S}^{\ell m}\Theta_{\ell m}^{(1,1)}(r) 
    &=M^2\kappa^{\frac{1}{2}}f(r)
    \left[g(r)+\chi Mm\left(h(r)+j(r)\partial_{r}\right)\right] \Psi^{\RW(0,1)}_{\ell m}(r)\,, \label{eq:SF_expand} \\
    \mathcal{H}_{A}^{\ell m}\Psi_{\ell m}^{\RW(1,1)}(r) 
    &=M^2\kappa^{\frac{1}{2}}f(r)
    \left[v(r)+\chi Mm\left(n(r)+p(r)\partial_{r}\right)\right]
    \Theta_{\ell m}^{(1,1)}(r)
    +V^{A(1,0)}_{\rm eff}(r,\chi)\Psi_{\ell m}^{\RW(0,1)}(r)\,,
    \label{eq:RW_expand} \\
    \mathcal{H}_{P}^{\ell m}\Psi_{\ell m}^{\ZM(1,1)}(r) 
    &=V^{P(1,0)}_{\rm eff}(r,\chi)\Psi_{\ell m}^{\ZM(0,1)}(r)\,,
    \label{eq:ZM_expand}
\end{align}
\end{subequations}
\end{widetext}
where we have introduced the differential operators
\begin{equation}
    \mathcal{H}_{i}^{\ell m}
    =f(r)^2\partial_{r}^2+\frac{2M}{r^2}f(r)\partial_{r}
    +\omega^2-V^{i(0,0)}_{\rm eff}(r,\chi)\,,
\end{equation}
with $i\in\left\{S,A,P\right\}$. Although $\Theta_{\ell m}(r)$ enters at $\mathcal{O}(\zeta^{1/2})$, we have followed the convention in \cite{Wagle:2021tam} to multiply a factor of $\zeta^{1/2}$ to Eq.~\eqref{eq:SF} and redefine $\zeta^{1/2}\Theta_{\ell m}(r)\rightarrow\Theta_{\ell m}(r)$ such that we can use the expansion scheme in Eq.~\eqref{eq:expansion_scalar}. 

The perturbed field equations describing the evolution of metric perturbations were also independently obtained in~\cite{Srivastava:2021imr}. Following~\cite{Srivastava:2021imr}, expanding the corresponding field equations for the scalar, axial, and polar perturbations, we have
\begin{widetext}
\begin{subequations}
\label{eqs:perturbed_final_Manu_expanded}
\begin{align}
    \mathcal{H}_{S}^{\ell m}\Theta_{\ell m}^{(1,1)}(r)
    & =M^2\kappa^{\frac{1}{2}}f(r) 
    \left[g(r)+\chi Mm\left(h(r)+j(r)\partial_{r}\right)\right]
    \Psi^{\RW(0,1)}_{\ell m}(r)\,, \label{eq:SFmanu_expand}	\\	
    \mathcal{H}_{A}^{\ell m}\Psi^{\RW(1,1)}_{\ell m}(r)
    & =M^2\kappa^{\frac{1}{2}}f(r)
    \left[v(r)+\chi Mm\left(n(r)+p(r)\partial_{r}\right)\right]
    \Theta_{\ell m}^{(1,1)}(r)
    +\chi m\kappa\left[A_2(r)-A_1(r)\partial_{r}\right]
    \Psi^{\RW(0,1)}_{\ell m}(r)\,, \label{eq:RWmanu_expand}\\	
    \mathcal{H}_{P}^{\ell m}\Psi^{\ZM(1,1)}_{\ell m}(r)
    & =\chi m\kappa\left[P_2(r)-P_1(r)\partial_{r}\right]
    \Psi^{\ZM(0,1)}_{\ell m}(r)\,, \label{eq:ZMmanu_expand}
\end{align}
\end{subequations}
\end{widetext}
where $A_1(r)$, $A_2(r)$, $P_1(r)$, and $P_2(r)$ are radial functions defined in~\cite{Srivastava:2021imr}. Although Eqs.~\eqref{eqs:perturbed_final_Wagle_expanded} and Eqs.~\eqref{eqs:perturbed_final_Manu_expanded} seem to hold different forms, we show in Appendix~\ref{appendix:equivalence_between_eqs} that these two sets of equations are equivalent after a redefinition of $\Psi^{\RW(1,1)}_{\ell m}(r)$ and $\Psi^{\ZM(1,1)}_{\ell m}(r)$. Since we have shown the equivalence between Eqs.~\eqref{eqs:perturbed_final_Wagle_expanded} and Eqs.~\eqref{eqs:perturbed_final_Manu_expanded}, here on, we will only use Eqs.~\eqref{eqs:perturbed_final_Wagle_expanded} obtained in~\cite{Wagle:2021tam} and the results would therefore extend to Eqs.~\eqref{eqs:perturbed_final_Manu_expanded} obtained in~\cite{Srivastava:2021imr}.

Note that in Eq.~\eqref{eq:SF_expand} we have dropped the term $V_{\rm eff}^{S(1,0)}(r,\chi)$ since it contributes at $\mathcal{O}(\zeta^{3/2})$ [or $\mathcal{O}(\zeta^2)$ if we absorb a factor of $\zeta^{1/2}$ into Eq.~\eqref{eq:SF}]. Similarly, when presenting the scalar field equation [i.e., Eqs.~\eqref{eq:scalar_radial_IRG_2_lm} and \eqref{eq:scalar_radial_ORG_lm}] in Sec.~\ref{sec:scalar_eqn}, the sources of which are constructed from the reconstructed metric in the IRG or the ORG, we have not included any $\mathcal{O}(\zeta^1)$ corrections to the potential of the scalar field equation as they contribute at $\mathcal{O}(\zeta^{3/2})$. Nonetheless, if one follows Eq.~\eqref{eq:QNM_slow_expansion_dCS} to expand the scalar QNM frequency $\omega_{\vartheta,\ell m}$, i.e.,
\begin{align}
    \omega_{\vartheta,\ell m}
    =& \;\omega_{\vartheta,\ell m}^{(0,0)}
    +\zeta\omega_{\vartheta,\ell m}^{(1,0)}+\mathcal{O}(\zeta^2) \nonumber\\
    =& \;\left(\omega_{\vartheta,\ell m}^{(0,0,0)}
    +\chi m\omega_{\vartheta,\ell 1}^{(0,1,0)}\right)
    +\zeta\left(\omega_{\vartheta,\ell m}^{(1,0,0)}
    +\chi m\omega_{\vartheta,\ell 1}^{(1,1,0)}\right) \nonumber\\
    &\; +\mathcal{O}(\zeta^2,\chi^2)\,,
\end{align}
one notices that $V_{\rm eff}^{S(1,0)}(r,\chi)$ contributes to $\omega_{\vartheta,\ell m}^{(1,0)}$, which are the frequency shifts of the scalar-led modes observed in \cite{Wagle:2021tam, Srivastava:2021imr}. Instead of calculating the odd-parity gravitational-led modes discussed in Sec.~\ref{sec:EVP_metric}, one could first solve Eq.~\eqref{eq:RW_expand} with $\Theta_{\ell m}^{(1,1)}(r)$ being the homogeneous solution $\Theta_{\ell m}^{H(1,1)}(r)$ of Eq.~\eqref{eq:SF_expand} and oscillating at $\omega_{\vartheta,\ell m}^{(0,0)}$. One can also ignore the second term in Eq.~\eqref{eq:RW_expand} since $\Psi_{\ell m}^{\RW(0,1)}(r)$ always oscillates at gravitational QNM frequencies. Then, one can expand Eq.~\eqref{eq:SF} to $\mathcal{O}(\zeta^{3/2})$, where $V_{\rm eff}^{S(1,0)}(r,\chi)$ now contributes, and insert $\Theta_{\ell m}^{H(1,1)}(r)$ and the particular solution $\Psi_{\ell m}^{\RW(1,1)}(r)$ of Eq.~\eqref{eq:RW_expand}, both at $\omega_{\vartheta,\ell m}^{(0,0)}$, into the source terms. The same inner product in Eqs.~\eqref{eq:inner_product} and \eqref{eq:inner_product_simplify} with $s=0$ can now be applied to this equation to obtain $\omega_{\vartheta,\ell m}^{(1,0)}$. On the other hand, to apply similar procedures to the modified Teukolsky equation in this paper, we need to reconstruct the metric at $\mathcal{O}(\zeta^1,\epsilon^1)$ [i.e., the contribution from $\left[R^*\!R\right]^{(1,1)}$ when expanding Eq.~\eqref{eq:EOM_theta} to $\mathcal{O}(\zeta^{3/2})$]. However, the corresponding metric reconstruction procedures have not been developed yet, so we will not compute $\omega_{\vartheta,\ell m}^{(1,0)}$ in this work and leave it to future work.

\subsection{Applying the EVP method to metric perturbation equations}
\label{sec:EVP_metric}

In this subsection, we apply the EVP method to Eqs.~\eqref{eqs:perturbed_final_Wagle_expanded} and \eqref{eqs:perturbed_final_Manu_expanded} to obtain the QNM spectra in dCS gravity. One can, in principle, use any numerical method to solve these equations, e.g., the direct integration method with shooting and/or the Wronskian technique \cite{Wagle:2021tam}. We here opt to implement the EVP method on these equations to allow for common ground on which to compare the results in Sec.~\ref{sec:QNM_freq} with the results in \cite{Wagle:2021tam, Srivastava:2021imr}.

In essence, the procedures for applying the EVP method to the metric perturbation equations remain similar to the procedures presented in Secs.~\ref{sec:EVP} and \ref{sec:EVP_dCS}. Following Sec.~\ref{sec:QNM_freq}, we perform the same expansion of $\omega_{\ell m}$ in Eq.~\eqref{eq:QNM_slow_expansion_dCS}. From Eq.~\eqref{eqs:perturbed_final_Wagle_expanded}, we find that the QNM shifts due to dCS gravity can be expressed as 
\begin{widetext}
\begin{subequations} \label{eq:QNM_shift_metric}
\begin{align} 
    & \omega_{\metric,\ell m}^{-(1,0)}
    =\frac{\left\langle V^{A(1,0)}_{\rm eff}(r,\chi)
    +M^2\kappa^{\frac{1}{2}}f(r)
    \left[v(r)+\chi Mm\left(n(r)+p(r)\partial_{r}\right)\right]
    \mathcal{O}^{S}\right\rangle_{\metric}}
    {\left\langle\partial_{\omega}\mathcal{H}_{A}^{\ell m}\right\rangle_{\metric}}\,, 
    \label{eq:QNM_shift_RW} \\
    & \omega_{\metric,\ell m}^{+(1,0)}
    =\frac{\left\langle V^{P(1,0)}_{\rm eff}(r,\chi)\right\rangle_{\metric}}
    {\left\langle\partial_{\omega}\mathcal{H}_{P}^{\ell m}\right\rangle_{\metric}}\,, 
    \label{eq:QNM_shift_ZM}
\end{align}    
\end{subequations}
\end{widetext}
where we use the subscript $\metric$ to label the QNM frequency shifts obtained from the metric perturbation equations in Eq.~\eqref{eqs:perturbed_final_Wagle_expanded}. In Eq.~\eqref{eq:QNM_shift_metric}, we have used the shorthand notation in Eq.~\eqref{eq:inner_product_simplify} but with ${}_{s}R_{\ell m}^{(0,1)}$ replaced by $\Psi^{\RW(0,1)}_{\ell m}(r)$ and $\Psi^{\ZM(0,1)}_{\ell m}(r)$ for $\omega_{\metric,\ell m}^{-(1,0)}$ and $\omega_{\metric,\ell m}^{+(1,0)}$, respectively. The inner product in Eq.~\eqref{eq:inner_product_simplify} is now defined as
\begin{equation} \label{eq:inner_product_RW}
    \langle\varphi_{1}(r)|\varphi_{2}(r)\rangle_{\metric}
    =\int_{\mathscr{C}}f(r)
    \varphi_{1}(r)\varphi_{2}(r) dr\,,
\end{equation}
with $\mathscr{C}$ being the same contour around the positive imaginary axis at the horizon in Eq.~\eqref{eq:contour}. The derivatives of the operators $\mathcal{H}_{A}^{\ell m}$ and $\mathcal{H}_{P}^{\ell m}$ about $\omega$ are given by
\begin{widetext}
\begin{subequations}
\begin{align}
    & \partial_{\omega}\mathcal{H}_{A}^{\ell m}
    =2\omega_{\ell m}
    +\chi m\left[\frac{24M^2(3r^2-13Mr+14M^2)}{\ell(\ell+1)\omega^2r^7}
    -\frac{4M^2}{r^3}\right]\,, \\
    & \partial_{\omega}\mathcal{H}_{P}^{\ell m}
    =2\omega_{\ell m}-\partial_\omega V_{\rm eff}^{P(0,0)}(r,\chi)\,,
\end{align} 
\end{subequations}
where $\partial_\omega V_{\rm eff}^{P(0,0)}(r,\chi)$ is too complicated to present here, and $V_{\rm eff}^{P(0,0)}(r,\chi)$ can be found in \cite{Wagle:2021tam, Srivastava:2021imr}. The operator $\mathcal{O}^{S}$ represents the solution to Eq.~\eqref{eq:SF_expand} to express $\Theta_{\ell m}^{(1,1)}(r)$ in terms of $\Psi^{\RW(0,1)}_{\ell m}(r)$, i.e.,
\begin{equation} \label{eq:O_S}
    \mathcal{O}^{S}
    =M^2\kappa^{\frac{1}{2}}\left(\mathcal{H}_S^{\ell m}\right)^{-1}
    \left\{f(r)\left[g(r)+\chi Mm\left(h(r)+j(r)\partial_{r}\right)\right]\right\}\,,
\end{equation}
\end{widetext}
where $\left(\mathcal{H}_S^{\ell m}\right)^{-1}$ can be constructed using Green's function as discussed in Sec.~\ref{sec:QNM_nonrotating}.

To evaluate Eq.~\eqref{eq:QNM_shift_metric}, we need to first solve Eq.~\eqref{eq:SF_expand} for the scalar field perturbation. Equation~\eqref{eq:SF_expand} is written in a form so that when one replaces $r$ with $r_{*}$, the left-hand side of Eq.~\eqref{eq:SF_expand} is the same as the left-hand side of Eq.~\eqref{eq:scalar_IRG_Schw_r*}, plus additional $\mathcal{O}(\chi^1)$ terms in the potential. Thus, one can directly use Eq.~\eqref{eq:scalar_sol} to solve for $\Theta_{\ell m}^{(1,1)}(r)$ after replacing $\mathcal{S}_{\vartheta}^{\ell m}(r)$ with the right-hand side of Eq.~\eqref{eq:SF_expand}. Furthermore, to evaluate the right-hand side of Eq.~\eqref{eq:SF_expand} and the inner products in Eq.~\eqref{eq:QNM_shift_metric}, we need to compute $\Psi^{\RW(0,1)}_{\ell m}(r)$ and $\Psi^{\ZM(0,1)}_{\ell m}(r)$. For non-rotating BHs in GR, one can directly use Leaver's method in \cite{Leaver:1985ax} to calculate $\Psi^{\RW(0,1)}_{\ell m}(r)$ and then apply the Chandrasekhar transformation in \cite{Chandrasekhar_1983} to obtain $\Psi^{\ZM(0,1)}_{\ell m}(r)$. For slowly-rotating BHs in GR, since Leaver's method has not been extended to the slow-rotation RW/ZM equation yet, as far as the authors know, we use the shooting method discussed in Sec.~\ref{sec:QNM_slow_rotation} to calculate $\Psi^{\RW(0,1)}_{\ell m}(r)$ and $\Psi^{\ZM(0,1)}_{\ell m}(r)$. We have confirmed that the GR QNM spectra obtained via this approach agree well with the known results in \cite{BertiRingdown, Berti:2005ys, Berti:2009kk}. After calculating $\Theta_{\ell m}^{(1,1)}(r)$, $\Psi^{\RW(0,1)}_{\ell m}(r)$, and $\Psi^{\ZM(0,1)}_{\ell m}(r)$, we then have all the ingredients to evaluate Eq.~\eqref{eq:QNM_shift_metric}. In the next subsection, we will compare these results to the ones obtained through the modified Teukolsky formalism in Sec.~\ref{sec:QNM_freq}.

\subsection{Comparison of the dCS QNM spectra}
\label{sec:comparison_QNM}

In this subsection, we apply the procedures described in Sec.~\ref{sec:EVP_metric} to calculate the slow-rotation dCS QNM spectra up to $\mathcal{O}(\zeta^1,\chi^1,\epsilon^1)$ and compare them to the results in Sec.~\ref{sec:QNM_freq}. We carefully follow steps similar to those described in Secs.~\ref{sec:QNM_nonrotating} and \ref{sec:QNM_slow_rotation} and compare the shifts for the non-rotating background case and the slowly-rotating background case separately. To ensure efficient comparisons, we consider the relative difference $\delta(\textrm{Re}[\omega_{0,\ell m}^{\pm (1,i,0)}], \textrm{Re}[\omega_{\metric,\ell m}^{\pm (1,i,0)}]) $ and $\delta(\textrm{Im}[\omega_{0,\ell m}^{\pm (1,i,0)}], \textrm{Im}[\omega_{\metric,\ell m}^{\pm (1,i,0)}])$, $i\in\{0,1\}$, as defined in Eq.~\eqref{eq:relerror}, where $\omega_{0,\ell m}^{\pm (1,i,0)}$ and $\omega_{\metric,\ell m}^{\pm (1,i,0)}$ are the QNM frequencies obtained from Eq.~\eqref{eq:master_eqn_Psi0_radial_simplify} and Eq.~\eqref{eqs:perturbed_final_Wagle_expanded}, respectively. For convenience, we use the shorthand notation
\begin{align} \label{eq:relerrorredef}
    \Delta\left(\rm{Re}\left[\omega_{\ell m}^{\pm (1,i,0)}\right]\right) 
    & \coloneqq\delta\left(\textrm{Re}\left[\omega_{0,\ell m}^{\pm(1,i,0)}\right],
    \textrm{Re}\left[\omega_{\metric,\ell m}^{\pm(1,i,0)}\right]\right) \,, \nonumber \\
    \Delta\left(\rm{Im}\left[\omega_{\ell m}^{\pm(1,i,0)}\right]\right) 
    & \coloneqq\delta\left(\textrm{Im}\left[\omega_{0,\ell m}^{\pm(1,i,0)}\right],
    \textrm{Im}\left[\omega_{\metric,\ell m}^{\pm(1,i,0)}\right]\right) \,.
\end{align}
The $\pm$ in the superscript denotes even- and odd-parity modes, respectively, and the superscripts $(1,0,0)$ and $(1,1,0)$ denote the QNM shifts at non-rotating and slowly-rotating orders in dCS gravity, respectively.

\subsubsection{Even-parity modes}
\label{sec:comparison_even}

The results presented in~\cite{Srivastava:2021imr} were obtained using the EVP method, which enables a direct comparison with our results derived from integrating the modified Teukolsky equation. However, a similar comparison cannot be made with the results reported in~\cite{Wagle:2021tam}. This is because the authors of~\cite{Wagle:2021tam} employed the direct integration method combined with the Wronskian technique without explicitly expanding the QNM frequency $\omega_{\ell m}$ perturbatively, as described in Eqs.~\eqref{eq:QNM_slow_expansion_dCS}. This results in higher-order terms in $\zeta$ and $\chi$ that may contribute to the QNM frequencies.

To facilitate a consistent comparison, we integrate the metric perturbation equations presented in Sec.~\ref{sec:perturb_eqn_metric} using the EVP method outlined in Sec.~\ref{sec:EVP_metric}. For the $(\ell=2,n=0)$ mode, the relative difference [i.e., Eq.~\eqref{eq:relerror}] between $\omega_{0,\ell m}^{+(1,1,0)}$ and $\omega_{\metric,\ell m}^{+(1,1,0)}$ is found to be $\sim10^{-11}$ for the real part and $\sim 10^{-10}$ for the imaginary part. The maximum relative difference for all the higher multipoles and overtones we studied was found to be $\sim 10^{-6}$ for the real part and $\sim 10^{-7}$ for the imaginary part of the $(\ell=4,n=2)$ QNM frequency. Recall that the even-parity sector does not exhibit any deviation from GR at $\mathcal{O}(\chi^0)$, therefore $\omega_{\metric,\ell m}^{+(1,0,0)}=0$. Furthermore, our results completely agree with those reported in~\cite{Srivastava:2021imr}.

A detailed tabulation of the deviations in the even-parity QNM shifts induced by dCS gravity is provided in Table~\ref{tab:QNM_results_rotating_even}. Notably, this work not only corroborates existing QNM results but also extends the literature by enabling the calculation of overtone frequencies in the QNM spectrum using the metric perturbation equations.

\subsubsection{Odd-parity modes}
\label{sec:comparison_odd}

Unlike the case of even-parity perturbations, our QNM frequencies of odd-parity perturbations cannot be compared directly with any previous works. Reference~\cite{Srivastava:2021imr} does not calculate the QNM frequencies for the odd-parity perturbations, whereas \cite{Wagle:2021tam} does calculate the QNM frequencies for the fundamental modes using the direct integration method, but the frequencies are not perturbatively expanded. Therefore, we evaluate the QNM frequencies using the EVP method for the metric perturbation equations~\eqref{eqs:perturbed_final_Wagle_expanded}. The calculated QNM frequency shifts, along with those obtained from the modified Teukolsky equation, are presented in Table~\ref{tab:QNM_results_nonrotating} for $\omega_{0/\metric,\ell m}^{+(1,0,0)}$ and in Table~\ref{tab:QNM_results_rotating_odd} for $\omega_{0/\metric,\ell m}^{+(1,1,0)}$. For this comparison, we will restrict ourselves to the QNM shifts obtained using the MTF in the IRG. 

The relative difference between the QNM shifts obtained using the modified Teukolsky equation and the metric perturbation equation for a non-rotating BH in dCS gravity is shown in Table~\ref{table:error_odd_RW_MTF}. We find that the relative difference for the fundamental mode of a non-rotating BH is $\sim 10^{-6}$ for the real part and $\sim 10^{-7}$ for the imaginary part. As higher multipoles and overtones are considered, the relative difference increases. However, even for the largest discrepancy observed, specifically, for the mode $(\ell=4, n=2)$, the relative difference remains small, at $0.005\%$ for the real part and $0.015\%$ for the imaginary part, therefore indicating strong agreement between the MTF and the metric perturbation approach.

At $\mathcal{O}(\chi^1)$, the relative difference is $0.014\%$ and $0.004\%$ for the real and the imaginary part of the $(\ell=2,n=0)$ mode. The maximum deviation in the real part is found for the multipole $(\ell= 4,n=2)$, reaching $0.5\%$ whereas the maximum difference in the imaginary part is found for the multipole $(\ell=4, n=0)$ at $4.6\%$. A detailed comparison for different multipoles is provided in Table~\ref{table:error_odd_RW_MTF}. These results combined with those from Sec.~\ref{sec:comparison_even} confirm that the MTF approach is in strong agreement with the results obtained by numerically integrating the metric perturbation equations previously studied in \cite{Wagle:2021tam, Srivastava:2021imr}. 

\begin{table*}[t]
    \begin{center}
    \begin{tabular}{c @{\hskip 0.1in} c @{\hskip 0.1in} c @{\hskip 0.1in} c @{\hskip 0.1in} c @{\hskip 0.1in} c}
    \hline 
    \hline
    $\ell$ & Overtones &
    $\Delta\left(\rm{Re}\left[\omega_{\ell m}^{-(1,0,0)}\right]\right)$ & $\Delta\left(\rm{Im}\left[\omega_{\ell m}^{-(1,0,0)}\right]\right)$ & 
    $\Delta\left(\rm{Re}\left[\omega_{\ell 1}^{-(1,1,0)}\right]\right)$ & 
    $\Delta\left(\rm{Im}\left[\omega_{\ell 1}^{-(1,1,0)}\right]\right)$  \\
    \hline
    & $n=0$ & $ 10^{-4} $ & $ 10^{-5} $ & $0.014$ & $0.004$ \\[0.2cm]
    $\ell=2$ & $n=1$ & $ 10^{-5} $ & $10^{-4}$ & $0.013$ & $0.045$ \\ [0.2cm]
    & $n=2$ & $ 0.001 $ & $0.003$ & $0.002$ & $0.194$ \\ [0.2cm]
    \hline
    & $n=0$ & $ 10^{-6} $  & $10^{-4}$ & $0.021$ & $0.220$ \\ [0.2cm]
    $\ell=3$ & $n=1$ & $ 10^{-4} $ & $0.002$ & $0.078$ & $0.128$  \\ [0.2cm]
    & $n=2$ & $ 0.003 $ & $0.001$ & $0.273$ & $0.310$  \\[0.2cm]
    \hline
    & $n=0$ & $ 10^{-4} $ & $0.002$ & $0.021$ & $4.645$ \\ [0.2cm]
    $\ell=4$ & $n=1$ & $ 10^{-4} $ & $0.009$ & $0.027$ & $0.649$ \\ [0.2cm]
    & $n=2$ & $ 0.005 $ & $0.015$ & $0.500$ & $0.044$  \\
    \hline
    \hline
    \\[-0.2cm]
    \end{tabular}
    \end{center}
    \caption{The relative difference [i.e., Eqs.~\eqref{eq:relerror} and \eqref{eq:relerrorredef}] in $\%$ between the odd-parity QNM frequency shifts in dCS gravity obtained from the MTF in the IRG [i.e., Eq.~\eqref{eq:master_eqn_Psi0_radial_simplify}] and the metric perturbation approach [i.e., Eq.~\eqref{eqs:perturbed_final_Wagle_expanded}].}
    \label{table:error_odd_RW_MTF}
\end{table*}

\section{Discussion}
\label{sec:conclusions}

In this paper, we computed the QNM spectrum of a slowly-rotating BH in dCS gravity to leading order in spin, where the BH spacetime is non-Ricci-flat but it remains perturbatively of  Petrov-type-D, including the perturbations of the dCS scalar field. We used the modified Teukolsky equation obtained in \cite{Wagle:2023fwl}, which describes the perturbations of the Weyl scalars $\Psi_0$ and $\Psi_4$. To compute the QNM frequencies from the modified Teukolsky equation, we applied the EVP method of \cite{Zimmerman:2014aha, Mark:2014aja, Hussain:2022ins}.

The modified Teukolsky equation in \cite{Wagle:2023fwl} is a set of second-order ordinary differential equations for $\Psi_{0,4}$ with the scalar field perturbation coupled to them. This set of equations was derived in two independent gauges, the IRG and the ORG, with $\Psi_0$ and $\Psi_4$ characterizing the GW perturbations in each gauge, respectively. Though the dCS correction to GWs $\Psi_{0,4}^{(1,1)}$ are sourced by both the scalar field perturbation and the GR gravitational perturbation $\Psi_{0,4}^{(0,1)}$, the scalar field perturbation is only sourced by $\Psi_{0,4}^{(0,1)}$ to leading order in the beyond-GR coupling. Exploiting this hierarchical structure, we solve for the scalar field first and eliminate its dependence in the modified Teukolsky equation, such that these equations are only driven by the GR Teukolsky functions. 

Since theories like dCS gravity admit breaking of isospectrality, i.e., even- and odd-parity modes carry different QNM frequencies \cite{Molina:2010fb, Wagle:2021tam, Srivastava:2021imr, Li:2023ulk}, we further divide the modified Teukolsky equation into even- and odd-parity components. To extract the modified Teukolsky equation for modes of different parity, we employ the analysis presented in \cite{Li:2023ulk}. We show that the scalar field perturbation only couples to the odd-parity part of $\Psi_0^{(1,1)}$ and $\Psi_4^{(1,1)}$, while the even-parity part of them remains completely uncoupled to the scalar field perturbation. This result is in line with previous studies in \cite{Molina:2010fb, Wagle:2021tam, Srivastava:2021imr}.

Due to the structure of the modified Teukolsky equation, where the higher-order $\mathcal{O}(\zeta^1,\epsilon^1)$ terms are sourced by the lower-order $\mathcal{O}(\zeta^0,\epsilon^1)$ terms, traditional methods for solving for the QNM spectrum cannot be directly applied, as these may lead to the growth of secular terms. Since the EVP method developed in \cite{Zimmerman:2014aha, Mark:2014aja, Hussain:2022ins} is explicitly designed to circumvent this problem, we have used it to calculate the QNM frequency shifts for slowly-rotating BHs in dCS gravity. Our results in the IRG using the modified Teukolsky equation for the $\Psi_0$ perturbation are consistent with the results obtained in the ORG using the modified Teukolsky equation for the $\Psi_4$ perturbation, demonstrating the self-consistency of this approach. We also compared our results to the ones obtained in previous studies \cite{Molina:2010fb, Wagle:2021tam, Srivastava:2021imr} via RW/ZM equations and find excellent consistencies for both even- and odd-parity modes. In addition, our approach allows us to calculate the spectrum of the first few overtones in a non-minimally, coupled, beyond-GR theory of gravity for the first time.

Combining our results with those of \cite{Wagle:2023fwl}, our work serves as the first working example of the modified Teukolsky formalism \cite{Li:2022pcy, Hussain:2022ins} in a non-minimally coupled beyond-GR theory, paving the path for future studies of QNMs and ringdown in general in a broad class of modified gravity models. One crucial drawback of approaches relying on the slow-rotation approximation is the inaccuracies it generates at higher values of spin. For instance, we estimate that the slow-rotation approximation would introduce an error $\gtrsim 1\%$ on the QNM frequencies at spins $\gtrsim 0.22M$. Current LIGO-Virgo-KAGRA observing runs \cite{ LIGOScientific:2021djp} show that remnant BHs of binary mergers usually spin at or beyond $a= 0.6M$ on average. Therefore, an approach that does not rely on the slow-rotation approximation would allow for more accurate ringdown tests of GR. This will be the topic of future work. 

In this paper, we limited our calculations to leading order in spin under the slow-rotation expansion so that we could directly compare our results to the ones in \cite{Cardoso:2009pk, Molina:2010fb, Pani:2011xj, Wagle:2021tam, Srivastava:2021imr}, therefore validating the modified Teukolsky method. Nevertheless, this new approach, in principle, works for BHs with any spin in dCS gravity, including rapidly-rotating ones, as demonstrated in \cite{Cano:2023tmv, Cano:2023jbk} in a different, higher-derivative gravity theory. Since the source terms of the modified Teukolsky equation in terms of the NP quantities computed in \cite{Wagle:2023fwl} work for any spin, and the background metric of a rapidly-rotating BH in dCS gravity can be computed with the method in \cite{Cano:2019ore}, the only challenge is to implement the metric reconstruction procedures for the full Kerr case. Fortunately, the CCK procedures for metric reconstruction have been widely implemented previously in the full Kerr case and in different scenarios \cite{Ori:2002uv, Yunes:2005ve, Keidl:2006wk, Keidl:2010pm, Pound:2021qin, Dolan:2021ijg, Ma:2024qcv}. Thus, we do not expect metric reconstruction to be an obstacle that prevents us from obtaining QNMs of rapidly-rotating BHs in dCS gravity, as we will demonstrate in our follow-up work. 

\section{Acknowledgements} 
\label{sec:acknowledgements}
\appendix

We thank Manu Srivastava for sharing notebooks of their work~\cite{Srivastava:2021imr}. Y. C. and D. L. acknowledge support from the Brinson Foundation, the Simons Foundation through Award No. 568762, and National Science Foundation (NSF) Grants No. PHY-2011961 and No. PHY-2011968. P.W. acknowledges funding from the Deutsche Forschungsgemeinschaft (DFG) - project number: 386119226. N. Y. and D. L. acknowledge support from the Simons Foundation through Award No. 896696, NSF Grant No. PHY-2207650, and the National Aeronautics and Space Administration through award 80NSSC22K0806. Some of our algebraic work used the package {\sc xAct}~\cite{xact} for Mathematica.

\section{Properties of the angular projection coefficients}
\label{appendix:projection_coeff}

In this appendix, we show some relations for the angular projection coefficients $\{\Lambda^{\ell_1\ell_2m}_{s_1s_2}, \Lambda^{\ell_1\ell_2m}_{s_1s_2c}, \Lambda^{\ell_1\ell_2m}_{s_1s_2s}\}$ and $\{\Lambda^{\dagger\ell_1\ell_2m}_{s_1s_2},\Lambda^{\dagger\ell_1\ell_2m}_{s_1s_2c},\Lambda^{\dagger\ell_1\ell_2m}_{s_1s_2s}\}$ that are used in the main body of this paper to simplify the equations in Secs.~\ref{sec:scalar_eqn} and \ref{sec:modified_teuk_eqn}. 

Following \cite{Campbell:1970ww}, let us first show that
\begin{subequations} \label{eq:relation_12}
\begin{align} 
    & \Lambda^{\ell\ell m}_{s_1s_1c}
    =m\Lambda^{\ell\ell 1}_{s_1s_1c}\,,
    \label{eq:relation_1} \\
    & \Lambda^{\ell\ell m}_{s_1s_1\pm1s}
    =m\Lambda^{\ell\ell 1}_{s_1s_1\pm1s}\,.
    \label{eq:relation_2}
\end{align}    
\end{subequations}
In \cite{Wagle:2023fwl}, we defined
\begin{subequations} \label{eq:def_Lambda}
\begin{align}
    & \Lambda^{\ell_1\ell_2 m}_{s_1 s_2 c}
    \equiv\int_{S^2}dS\;\cos{\theta}
    \,{}_{s_1}Y_{\ell_1 m}
    \;{}_{s_2}\bar{Y}_{\ell_2 m}\,,
    \label{eq:def_Lambda_c} \\
    & \Lambda^{\ell_1\ell_2 m}_{s_1 s_2 s}
    \equiv\int_{S^2}dS\;\sin{\theta}
    \,{}_{s_1}Y_{\ell_1 m}
    \;{}_{s_2}\bar{Y}_{\ell_2 m}\,,
    \label{eq:def_Lambda_s}
\end{align}  
\end{subequations}
where $dS$ is the solid angle element, and the integration is over the entire 2-sphere. Using that
\begin{equation}
    {}_{0}Y_{10}(\theta,\phi)=\sqrt{\frac{3}{4\pi}}\cos{\theta}\,,\quad
    {}_{\pm 1}Y_{10}(\theta,\phi)=\pm\sqrt{\frac{3}{8\pi}}\sin{\theta}\,,
\end{equation}
we can write $\Lambda^{\ell \ell m}_{s_1s_1c}$ and $\Lambda^{\ell \ell m}_{s_1s_1\pm1s}$ as
\begin{subequations} \label{eq:Lambda_1}
\begin{align}
    & \Lambda^{\ell \ell m}_{s_1s_1c}
    =\sqrt{\frac{4\pi}{3}}\int_{S^2}dS\;{}_{s_1}\bar{Y}_{\ell m}\,
    {}_{s_1}Y_{\ell m}\,{}_{0}Y_{10}\,, 
    \label{eq:Lambda_c_1} \\
    & \Lambda^{\ell \ell m}_{s_1s_1\pm 1s}
    =\pm\sqrt{\frac{8\pi}{3}}\int_{S^2}dS\;{}_{s_1\pm 1}\bar{Y}_{\ell m}\,
    {}_{s_1}Y_{\ell m}\,{}_{\pm1}Y_{10}\,.
    \label{eq:Lambda_s_1}
\end{align}    
\end{subequations}
Using the relation between the spin-weighted spherical harmonics ${}_{s}Y_{\ell m}(\theta,\phi)$ and the Wigner rotation matrices $D_{sm}^{\ell}(\phi,\theta,\gamma)$ in \cite{Goldberg:1966uu},
\begin{equation} \label{eq:Wigner}
    {}_sY_{\ell m}(\theta,\phi)e^{-is\gamma}
    =\sqrt{\frac{2\ell+1}{4\pi}}D_{-sm}^{\ell}(\phi,\theta,\gamma)\,,
\end{equation}
we can reduce Eq.~\eqref{eq:Lambda_1} to
\begin{subequations} \label{eq:Lambda_2}
\begin{align}
    & \Lambda^{\ell \ell m}_{s_1s_1c}
    =\frac{2\ell+1}{8\pi^2}
    \int^{2\pi}_{0}d\gamma\int_{S^2}dS\;
    \bar{D}_{-s_1m}^{\ell}\,D_{-s_1m}^{\ell}
    \,D_{00}^{1}\,, \label{eq:Lambda_c_2} \\
    & \Lambda^{\ell \ell m}_{s_1s_1\pm1s}
    =\pm\frac{2\ell+1}{4\sqrt{2}\pi^2}
    \int^{2\pi}_{0}d\gamma\int_{S^2}dS\;
    \bar{D}_{-s_1\mp\,1m}^{\ell}\,D_{-s_1 m}^{\ell}
    \,D_{\mp10}^{1}\,, \label{eq:Lambda_s_2}
\end{align}    
\end{subequations}
where we have added an additional integral in $\gamma$ and a factor of $1/(2\pi)$. This trick uses the fact that the integrands in Eq.~\eqref{eq:Lambda_2} do not really depend on $\gamma$ since all the factors of $e^{-is\gamma}$ cancel out after using Eq.~\eqref{eq:Wigner}. Now, using that \cite{Campbell:1970ww}
\begin{widetext}
\begin{equation}
    \int^{2\pi}_{0}d\gamma\int_{S^2}dS\; 
    \bar{D}_{s_3m_3}^{\ell_3}D_{s_2m_2}^{\ell_2}D_{s_1m_1}^{\ell_1}
    =\frac{8\pi^2}{2\ell_3+1}\delta_{s_1+s_2,s_3}\delta_{m_1+m_2,m_3}
    C_{\ell_1 m_1 \ell_2 m_2}^{\ell_3 m_3}
    C_{\ell_1 s_1 \ell_2 s_2}^{\ell_3 s_3}\,,
\end{equation}    
\end{widetext}
where $C_{\ell_1 m_1 \ell_2 m_2}^{\ell_3 m_3}$
are the Clebsch-Gordan coefficients, we find that
\begin{subequations} \label{eq:Lambda_3}
\begin{align}
    & \Lambda^{\ell \ell m}_{s_1s_1c}
    =C_{1 0 \ell m}^{\ell m}
    C_{1 0 \ell(-s_1)}^{\ell(-s_1)}\,, 
    \label{eq:Lambda_c_3} \\
    & \Lambda^{\ell \ell m}_{s_1s_1\pm1s}
    =\pm\sqrt{2}C_{1 0 \ell m}^{\ell m}
    C_{1(\mp1)\ell(-s_1)}^{\ell(-s_1\mp1)}\,, 
    \label{eq:Lambda_s_3}
\end{align}    
\end{subequations}
Using the relation in \cite{NIST:DLMF} that
\begin{equation}
    C_{1 0 \ell m}^{\ell m}
    =m C_{1 0 \ell 1}^{\ell 1}\,,
\end{equation}
we obtain Eq.~\eqref{eq:relation_12}. 

Next, let us show that
\begin{subequations} \label{eq:relation_34}
\begin{align} 
    & \Lambda^{\ell\ell m}_{-s_1-s_1c}
    =-\Lambda^{\ell\ell m}_{s_1s_1c}\,,
    \label{eq:relation_3} \\
    & \Lambda^{\ell\ell m}_{-s_1-s_1\mp 1s}
    =\Lambda^{\ell\ell m}_{s_1s_1\pm 1s}\,.
    \label{eq:relation_4}
\end{align}    
\end{subequations}
First, using Eqs.~\eqref{eq:angular_transformation} and \eqref{eq:def_Lambda}, we find that
\begin{subequations} \label{eq:Lambda_minus_1}
\begin{align}
    & \Lambda^{\ell_1\ell_2 m}_{-s_1 -s_2 c}
    =(-1)^{s_1+s_2}\int_{S^2}dS\;\cos{\theta}
    \,{}_{s_1}\bar{Y}_{\ell_1 -m}
    \;{}_{s_2}Y_{\ell_2 -m}\,, 
    \label{eq:Lambda_c_minus_1} \\
    & \Lambda^{\ell_1\ell_2 m}_{-s_1 -s_2 s}
    =(-1)^{s_1+s_2}\int_{S^2}dS\;\sin{\theta}
    \,{}_{s_1}\bar{Y}_{\ell_1 -m}
    \;{}_{s_2}Y_{\ell_2 -m}\,.
    \label{eq:Lambda_s_minus_1}
\end{align}    
\end{subequations}
Since the integrands in Eq.~\eqref{eq:Lambda_minus_1} are real, we can move the complex conjugate of ${}_{s_1}\bar{Y}_{\ell_1 -m}$ to ${}_{s_2}Y_{\ell_2 -m}$. Thus, we obtain
\begin{subequations} \label{eq:Lambda_minus_2}
\begin{align}
    & \Lambda^{\ell_1\ell_2 m}_{-s_1 -s_2 c}
    =(-1)^{s_1+s_2}\Lambda^{\ell_1\ell_2 -m}_{s_1 s_2 c}\,,
    \label{eq:Lambda_c_minus_2}\\
    & \Lambda^{\ell_1\ell_2 m}_{-s_1 -s_2 s}
    =(-1)^{s_1+s_2}\Lambda^{\ell_1\ell_2 -m}_{s_1 s_2 s}\,.
    \label{eq:Lambda_s_minus_2}
\end{align}    
\end{subequations}
Using Eq.~\eqref{eq:relation_12} with Eq.~\eqref{eq:Lambda_minus_2}, we obtain Eq.~\eqref{eq:relation_34}.

Finally, let us show that
\begin{subequations} \label{eq:relation_56}
\begin{align} 
    & \Lambda^{\dagger\ell\ell m}_{-s_1s_2c}
    =(-1)^{m+s_1}\Lambda^{\ell\ell m}_{s_1s_2c}\,,
    \label{eq:relation_5} \\
    & \Lambda^{\dagger\ell\ell m}_{-s_1s_2s}
    =(-1)^{m+s_1}\Lambda^{\ell\ell m}_{s_1s_2s}\,.
    \label{eq:relation_6}
\end{align}    
\end{subequations}
In \cite{Wagle:2023fwl}, we defined
\begin{subequations} \label{eq:def_Lambda_dagger}
\begin{align}
    & \Lambda^{\dagger\ell_1\ell_2 m}_{s_1 s_2 c}
    \equiv\int_{S^2}dS\;\cos{\theta}
    \,{}_{s_1}\bar{Y}_{\ell_1 -m}
    \;{}_{s_2}\bar{Y}_{\ell_2 m}\,,
    \label{eq:def_Lambda_c_dagger} \\
    & \Lambda^{\dagger\ell_1\ell_2 m}_{s_1 s_2 s}
    \equiv\int_{S^2}dS\;\sin{\theta}
    \,{}_{s_1}\bar{Y}_{\ell_1 -m}
    \;{}_{s_2}\bar{Y}_{\ell_2 m}\,.
    \label{eq:def_Lambda_s_dagger}
\end{align}   
\end{subequations}
Using Eq.~\eqref{eq:angular_transformation}, we find
\begin{subequations} \label{eq:Lambda_dagger_1}
\begin{align}
    & \Lambda^{\dagger\ell_1\ell_2 m}_{-s_1 s_2 c}
    =(-1)^{m+s_1}\int_{S^2}dS\;\cos{\theta}
    \,{}_{s_1}Y_{\ell_1 m}
    \;{}_{s_2}\bar{Y}_{\ell_2 m}\,, 
    \label{eq:Lambda_c_dagger_1} \\
    & \Lambda^{\dagger\ell_1\ell_2 m}_{-s_1 s_2 s}
    =(-1)^{m+s_1}\int_{S^2}dS\;\sin{\theta}
    \,{}_{s_1}Y_{\ell_1 m}
    \;{}_{s_2}\bar{Y}_{\ell_2 m}\,,
    \label{eq:Lambda_s_dagger_1}
\end{align}    
\end{subequations}
which gives us Eq.~\eqref{eq:relation_56} when comparing Eq.~\eqref{eq:Lambda_dagger_1} to Eq.~\eqref{eq:def_Lambda}.

\section{List of radial operators}
\label{appendix:radial_operators}

In this appendix, we provide the explicit expressions of the radial operators 
$\{\hat{A}_i^{\ell m},\hat{B}_i^{\ell m},\hat{\tilde{B}}_i^{\ell m}\}$ and $\{\hat{\mathscr{A}}_i^{\ell m},\hat{\mathscr{B}}_i^{\ell m},\hat{\tilde{\mathscr{B}}}_i^{\ell m}\}$ in terms of the auxiliary radial functions in \cite{Wagle:2023fwl}.

First, the radial operators $\{\hat{A}_i^{\ell m},\hat{B}_i^{\ell m},\hat{\tilde{B}}_i^{\ell m}\}$ used in the master equation for $\Psi_0^{(1,1)}$ in the IRG [Eqs.~\eqref{eq:master_eqn_Psi0_radial_simplify} and \eqref{eq:source_simplified_IRG}] are defined via
\begin{subequations} \label{eq:A_i_list}
\begin{align}
    \hat{A}_1^{\ell m}
    =& \;i\left(k_{1}^{\ell m}(r)+k_{2}^{\ell m}(r)\partial_r\right)
    \mathcal{H}_{\vartheta}^{-1}\mathcal{V}^{\ell m} \nonumber\\
    & \;+\frac{1}{1-\bar{\eta}_{\ell m}}
    \left(k_{3}^{\ell m}(r)+k_{4}^{\ell m}(r)\partial_r\right)\,, \\
    \hat{A}_2^{\ell m}
    =& \;\left(k_{5}^{\ell m}(r)+k_{6}^{\ell m}(r)\partial_r\right)
    \mathcal{H}_{\vartheta}^{-1}\mathcal{V}^{\ell m} \nonumber\\
    & \;-\frac{i}{1-\bar{\eta}_{\ell m}}
    \left(k_{7}^{\ell m}(r)+k_{8}^{\ell m}(r)\partial_r\right)\,,  \\
    \hat{A}_3^{\ell m}
    =& \;\left(k_{9}^{\ell m}(r)+k_{10}^{\ell m}(r)\partial_r\right)
    \mathcal{H}_{\vartheta}^{-1}\mathcal{V}^{\ell m} \nonumber\\
    & \;-\frac{i}{1-\bar{\eta}_{\ell m}}
    \left(k_{11}^{\ell m}(r)+k_{12}^{\ell m}(r)\partial_r\right)\,, \\
    \hat{B}_1^{\ell m}
    =& \;-i\left(q_{1}^{\ell m}(r)
    +q_{2}^{\ell m}(r)\partial_r\right)\,, \\
    \hat{B}_2^{\ell m}
    =& \;-i\left(q_{3}^{\ell m}(r)
    +q_{4}^{\ell m}(r)\partial_r\right)\,, \\
    \hat{B}_3^{\ell m}
    =& \;-i\left(q_{5}^{\ell m}(r)
    +q_{6}^{\ell m}(r)\partial_r\right)\,,\\
    \hat{\tilde{B}}_1^{\ell m}
    =& \;-i\left(\tilde{q}_{1}^{\ell m}(r)
    +\tilde{q}_{2}^{\ell m}(r)\partial_r\right)\,, \\
    \hat{\tilde{B}}_2^{\ell m}
    =& \;-i\tilde{q}_{3}^{\ell m}(r)\,,
\end{align}
\end{subequations}
where $\mathcal{H}_{\vartheta}^{-1}$ is the Green's function corresponding to the left-hand side of Eq.~\eqref{eq:scalar_radial_IRG_2_lm}, which can be retrieved from Eq.~\eqref{eq:scalar_sol}, and the operator $\mathcal{V}^{\ell m}$ is
\begin{align}
    \mathcal{V}^{\ell m}
    =& \;g_1^{\ell m}(r)+g_2^{\ell m}(r)\partial_r
    -i\chi m\Lambda^{\ell\ell 1}_{10s}
    \left[g_3^{\ell m}(r)+h_1^{\ell m}(r)\right. \nonumber \\
    & \;\left.+\left(g_4^{\ell m}(r)+h_2^{\ell m}(r)\right)
    \partial_r\right]\,.
\end{align}
Note that there is an extra factor of $1/(1-\bar{\eta}_{\ell m})$ in several terms of Eq.~\eqref{eq:A_i_list} since we have extracted a factor of $1-\bar{\eta}_{\ell m}$ in Eq.~\eqref{eq:A_i_m_simplified} when defining $\hat{A}_i^{\ell m}$. The definition of all the radial functions $k_i^{\ell m}(r)=k_i^{\ell m}(r,\omega,M)$, $q_i^{\ell m}(r)=q_i^{\ell m}(r,\omega,M)$, and $\tilde{q}_i^{\ell m}(r)=\tilde{q}_i^{\ell m}(r,\omega,M)$ can be found in \cite{Wagle:2023fwl} and the supplementary Mathematica notebook \cite{Pratikmodteuk}. The radial functions $g_i^{\ell m}(r)$ and $h_i^{\ell m}(r)$ follow the redefinition in Eq.~\eqref{eq:redefine_g_h}.

Next, the radial operators $\{\hat{\mathscr{A}}_i^{\ell m},\hat{\mathscr{B}}_i^{\ell m},\hat{\tilde{\mathscr{B}}}_i^{\ell m}\}$ used in the master equation for $\Psi_4^{(1,1)}$ in the ORG
[Eqs.~\eqref{eq:master_eqn_Psi4_radial_simplify} and \eqref{eq:source_simplified_ORG}] are 
\begin{subequations}
\begin{align}
    \hat{\mathscr{A}}_1^{\ell m}
    =& \;i\left(\textgoth{k}_{1}^{\ell m}(r)
    +\textgoth{k}_{2}^{\ell m}(r)\partial_r\right)
    \mathcal{H}_{\vartheta}^{-1}\mathcal{U}^{\ell m} \nonumber\\
    & \;+\frac{1}{1-\bar{\eta}_{\ell m}}
    \left(\textgoth{k}_{3}^{\ell m}(r)
    +\textgoth{k}_{4}^{\ell m}(r)\partial_r\right)\,, \\
    \hat{\mathscr{A}}_2^{\ell m}
    =& \;\left(\textgoth{k}_{5}^{\ell m}(r)
    +\textgoth{k}_{6}^{\ell m}(r)\partial_r\right)
    \mathcal{H}_{\vartheta}^{-1}\mathcal{U}^{\ell m} \nonumber\\
    & \;-\frac{i}{1-\bar{\eta}_{\ell m}}
    \left(\textgoth{k}_{7}^{\ell m}(r)
    +\textgoth{k}_{8}^{\ell m}(r)\partial_r\right)\,,  \\
    \hat{\mathscr{A}}_3^{\ell m}
    =& \;\left(\textgoth{k}_{9}^{\ell m}(r)
    +\textgoth{k}_{10}^{\ell m}(r)\partial_r\right)
    \mathcal{H}_{\vartheta}^{-1}\mathcal{U}^{\ell m} \nonumber\\
    & \;-\frac{i}{1-\bar{\eta}_{\ell m}}
    \left(\textgoth{k}_{11}^{\ell m}(r)
    +\textgoth{k}_{12}^{\ell m}(r)\partial_r\right)\,, \\
    \hat{\mathscr{B}}_1^{\ell m}
    =& \;-i\left(\textgoth{q}_{1}^{\ell m}(r)
    +\textgoth{q}_{2}^{\ell m}(r)\partial_r\right)\,, \\
    \hat{\mathscr{B}}_2^{\ell m}
    =& \;-i\left(\textgoth{q}_{3}^{\ell m}(r)
    +\textgoth{q}_{4}^{\ell m}(r)\partial_r\right)\,, \\
    \hat{\mathscr{B}}_3^{\ell m}
    =& \;-i\left(\textgoth{q}_{5}^{\ell m}(r)
    +\textgoth{q}_{6}^{\ell m}(r)\partial_r\right)\,, \\
    \hat{\tilde{\mathscr{B}}}_1^{\ell m}
    =& \;-i\left(\tilde{\textgoth{q}}_{1}^{\ell m}(r)
    +\tilde{\textgoth{q}}_{2}^{\ell m}(r)\partial_r\right)\,, \\
    \hat{\tilde{\mathscr{B}}}_2^{\ell m}
    =& \;-i\tilde{\textgoth{q}}_{3}^{\ell m}(r)\,,
\end{align}
\end{subequations}
where
\begin{align}
    \mathcal{U}^{\ell m}
    =& \;\textgoth{g}_1^{\ell m}(r)
    +\textgoth{g}_2^{\ell m}(r)\partial_r
    -i\chi m\Lambda^{\ell\ell 1}_{10s}
    \left[\textgoth{g}_3^{\ell m}(r)
    +\textgoth{h}_1^{\ell m}(r)\right. \nonumber \\
    & \;\left.+\left(\textgoth{g}_4^{\ell m}(r)
    +\textgoth{h}_2^{\ell m}(r)\right)
    \partial_r\right]\,.
\end{align}
The definition of all the radial functions $\textgoth{k}_i^{\ell m}(r)=\textgoth{k}_i^{\ell m}(r,\omega,M)$, $\textgoth{q}_i^{\ell m}(r)=\textgoth{q}_i^{\ell m}(r,\omega,M)$, and $\tilde{\textgoth{q}}_i^{\ell m}(r)=\tilde{\textgoth{q}}_i^{\ell m}(r,\omega,M)$ can be found in \cite{Wagle:2023fwl} and the supplementary Mathematica notebook \cite{Pratikmodteuk}. The radial functions $\textgoth{g}_i^{\ell m}(r)$ and $\textgoth{h}_i^{\ell m}(r)$ follow the redefinition in Eq.~\eqref{eq:redefine_g_h}.

\section{Equivalence between master equations obtained using metric perturbation approach in \cite{Wagle:2021tam} and \cite{Srivastava:2021imr}}
\label{appendix:equivalence_between_eqs}

In this appendix, we show that the master equations obtained in~\cite{Wagle:2021tam} and \cite{Srivastava:2021imr} are equivalent. The master equations describing the gravitational and scalar perturbations for a slowly-rotating BH in dCS gravity were found in~\cite{Wagle:2021tam} as shown in Eq.~\eqref{eqs:perturbed_final_Wagle}. For the same system, another set of differential equations were obtained in~\cite{Srivastava:2021imr}, and they are
\begin{widetext}
\begin{subequations}
\label{eqs:perturbed_final_Manu-2}
\begin{align}
    & \left[f(r)^2\partial_{r}^2+\left(\frac{2M}{r^2}f(r)
    +\zeta\chi m\kappa T_1(r)\right)\partial_r
    +\omega^2-V^\textrm{Smod}_{\rm eff}(r,\chi,\zeta)\right]
    \Theta_{\ell m}(r)\nonumber \\ 
    & =\zeta^{\frac{1}{2}}M^2\kappa^{\frac{1}{2}}
    f(r)\left[g(r)+\chi Mm\left(h(r)+j(r)\partial_r\right)\right]
    \Psi^{\RWmod}_{\ell m}(r)\,, \label{eq:SF2}\\
    & \left[f(r)^2\partial_{r}^2+\left(\frac{2M}{r^2}f(r)
    +\zeta\chi m\kappa A_1(r)\right)\partial_r
    +\omega^2-V^\textrm{Amod}_{\rm eff}(r,\chi,\zeta)\right]
    \Psi^{\RWmod}_{\ell m}(r)\nonumber \\ 
    & =\zeta^{\frac{1}{2}}M^2\kappa^{\frac{1}{2}}f(r) 
    \left[v(r)+\chi Mm\left(n(r)+p(r)\partial_{r}\right)\right]
    \Theta_{\ell m}(r)\,, \label{eq:RW2}\\
    & \left[f(r)^2\partial_{r}^2+\left(\frac{2M}{r^2}f(r)
    +\zeta\chi m\kappa P_1(r)\right)\partial_r  
    +\omega^2-V^\textrm{Pmod}_{\rm eff}(r,\chi,\zeta)\right]
    \Psi^{\ZMmod}_{\ell m}(r)=0 \,,\label{eq:ZM2}
\end{align}
\end{subequations}
\end{widetext}
where evidently these equations are different from those presented in Eq.~\eqref{eqs:perturbed_final_Wagle} as the potential terms are different in Eqs.~\eqref{eqs:perturbed_final_Wagle} and \eqref{eqs:perturbed_final_Manu-2}. Moreover, Eq.~\eqref{eqs:perturbed_final_Manu-2} contains extra terms (i.e., $T_1(r)$, $A_1(r)$, and $P_1(r)$) in the prefactor of the first derivative of the perturbation function. These differences lie in the definition of the master functions in terms of the metric perturbation. Though the scalar field perturbation is defined identically in \cite{Wagle:2021tam} and \cite{Srivastava:2021imr}, the gravitational master functions are not. However, we can define a transformation between the master functions, such that
\begin{align} \label{eq:Manu_to_Wagle}
    \Psi^{\RWmod}_{\ell m} 
    &=\Psi^{\RW}_{\ell m}\left(1-\delta\Psi^{\RW}_{\ell m}\right)\,, \nonumber \\
    \Psi^{\ZMmod}_{\ell m} 
    &=\Psi^{\ZM}_{\ell m}\left(1-\delta\Psi^{\ZM}_{\ell m}\right)\,,
\end{align}
where
\begin{widetext}
\begin{subequations}
\begin{align}
    \delta\Psi^{\RW}_{\ell m} 
    =& \;-\frac{1}{14\ell(\ell+1)Mr^9\omega}
    \left[\pi m\left(63\left(83\ell^2+83\ell-640\right)M^3r
    +32M^2r^2\left(-5\ell^2-5\ell+63r^2\omega^2-35\right)
    \right.\right. \nonumber\\ 
    & \;\left.\left.+70Mr^3\left(-3\ell^2-3\ell+12r^2\omega^2-4\right)
    +140r^4\left(-3\ell^2-3\ell+2r^2\omega^2+12\right)
    +60480 M^4\right)\right]\,, \\
    \delta\Psi^{\ZM}_{\ell m} 
    =& \;\frac{1}{14\ell(\ell+1)Mr^9\omega
    \left(\left(\ell^2+\ell-2\right)r+6M\right)^2}
    \left[\pi m\left(-144\left(2730\ell^2+2730\ell-11603\right)M^5r
    \right.\right. \nonumber\\ 
    & \;\left.\left.+70\left(\ell^2+\ell-2\right)Mr^5
    \left(3\ell^2\left(4r^2\omega^2+3\right)
    +3\ell\left(4r^2\omega^2+3\right)
    +4\left(5r^2\omega^2-42\right)\right)\right.\right. \nonumber\\ 
    & \;\left.\left.+280\left(\ell^2+\ell-2\right)^2r^6
    \left(r^2\omega^2-3\right)
    +6M^4r^2\left(-5439\ell^4-10878\ell^3+55307\ell^2
    +60746\ell+11340r^2\omega^2-96536\right)
    \right.\right. \nonumber\\ 
    & \;\left.\left.+3M^3r^3\left(6479\ell^4+12958\ell^3+\ell^2
    \left(7812r^2\omega^2-17747\right)
    +\ell\left(7812r^2\omega^2-24226\right)
    -6504r^2\omega^2+37096\right)\right.\right. \nonumber\\ 
    & \;\left.\left.+12M^2r^4\left(3\ell^4\left(56r^2\omega^2+15\right)
    +6\ell^3\left(56r^2\omega^2+15\right)
    +\ell^2\left(296r^2\omega^2+625\right)
    +4\ell\left(32r^2\omega^2+145\right)\right.\right.\right. \nonumber\\ 
    & \;\left.\left.\left.-4\left(57r^2\omega^2+1280\right)\right)
    -1469664M^6\right)\right]\,.
\end{align}
\end{subequations}
\end{widetext}
Applying the transformation in Eq.~\eqref{eq:Manu_to_Wagle} to Eq.~\eqref{eqs:perturbed_final_Manu-2}, we obtain the gravitational perturbed equations~\eqref{eqs:perturbed_final_Wagle} exactly. To show equivalence of the scalar field perturbation equation, we also need to use the scalar field equation and the RW equation in GR~\cite{Regge:1957td, Pani:2013pma} to eliminate all higher order terms. Through this analysis, we conclude that the perturbation equations obtained in \cite{Wagle:2021tam} and \cite{Srivastava:2021imr} are equivalent.

\section{GR values for $\omega_{\ell m}^{(0,0,0)}$ and $\omega_{\ell m}^{(0,1,0)}$}
\label{appendix:GR_values}

In this appendix, we present the values of $\omega_{\ell m}^{(0,0,0)}$ and $\omega_{\ell 1}^{(0,1,0)}$ at $\ell=2,3,4$ and $n=0,1,2$ in Table~\ref{tab:QNM_results_GR} for completeness. The values of $\omega_{\ell m}^{(0,0,0)}$ are obtained via Leaver's method \cite{Leaver:1985ax}, while the values of $\omega_{\ell 1}^{(0,1,0)}$ are obtained via the EVP method prescribed in Sec~\ref{sec:EVP}.

\begin{table*}[]
    \centering
    \begin{tabular}{cccc}
        \hline\hline
        $\ell$ & Overtones & $\omega_{\ell m}^{(0,0,0)}$ 
        & $\omega_{\ell 1}^{(0,1,0)}$ \\
        \hline
        & $n=0$ &  $0.7473434-0.1779246i$ & $0.1257661 +  0.0019958i$ \\
        $\ell=2$ & $n=1$ & $0.6934220-0.5478298i$ & $0.1438801 + 0.0127691i$ \\
        & $n=2$ & $0.6021069-0.9565540i$ & $0.1720698 + 0.0448677i$ \\
        \hline
        & $n=0$ & $1.1988866-0.1854061i$ & $0.1347096 + 0.0013041i $ \\
        $\ell=3$ & $n=1$ & $1.1652876-0.5625962i$ & $0.1428287 + 0.0057983i$ \\
        & $n=2$ & $1.1033698-0.9581855i$ & $0.1575689 + 0.0161148i$ \\
        \hline
        & $n=0$ & $1.6183568-0.1883279i$ & $0.1396268 + 0.0006829i $ \\
        $\ell=4$ & $n=1$ & $1.5932631-0.5686687i$ & $0.1441972 + 0.0028554i$ \\
        & $n=2$ & $1.5454191-0.9598164i$ & $0.1528679 + 0.0075248i$ \\
        \hline\hline
    \end{tabular}
    \caption{The values of $\omega_{\ell m}^{(0,0,0)}$ and $\omega_{\ell 1}^{(0,1,0)}$ at $\ell=2,3,4$ and $n=0,1,2$ in GR. We have set $M=1/2$ in this table.}
    \label{tab:QNM_results_GR}
\end{table*}

\bibliographystyle{apsrev4-1}
\bibliography{reference_new}
    
\end{document}